%% file: Article_KS.tex
\documentclass{jfm}
\usepackage{graphicx}
\usepackage{epstopdf,epsfig}
\usepackage{amsfonts,amssymb,amsmath,mathbbol} 
\usepackage{caption}
\usepackage{subcaption}

\usepackage{xcolor} %new package
\usepackage{color} %new package
\usepackage{float}
\usepackage{natbib}
\usepackage{rotating}
\usepackage{overpic}
\usepackage{lineno}
\usepackage{multirow}
\usepackage{array}

\newcommand{\RomanNumeralCaps}[1]

\shorttitle{ film boiling} %for header on odd pages
\shortauthor{Haotao et al.} %for header on even pages

\newcommand{\rb}{\textcolor{black}} % for commenting
\newcommand{\ra}{\textcolor{black}} 
\newcommand{\add}{\textcolor{black}} 

\title{Three-dimensional simulation of film boiling on a horizontal surface with magnetic field}

\author{
Hao-Tao Gu\aff{1},
Kirti Chandra Sahu\aff{2},
Jie Zhang\aff{1}
\corresp{\email{j\_zhang@xjtu.edu.cn}}, 
\and Ming-Jiu Ni\aff{1,3}
}

\affiliation
{
\aff{1}
State Key Laboratory for Strength and Vibration of Mechanical Structures, School of Aerospace, Xi’an Jiaotong University, Xi’an, Shaanxi 710049, PR China
\aff{2}
Department of Chemical Engineering, Indian Institute of Technology Hyderabad, Sangareddy 502 284, Telangana, India
\aff{3}
 School of Engineering Science, University of Chinese Academy of Sciences, Beijing 101408, PR China
}

\begin{document}
%\linenumbers
\maketitle
	
\begin{abstract}
This study conducts a numerical investigation into the three-dimensional film boiling of liquid under the influence of external magnetic fields. The numerical method incorporates a sharp phase-change model based on the volume-of-fluid approach to track the liquid-vapor interface. Additionally, a consistent and conservative scheme is employed to calculate the induced current densities and electromagnetic forces. We investigate the magnetohydrodynamic effects on film boiling, particularly examining the pattern transition of the vapor bubble and the evolution of heat transfer characteristics, exposed to either a vertical or horizontal magnetic field. In single-mode scenarios, film boiling under a vertical magnetic field displays an isotropic flow structure, forming a columnar vapor jet at higher magnetic field intensities. In contrast, horizontal magnetic fields result in anisotropic flow, creating a two-dimensional vapor sheet as the magnetic strength increases. In multi-mode scenarios, the patterns observed in single-mode film boiling persist, with the interaction of vapor bubbles introducing additional complexity to the magnetohydrodynamic flow. More importantly, our comprehensive analysis reveals how and why distinct boiling effects are generated by various orientations of magnetic fields, which induce directional electromagnetic forces to suppress flow vortices within the cross-sectional plane.
\end{abstract}

\section{Introduction}
Boiling is widely employed in industrial heat-related processes due to its exceptional heat transfer efficiency \citep{dhir1,Tomar1,guion2018simulations,lavino2021surface}. Depending on the overheat level, boiling can be categorized into three stages: nucleate boiling, transition boiling, and film boiling. The phenomenon of film boiling is especially common under conditions of significant overheat and is a crucial process in industrial applications. %applications such as fusion nuclear reactors \citep{dhir1}. 
In the context of magnetic confinement fusion, a method used for controllable nuclear fusion, liquid metal lithium can serve as a heat remover within a liquid blanket, facilitating the energy conversion process. Film boiling of this liquid metal occurs under conditions of extremely high heat flux and overheat generated by nuclear fusion. The presence of a magnetic field (MF) induces magnetohydrodynamics (MHD) effects that influence the entire process. Considering this perspective, investigating the film boiling of liquid metal under the influence of a magnetic field emerges as a noteworthy research topic.

The initial studies of film boiling on a horizontal surface were predominantly based on theoretical and experimental investigations. Pioneering work by \citet{Chang1,Chang2} involved analyzing the mechanism of film boiling on a horizontal surface using the theory proposed by \citet{Taylor1} for the most unstable wavelength. \citet{Zuber1} extended this theory to derive a model to predict the minimal heat flux during film boiling and demonstrated that Taylor instability predominantly governed the interface with the wavelength satisfying $\lambda_c \leq \lambda \leq \lambda_d$, where $\lambda_c(=2\pi\sqrt{\sigma/(\rho_l-\rho_v)g})$ represents the most critical wavelength, and $\lambda_d (=\sqrt{3}\lambda_c)$ denotes the most dangerous wavelength. Subsequently,  \citet{Klimenko1} and \citet{Klimenko2} introduced a new correlation with an accuracy of approximately $\pm 25$\%. This theoretical model has been widely acknowledged as a foundational basis for subsequent research on film boiling.

The emergence of powerful supercomputers, advancements in numerical simulation techniques, and the development of precise interface tracking methods have steered contemporary research on film boiling over a horizontal surface towards direct numerical simulations. However, many numerical simulations of film boiling have still focused on two-dimensional (2D) simulations and single-mode models, with the computational domain width close to the most dangerous wavelength $\lambda_d$. These 2D single-mode film boiling numerical simulations were mainly used to investigate the influence of different physical quantities (e.g., Jakob number, Morton number, density ratio) on morphology and heat transfer in film boiling. For high Jakob numbers, the rapid formation of vapor makes bubble detachment more difficult, resulting in the formation of tall and stable vapor columns \citep{Son1,Juric1,guo1}. For low Morton numbers or high-density ratios, it becomes more challenging for bubbles to pinch off \citep{Juric1,akhtar1,mortazavi1}. Some studies \citep{Agarwal1,Tomar1, guo1} suggested the periodic nature of bubble growth at nodes and anti-nodes in accordance with wavelength estimated based on Taylor's theory. Some other researchers have performed multi-mode film boiling simulations, considering factors like the interaction between bubbles and the initial interface. A few studies \citep{esmaeeli2,hens1,ningegowda1} have explored the impact of the Jakob number on multi-mode film boiling, revealing that, at high Jakob numbers, boiling occurs in the form of vapor columns. The proximity between vapor columns induces mutual influence and potential merging due to the instability of these columns, ultimately resulting in a chaotic process. Several researchers have highlighted the shortcomings in 2D film boiling simulations \citep{gibou1, akhtar1, zhang1}. These studies observed that bubbles did not detach from the vapor film in their 2D simulations. This issue was attributed to the curvature of the stem connecting the thin film and the bubble approaching zero, thus neglecting the surface tension effect.

Earlier studies on film boiling with large 2D domains have consistently confirmed that the distribution of bubbles corresponds to the most dangerous wavelength in film boiling on a horizontal surface \citep{esmaeeli2, Tomar1, hens1, akhtar2, ningegowda1}. The three-dimensional (3D) numerical simulations of film boiling employing simplified liquid/vapor parameter models primarily focus on single-mode scenarios \citep{Esmaeeli1, tsui1, zhang1, kumar1}. Moreover, limited research has been devoted to exploring the underlying physical mechanisms, with much of the existing work replicating earlier 2D studies. These investigations often revisit aspects like the impact of overheat, Grashof number, and density ratio \citep{khorram1}. Furthermore, the scope of single-mode film boiling simulations is constrained by computational domain limitations, preventing the complete visualization of bubble generation throughout the boiling process. Notably, three-dimensional (3D) multi-modal film boiling simulations in existing literature have typically employed computational domains of sizes not exceeding 2$\lambda_d$ \citep{esmaeeli2, khorram1, shin1}, thus inadequately capturing the actual distribution of bubbles.

From the above discussion, it is evident that research on conventional film boiling has indeed reached a profound level of depth. However, there are still areas where further exploration is warranted, particularly in understanding the underlying physical mechanisms, expanding beyond single-mode simulations, and overcoming computational limitations to capture the full dynamics of multi-modal film boiling. Film boiling coupled with multiphysics, particularly electrohydrodynamics (EHD), has recently emerged as the forefront of film boiling investigation. Extensive experimental studies \citep{Ogata1,Wang1,Pearson1,Patel1,Patel2} and numerical simulations \citep{Tomar2,Tomar3,pandey2,pandey1,pandey3,rouzbahani1,wangqi1} have shown that electric fields aid in accelerating bubble detachment during the boiling process, thereby enhancing heat transfer efficiency. Furthermore, numerical simulations focusing on 2D multi-model film boiling under electric fields have found that the electric field promotes the occurrence of Rayleigh-Taylor instability at the vapor-liquid interface and increases the number of bubble nucleation sites. However, as another crucial physical phenomenon, magnetic fields offer considerable research potential, particularly in confined nuclear fusion. The effect of MF on fluids can be divided into magnetohydrodynamics (MHD) and ferrohydrodynamics (FHD), where MHD effects result from the action of electromagnetic forces or Lorentz forces, and FHD effects result from Kelvin forces. Regarding the study of MF influence on boiling heat transfer, both experimental and numerical research have focused more extensively on FHD. This is mainly because ferrofluids, compared to magnetic fluids such as liquid lithium, have decent transparency, which brings convenience to experimental investigations. A plethora of experimental \citep{lee1,shojaeian1,abdollahi1,zonouzi1} and numerical \citep{aminfar1,aminfar2,malvandi1,guo2} studies have demonstrated that MFs enhance the heat transfer characteristics of ferrofluid boiling, while also increasing the number of bubble nucleation sites during the boiling process.

In contrast, research on magnetohydrodynamic (MHD) effects on film boiling, even today, primarily relies on early theoretical and experimental studies. \citet{faber1} investigated the effect of a vertical MF on the subcooled boiling of mercury and found that a strong MF suppresses natural convection but promotes nucleate boiling. \citet{fraas1} studied the effect of a vertical MF on boiling in potassium salts and metals and concluded that the MF had an insignificant effect. \citet{wagner1} studied the influence of a horizontal magnetic field on nucleate boiling of liquid metallic mercury, finding that the horizontal magnetic field reduced the heat transfer efficiency of boiling and promoted the transition from nucleate to film boiling. \citet{takahashi1} examined the impact of a vertical MF on saturated nucleate boiling of mercury on a horizontal surface. It was shown that there was a decrease in the Nusselt number and heat transfer efficiency with increasing MF. Experiments conducted by \citet{de1} showed that the magnetic field slowed down the frequency of bubble release during nucleate boiling. The theoretical research conducted by \citet{arias1,arias2} corroborated the findings from the aforementioned experimental investigations. However, theoretical investigation of \citet{arias3} within the Taylor–Helmholtz instability framework indicated that the MF did not influence film boiling. The aforementioned literature review suggests that the impact of magnetic fields on boiling does not yield consistent results, and some findings are even contradictory. Additionally, due to the opacity of liquid metals, experimental studies were unable to capture the dynamics of bubbles during the boiling process accurately. Consequently, numerical simulation becomes the only feasible method to explore the mechanisms underlying the influence of magnetic fields on film boiling with liquid metals, a subject that, to our knowledge, has not been investigated previously.

Thus, the present study aims to comprehend the behavior of film boiling under the influence of applied magnetic fields in different directions. The scarcity of research in this area can be attributed to the increased significance of magnetohydrodynamic (MHD) effects in 3D models. Additionally, the flow and thermal boundary layers of liquid metals are extremely thin, necessitating grids with fine resolution and consequently leading to substantial computational demands. Furthermore, numerical simulations involving incompressible multiphase magnetohydrodynamics with phase change impose high requirements on the robustness of the employed numerical solver. The phase change model utilized in this study was developed in our earlier work \citep{zhao1}, which employed a sharp interface method to enforce the temperature and concentration conditions at the liquid/vapor interface, integrating a volume-of-fluid method and an embedded boundary approach for sharp scheme construction. \citet{zhao1} extensively validated the phase change model against various benchmark problems and rigorous numerical tests, showcasing its exceptional accuracy and robustness. In addition to this phase change model, we integrate a consistent and conservative scheme \citep{zhang2} for discretizing the magnetohydrodynamic (MHD) equations, enabling the simulation of 3D film boiling on a horizontal surface in the presence of an applied magnetic field. Our investigation comprises two primary components. Firstly, we focus on single-mode film boiling under vertical and horizontal magnetic fields in three dimensions. In contrast to previous studies, we significantly extend the simulation duration to observe the impact of the magnetic field after the complete development of the boiling flow. Additionally, we increase the computational domain height to explore a wider range of boiling dynamics. Secondly, we extend the computational domain size to $4\lambda_d\times4\lambda_d\times4\lambda_d$ and analyze the behavior of multi-mode film boiling under both vertical and horizontal magnetic fields. 

The rest of the manuscript is structured as follows. In \S\ref{sec:problem}, we present the physical model and governing equations and introduce the phase change model developed by \citet{zhao1}. The results are discussed in \S\ref{sec:dis}, where we analyze the influence of horizontal and vertical magnetic fields on both single-mode and multi-mode film boiling phenomena in \S\ref{sec:single-mode} and \S\ref{sec:multi-mode}, respectively. Finally, concluding remarks are provided in \S\ref{sec:conclusion}.

\section{Problem formulation and numerical method} \label{sec:problem}
\subsection{Physical model}

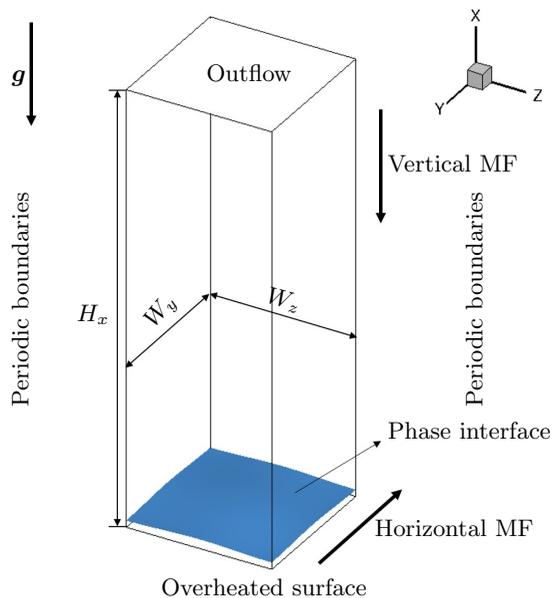
\begin{figure}
	\centering
	\begin{overpic}
			[scale=0.45,tics=5]{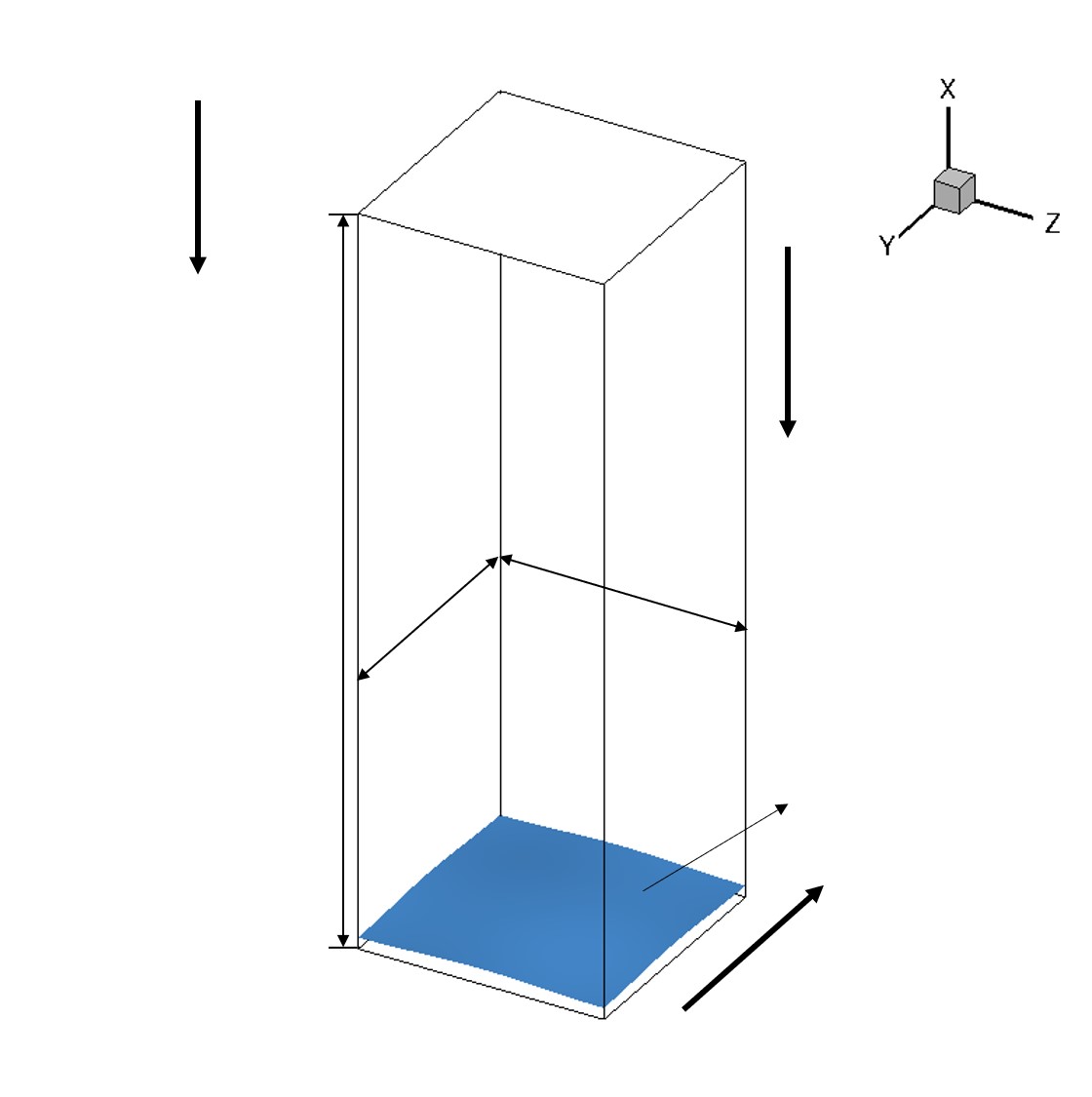}
			\put(15,82){$\boldsymbol{g}$}
			\put(35,44){\begin{turn}{50}$W_y$\end{turn}}
			\put(54,48){\begin{turn}{-15}$W_z$\end{turn}}
			\put(25,45){$H_x$}
			\put(45,82){Outflow}
			\put(38,3){Overheated surface}
			\put(73,27){Phase interface}
			\put(71,12){Horizontal MF}
			\put(73,68){Vertical MF}
			\put(15,32){\begin{turn}{90}Periodic boundaries\end{turn}}
			\put(85,32){\begin{turn}{90}Periodic boundaries\end{turn}}
	\end{overpic}
	\caption{Schematic diagram of the film boiling configuration under the applied MFs. The boundary conditions employed along the boundaries of the computational domain are also specified.}
	\label{fig1}
\end{figure}

We examine the film boiling phenomenon of liquid under horizontally and vertically applied MFs by performing 3D numerical simulations. Firstly, based on previous studies on film boiling without an imposed MF, a brief description of the entire physical process is provided: during film boiling, there exists a layer of vapor film between the overheated wall and the upper liquid layer, preventing direct contact between the upper fluid and the overheated wall. As the evaporation proceeds, the vapor film gradually thickens, leading to instability at the interface. Eventually, the influence of gravity triggers the Rayleigh-Taylor instability, giving rise to the formation of bubbles. According to this physical process, the computational model set up for this study is illustrated in Fig. \ref{fig1}. The size of the computational domain is $H_x \times W_y \times W_z$, where $H_x, W_y$, and $W_z$ are multiples of $\lambda_d$. This choice aligns with the observation by \citet{tsui1} that setting the computational domain width as $\lambda_d$ in 3D simulations adequately captures the physical phenomenon. A Cartesian coordinate system $(x,y,z)$ is employed, with $x$ representing the vertically upward direction and $y$ and $z$ indicating the horizontal and spanwise directions, respectively. The center of the coordinate system is aligned with the center of the bottom plane $(yz)$. The gravity $\boldsymbol{g}$ acts in the negative $x$ direction. An external uniform magnetic field is applied either in the vertical ($x$) or horizontal ($y$) direction.

We employ the following boundary conditions: no-slip and no-through-flow conditions at the bottom (overheated surface), periodic boundary conditions at the four vertical sides (front/back/left/right), and outflow conditions at the top. The phase interface and the liquid are maintained at the saturation temperature ($T_{\text{sat}}$), which is a valid assumption except when exposed to extremely high heat flux \citep{Juric1, akhtar1}. Additionally, the temperature of the overheated surface is kept constant at $T_{\text{wall}}$ ($T_{\text{wall}} = T_{\text{sat}} + \Delta T$). Initially (at time $t=0$), the interface is defined using sinusoidal functions, and the temperature in the vapor phase linearly decreases from the overheated surface to the phase interface.

%Fig 1
	
\subsection{Governing equations}

The viscous incompressible MHD two-phase flow, incorporating liquid-vapor phase change, is governed by the mass, momentum, and energy conservation equations. They are expressed as follows:
    \begin{equation}
		\nabla \cdot \boldsymbol{u}=\ddot{m} \left(\frac{1}{\rho_{v}}-\frac{1}{\rho_{l}}\right), 
		\label{eq1}
	\end{equation}
	\begin{equation}
		\rho\left(\frac{\partial \boldsymbol{u}}{\partial t}+\boldsymbol{u} \cdot \nabla \boldsymbol{u}\right)=-\nabla p+\nabla \cdot\left(\mu\left(\nabla \boldsymbol{u}+\nabla \boldsymbol{u}^{T}\right)\right)+\rho \boldsymbol{g}+\sigma \kappa \delta_{s} \boldsymbol{n}+ \boldsymbol{F_l},
		\label{eq2}
	\end{equation}
	\begin{equation}
		\frac{\partial T}{\partial t}+\boldsymbol{u} \cdot \nabla T=\nabla \cdot\left(\frac{k}{\rho c_{p}} \nabla T\right)+\Phi.
		\label{eq3}
	\end{equation}
Here, $\boldsymbol{u}$ signifies the fluid velocity, $\ddot{m}$ represents the volume mass transfer rate, $\rho$ denotes the density, and subscripts $l$ and $v$ designate the physical properties for the liquid and vapor phases, respectively. In Eq. (\ref{eq2}), $p$ stands for pressure, $\mu$ is the dynamic viscosity, and $\sigma \kappa \delta_{s} \boldsymbol{n}$ represents the surface tension force, wherein, $\sigma$ is the surface tension coefficient, assumed to be a temperature-independent constant. The interface curvature is denoted by $\kappa$, $\delta_{s}$ is the Dirac distribution function (whose value is equal to 1 at the interface and zero elsewhere), and $\boldsymbol{n}$ denotes the unit normal vector at the local interface. 

Besides, $\boldsymbol{F_l}$ is the Lorentz force arising from the electromagnetic induction, which is given by
\begin{equation}
	\boldsymbol{F_l}=\boldsymbol{J} \times \boldsymbol{B},
\end{equation}
where $\boldsymbol{J}$ is the induced current density, $\boldsymbol{B}$ represents the applied MF. According to the Ohm’s law, the current density is given by
\begin{equation}
	\boldsymbol{J}=\sigma_{e}(-\nabla \varphi+\boldsymbol{u} \times \boldsymbol{B}),
	\label{eq5}
\end{equation}
where $\sigma_{e}$ denotes the electric conductivity and $\varphi$ represents the induced electric potential. In Eq. (\ref{eq5}), the induced magnetic intensity is neglected because it is significantly smaller than the external field, acknowledging that the magnetic Reynolds number ($Re_m = \eta \sigma_e L u'$, with $\eta$ the magnetic permeability, $L$ the characteristic length and $u'$ the characteristic velocity) is much smaller than 1 for the MHD flow considered in the present study. In order to satisfy the charge conservation, we employ
\begin{equation}
	\nabla \cdot \boldsymbol{J}=0.
	\label{eq6}
 \end{equation}
By combining Eqs. (\ref{eq5}) and (\ref{eq6}), the electric potential Poisson equation is derived as: 
\begin{equation}
	\nabla \cdot\left(\sigma_{e} \nabla \varphi\right)=\nabla \cdot\left(\sigma_{e} \boldsymbol{u} \times \boldsymbol{B}\right).
\end{equation}
It is to be noted that solving this Poisson equation allows for the determination of the electric potential $\varphi$. Subsequently, the current density $\boldsymbol{J}$ can be calculated using Eq. (\ref{eq5}). Finally, the Lorentz force $\boldsymbol{F_l}$ can be determined. In Eq. (\ref{eq3}) governing the temperature transport, $c_{p}$ represents the volumetric heat capacity, $k$ is the thermal conductivity, $T$ denotes the temperature, and $\Phi$ stands for the internal dissipation term, encompassing viscous dissipation and Joule dissipation. \rb{However, in this study, both dissipation effects are considered negligible compared to the temperature terms, a simplification consistent with other studies \citep{burr2002rayleigh, vogt2018transition, chen2024flow}.}

The presence of phase change necessitates that the above equations must satisfy the following interfacial jump conditions:
\begin{equation}
    \dot{m}=\rho_{l}\left(\boldsymbol{u}_{\Gamma}-\boldsymbol{u}_{l}\right) \cdot \boldsymbol{n}=\rho_{v }\left(\boldsymbol{u}_{\Gamma}-\boldsymbol{u}_{v}\right) \cdot \boldsymbol{n}, 
\label{eq8}
\end{equation}
\begin{equation}
	\dot{m}=\frac{\left(\boldsymbol{q}_{l}-\boldsymbol{q}_{v}\right) \cdot \boldsymbol{n}}{h_{lg}}=\frac{\left((k \nabla T)_{l}-(k \nabla T)_{\text {v}}\right) \cdot \boldsymbol{n}}{h_{lg}}.
	\label{eq10}
\end{equation}
Eq. (\ref{eq8}) describes the mass transfer rate from liquid to vapor in accordance with the mass conservation law. Note that $\boldsymbol{u}_{\Gamma}$ is the propagation velocity of the interface, which does not equal to the local fluid velocity $\boldsymbol{u}_{(l, v)}$ in the presence of phase change. Eq. (\ref{eq10}) elucidates the source of the mass transfer rate generated by the thermal flux across the interface. Here, $\dot{m}$ denotes the area mass transfer rate, $\boldsymbol{q}$ is the thermal flux at the respective side of liquid and vapor, and $h_{lg}$ is the latent heat. Then the velocity jump condition (Eq.~\ref{eq8}) leads to the following divergence-constraint condition on the velocity field
\begin{equation}
	\nabla\cdot\boldsymbol{u} = (\frac{1}{\rho_v} - \frac{1}{\rho_l})\ddot{m}.
	\label{eq8.2}
\end{equation}

The relationship between $\dot{m}$ and $\ddot{m}$ is given by
\begin{equation}
	V \ddot{m}=\dot{m} S_{\Gamma},
\end{equation}
where $V$ denotes the volume of the discretized cell containing the interface and $S_{\Gamma}$ is the area of that interface. It is to be noted that both $\ddot{m}$ and $\dot{m}$ are exclusively estimated at the interface. In this work, we assume the interface temperature remains constant at the saturation temperature:
\begin{equation}
	T_{\Gamma}=T_{sat},
\end{equation}
which effectively serves as a Dirichlet-like boundary condition for the solution of the temperature equation in the separated liquid and vapor phases.

\subsection{Numerical method}

In the current study, our MHD-phase change solver is implemented within the framework of the open-source platform \textit{Basilisk} \citep{Popinet1, Popinet2}, well-regarded for its capability in simulating complex multiphase flows. A staggered-in-time approximate projection method is employed to discretize the incompressible Navier–Stokes equations, while the volume-of-fluid (VOF) method \citep{weymouth1,scardovelli1999direct} accurately captures and advances the phase interface. The Bell-Colella-Glaz (BCG) second-order scheme \citep{Bell1} discretizes the advection term, and a fully implicit scheme is applied to discretize the diffusion term. Additionally, a quad/octree-based adaptive mesh refinement (AMR) technique is utilized for spatial discretization, enabling grid refinement primarily near the interface to optimize computational resources.

In our previous study \citep{zhao1}, we developed a numerical method for the phase change model capable of simulating boiling and evaporating flows. \rb{This method imposes jump conditions at the interface using a sharp scheme while exactly conserving both mass and energy. The relative conservation error is found to be consistently smaller than $10^{-3}$.} In the present study, we adopt this numerical method to address the phase-change model for film boiling. Additionally, in the presence of an external magnetic field (MF), we utilize a consistent and conservative scheme, which was also developed earlier \citep{zhang2, Zhang3}, for computing electric currents and Lorentz forces. Detailed information about the numerical technique and validations regarding the MHD-phase change solver can be found in previous works \citep{zhao1, zhang2, Zhang3}. In view of this, we will only present some crucial information featuring the progression of the numerical solution from time level $n$ to level $n+1$:
\begin{enumerate}
\item A split advection method proposed by \citet{weymouth1} is employed to discretize the VOF advection equation, and the volume fraction $C^{n+\frac{1}{2}}$ is updated accordingly. However, the presence of phase change alters the form of the VOF advection equation to
\begin{equation}
	\frac{\partial C}{\partial t}+\nabla \cdot\left(\boldsymbol{u} C \right)=-\frac{\ddot{m}}{\rho_l},
\end{equation}
which is consistent with the formulation of the mass conservation equation. Due to the discontinuity in velocity at the interface, it is not possible to solve the VOF advection equation directly, as is the case with ordinary two-phase flows. To address this, an extended velocity field $\boldsymbol{u}_E$ is introduced, $\boldsymbol{u}_E$=$\boldsymbol{u}$-$\boldsymbol{u}_J$. $\boldsymbol{u}_J$ is the velocity jump, which could be obtained by solving for the velocity potential $\psi$ using the following equations:
\begin{equation}
	\left\{\begin{array}{l}
		\nabla^{2} \psi=\left(\frac{1}{\rho_{v }}-\frac{1}{\rho_{l}}\right) \ddot{m}, \\
		\boldsymbol{u}_{J}=\nabla \psi.
	\end{array}\right.
\end{equation}
To avoid the negative effect of the source term, the interfacial velocity ($\boldsymbol{u}_{\Gamma}$) is chosen to advect the VOF advection equation:
\begin{equation}
	\frac{\partial C}{\partial t} + \boldsymbol{u}_{\Gamma} \cdot \nabla C =0,
 \label{eq:VOF}
\end{equation}
where $\boldsymbol{u}_{\Gamma}$=$\boldsymbol{u}_E$- $\frac{\ddot{m}}{\rho_l}\boldsymbol{n}$, wherein $\boldsymbol{n}$ directs towards the vapor phase in normal direction at the interface.
\item The physical properties $\phi^{n+\frac{1}{2}}$ are updated using
\begin{equation}
	\phi^{n+\frac{1}{2}}=C^{n+\frac{1}{2}} \phi_{l}+\left(1-C^{n+\frac{1}{2}}\right) \phi_{v}, \quad {\rm where} ~~ \phi=\rho, ~\mu, ~\sigma_{e}.
\end{equation}
\item The temperature $T^{n+\frac{1}{2}}$ is then updated. Note that the temperature fields of the liquid and vapor phases are solved separately to ensure sharp temperature gradients at the interface for accurate heat flux calculations. Specifically, the temperature equation is solved only in the vapor phase for saturated film boiling, maintaining the liquid temperature at $T_\Gamma$. The interfacial temperature ($T_{\Gamma}$) is imposed as the interfacial boundary condition, which is realized via an embedded boundary method. Besides, the advection term of the temperature is discretized by using a geometrical scheme, which is consistent with that applied in Eq.~(\ref{eq:VOF}).
\begin{equation}
	\frac{T^{n+\frac{1}{2}}-T^{n-\frac{1}{2}}}{\Delta t}+\nabla \cdot(T \boldsymbol{u})^{n+\frac{1}{2}}=\nabla \cdot\left(\left(\frac{\lambda}{\rho c_{p}}\right) \nabla_{f} T\right)^{n+\frac{1}{2}}+\left(\frac{\ddot{m}}{\rho} T_{\Gamma}\right)^{n+\frac{1}{2}}.
	\end{equation}
\item The Lorentz force $\boldsymbol{F_l}^{n+\frac{1}{2}}$ is updated by solving the electric potential Poisson equation and Ohm's law using
\begin{equation}
	\nabla \cdot\left(\sigma_{e} \nabla \varphi \right)^{n+\frac{1}{2}}=\nabla \cdot\left(\sigma_{e}(\boldsymbol{u} \times \boldsymbol{B}) \right)^{n+\frac{1}{2}},
	\end{equation}
\begin{equation}
	\boldsymbol{J}^{n+\frac{1}{2}}=\sigma_{e}^{n+\frac{1}{2}}\left(-\nabla \varphi+\boldsymbol{u} \times \boldsymbol{B}\right)^{n+\frac{1}{2}},
\end{equation}
\begin{equation}
	\boldsymbol{F_l}^{n+\frac{1}{2}}=\boldsymbol{J}^{n+\frac{1}{2}} \times \boldsymbol{B}^{n+\frac{1}{2}}.
\end{equation}
\item The mass transfer rate $\dot{m}^{n+\frac{1}{2}}$ is updated according to Eq. (\ref{eq10}), while the embedded boundary method is used to calculate the temperature gradient, ensuring that the interpolation template only uses those cell values located in the vapor phase. This method avoids errors caused by artificially thickening the interface and ensuring accurate mass transfer rate calculations.
\item Finally, the pressure field ${p}^{n+1}$ and the velocity field $\boldsymbol{u}^{n+1}$ are updated using the approximate projection method, which has been incorporated in \textit{Basilisk} \citep{Popinet1, Popinet2}.
\end{enumerate}

\subsection{Dimensionless parameters}
In this study, the length, velocity and time scales are defined as $\lambda'$=$\sqrt{\sigma/(\rho_l-\rho_v)g}$, $t'$=$\sqrt{\lambda'/g}$ and $u'$=$\sqrt{\lambda'g}$. The critical wavelength is given by $\lambda_c$=$2\pi$$\lambda'$. The most dangerous wavelength is given by $\lambda_d$=$\sqrt{3}$$\lambda_c$. The relevant dimensionless parameters in this problem are listed below. The Grashof number $(Gr=\rho_v(\rho_l-\rho_v)g\lambda'^3/\mu_v^2)$ describes the ratio of the buoyancy to viscous force. The Prandlt number $(Pr=c_{p,v}\mu_v/k_v)$ represents the ratio of the momentum diffusion to the thermal diffusion in vapor. The Jakob number $(Ja=c_{p,v}\Delta T/h_{lg})$ represents the ratio of overheat to latent heat.

To investigate the heat transfer characteristics in film boiling studies, the Nusselt number $(Nu)$ is the most relevant dimensionless number, representing the ratio between heat convection and heat conduction. It measures the strength of boiling heat transfer during film boiling. In the current study, the space-averaged Nusselt number is defined as
\begin{equation}
	Nu=\left.\frac{1}{W_y} \frac{1}{W_z}  \int_{0}^{W_y} \int_{0}^{W_z} \frac{\lambda^{\prime}}{T_{\text {wall }}-T_{\text {sat }}} \frac{\partial T}{\partial n}\right|_{\text {wall }} \mathrm{~d} y \mathrm{~d} z. 
	\label{nu}
\end{equation}
Further, the space- and time-averaged Nusselt number can be calculated as
\begin{equation}
	\overline{Nu}=\frac{1}{t_{1}-t_{0}} \int_{t_{0}}^{t_{1}} Nu \mathrm{~d} t,
	\label{time-nu}
\end{equation}
where $t_{0}$ is the initial time and $t_{1}$ is the end time.

For MHD flow, the interaction number ($N$) describes the ratio between the Lorentz and gravitational forces. This is represented as
\begin{equation}
N=\sigma_{e(l)}B^2\lambda'/\rho_l\sqrt{g\lambda'}.
\label{nn}
\end{equation}

\ra{Additionally, we note that another dimensionless parameter, the Hartmann number ($Ha$), is used to describe the magnitude of the Lorentz force. It is expressed as $Ha = \sqrt{ReN}$, where $Re$ is the Reynolds number, defined as $Re = \rho_l u'\lambda'/\mu_l$. In this study, we primarily use the interaction number ($N$) to characterize the strength of the external magnetic field (MF), while also providing the corresponding $Ha$ values.}

In the present study, we explore the film-boiling phenomenon based on the liquid metal lithium.
%, the physical properties of which are listed in Table \ref{table1}. 
However, in the case of liquid metal lithium, both the thermal and flow boundary layers are very thin, posing a significant challenge to achieving the required high spatial resolution for resolving the extremely thin vapor film which has a thickness of $\delta \propto \left[ \frac{\mu_v k_v \Delta T}{h_{lg} \rho_v (\rho_l - \rho_v) g} \left( \frac{\sigma}{(\rho_l - \rho_v) g} \right)^{1/2} \right]^{1/4}$ \citep{Berenson1, Kim1}. Typically, we found that when the computational domain is confined to $H_x \times W_y \times W_z = 3\lambda_d \times \lambda_d \times \lambda_d$, the minimum vapor film thickness cannot be resolved even when grid resolution is 512 grids within a most dangerous wavelength, which represents the diameter of the vapor bubble in the single-mode film boiling. It implies more than fifty million grids are required even if we use adaptive mesh refinement technique to reduce the computational cost. More seriously, the time step is highly restrained by the size of the smallest mesh. To solve this problem, we manually adjust some of the physical properties of liquid lithium in our numerical simulations, especially the viscosity and conductivity of the vapor phase so that the vapor layer is now almost eight times thicker than that of the original liquid lithium. The physical properties of the fluid under investigation are listed in Table \ref{table2}, and we will show that the grid resolution of $\lambda_d/128$ can accurately resolve the thin vapor layer. \rb{Furthermore, in Appendix~A, we demonstrate that even when using more realistic physical properties, the MHD effects on the rising behavior of the vapor bubble remain very similar. This supports the notion that adjustments to the fluid properties do not have a qualitative impact on the conclusions drawn in the present study.}

According to Table \ref{table2}, the values of characteristic scales and dimensionless parameters are presented in Table \ref{table3}. The physical quantities described in the rest of the manuscript have been non-dimensionalized using the characteristic scales. And the temperature field is normalized based on the saturation temperature $T_{\mathrm{sat}}$, given that $(T-T_{\mathrm{sat}})/(T_{\mathrm{wall}} - T_{\mathrm{sat}})$ standardizing the relative magnitude of the vapor phase temperature field to values between 0 and 1. As a result, the temperature of the liquid phase is set to 0, while the temperature of the overheated wall is fixed at 1.
	
%	\begin{table}
%    	\begin{center}
%			\def~{\hphantom{0}}
%		\begin{tabular}{lcccccc}
%			& $\rho$(kg/m$^{3}$) & $\mu$(Ps$\cdot$s) & $k$(W/(m$\cdot$K)) & $c_{p}$(J/(kg$\cdot$K))  & $h_{lg}$(J/kg) &  $\sigma$(N/m) \\
%			vapor & 0.0546 & 1.442$\times$$10^{-4}$ & 0.192 & 9259 &\multirow{2}{*}{1.933$\times$ $10^7$} & \multirow{2}{*}{0.226}  \\
%			Liquid  &   401.3 & 1.59$\times$$10^{-4}$ & 68.5 & 4250 &  &                
%		\end{tabular}
%			\caption{Physical properties of liquid metal lithium at $T_{sat}$=1615K.}
%		\label{table1}
%		\end{center}
%	\end{table}

	\begin{table}
		\begin{center}
			\def~{\hphantom{0}}
			\begin{tabular}{lccccccc}
				& $\rho$(kg/m$^{3}$) & $\mu$(Ps$\cdot$s) & $k$(W/(m$\cdot$K)) & $c_{p}$(J/(kg$\cdot$K))  & $h_{lg}$(J/kg) &  $\sigma$(N/m) & $\Delta$$T$(K) \\
				Vapor & 0.546 & 0.049 & 6.528 & 9259 &\multirow{2}{*}{1.933$\times$ $10^6$} & \multirow{2}{*}{0.23} & \multirow{2}{*}{20}  \\
				Liquid  &   401.3 & 0.159 & 68.5 & 4250 &  &  &              
			\end{tabular}
			\caption{Physical properties of fluids used in the present study.}
			\label{table2}
		\end{center}
	\end{table}
	
	\begin{table}
		\begin{center}
			\def~{\hphantom{0}} 
			\begin{tabular}{cccccccc}
			$\lambda'$ & $t'$ & $u'$ & $\lambda_c$ & $\lambda_d$ & $Gr$ & $Pr$ & $Ja$  \\ [3pt]
			0.0076 m &0.0279 s&0.273 m/s& 0.048 m &0.083 m &	0.392 &69.499 & 0.096  \\ 
			\end{tabular}	
			\caption{Characteristic scales and values of the dimensionless numbers used in the present study.}
			\label{table3}		
		\end{center}
	\end{table}

\section{Results and discussion} \label{sec:dis}
\subsection{Single-mode film boiling} \label{sec:single-mode}

\begin{figure}
	\centering
   \scalebox{0.9}{\input{figure/grid-test.tex}}	
\caption{The grid independence test performed at different grid resolutions. The bubble shapes on the vertical cross-section ($XOY$ plane) at (a) $t=14.8$ and (b) $t=16.7$ for different grid resolutions. (c) The temporal evolution of the space-averaged Nusselt number ($Nu$) obtained using different grid resolutions.}
\label{grid-test}
\end{figure}
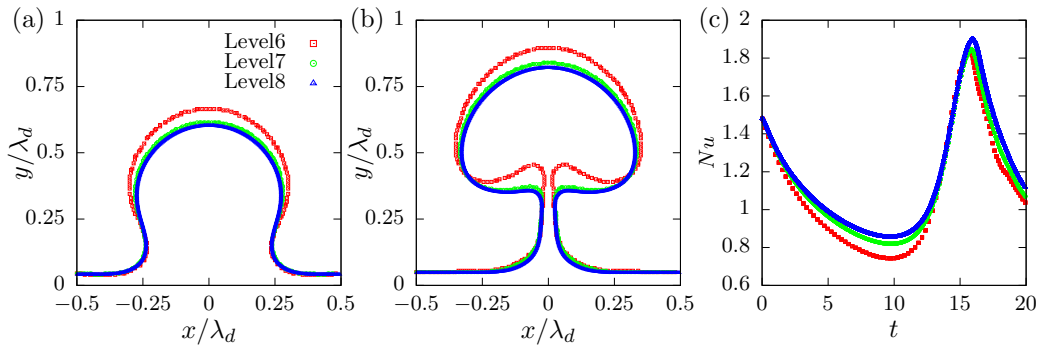

In the case of single-mode film boiling with and without MFs, a computational domain with dimensions $3\lambda_d \times \lambda_d \times \lambda_d$ is used. An initial perturbation in the following form is applied to the phase interface:
\begin{equation}
	x=\frac{\lambda_d}{128} \left( 8.0 + \mathrm{cos}\left( \frac{2 \pi y}{\lambda_d} \right)+ \mathrm{sin}\left( \frac{2 \pi z}{\lambda_d} \right)  \right).
\end{equation}
A grid convergence test is conducted for film boiling without MF, considering three different grids: $192 \times 64 \times 64$ (referred to as grid-level 6), $384 \times 128 \times 128$ (grid-level 7), and $768 \times 256 \times 256$ (grid-level 8). The vapor bubble shapes on the vertical cross-section ($XOY$ plane) at $t=14.8$ and $t=16.7$ are illustrated in Fig. \ref{grid-test}(a) and Fig. \ref{grid-test}(b), respectively. The temporal evolution of the space-averaged Nusselt number ($Nu$) is also depicted in Fig. \ref{grid-test}(c). It is evident that the grid resolution at grid-level 7 is adequate to achieve the desired computational accuracy. Consequently, this grid size is adopted for the remainder of our study. 

\rb{The computational results presented in this study have been obtained using workstations equipped with two AMD$^\circledR$ EPYC 7742 CPUs, each with 64 cores operating at 2.20 GHz and 256 GB of RAM. For single-mode boiling flows without an external magnetic field, as discussed in \S 3.1, a typical run with a grid refined to Level 7 and covering a dimensionless time span of $0 < t < 75$ required more than two months using 24 processors. When a vertical magnetic field was applied at $N = 6.3$, the computational time extended to over three months due to the additional Poisson equation that needed to be solved. For multi-mode film boiling simulations discussed in \S 3.2, the computations were carried out using 128 AMD processors. In this case, with a grid refined to Level 7 and a dimensionless computational time span of $0 < t < 100$ for $N = 11.1$, each simulation took nearly six months to complete. Moreover, test cases indicated that refining the grid from Level 7 to Level 8 would increase the computational time by approximately threefold, primarily due to the increased time constraints imposed by CFL condition. For example, the dimensionless time step was reduced from 2.7 $\times 10^{-4}$ to 1.0 $\times 10^{-4}$. Based on a comprehensive evaluation of computational efficiency and accuracy, we concluded that a spatial resolution of Level 7 is the most practical choice for our numerical simulations.}

%Fig 2
 
\subsubsection{Film boiling without a MF}

%Fig 3
\begin{figure}
	\centering
     \scalebox{1}{\input{figure/nu-b0.tex}}	
	\caption{Temporal evolution of the space-averaged $Nu$ for the single-mode film boiling without a MF. The dashed line represents the theoretical results of \citet{Klimenko1}.} 
	\label{nu-b0}
\end{figure}
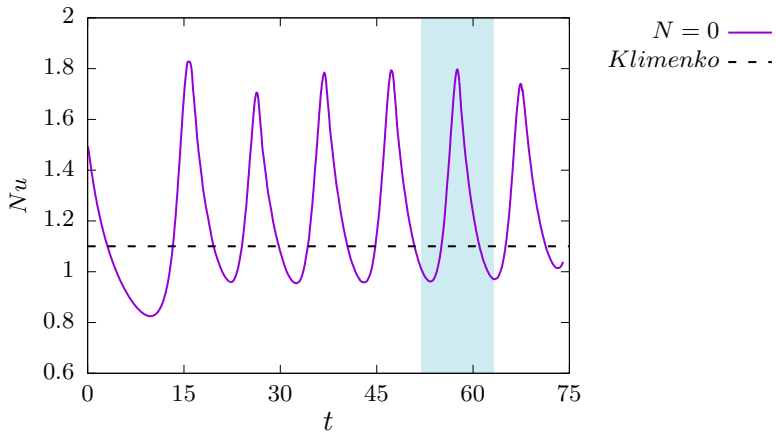

\begin{figure}
	\centering
    \includegraphics[height=4.8cm]{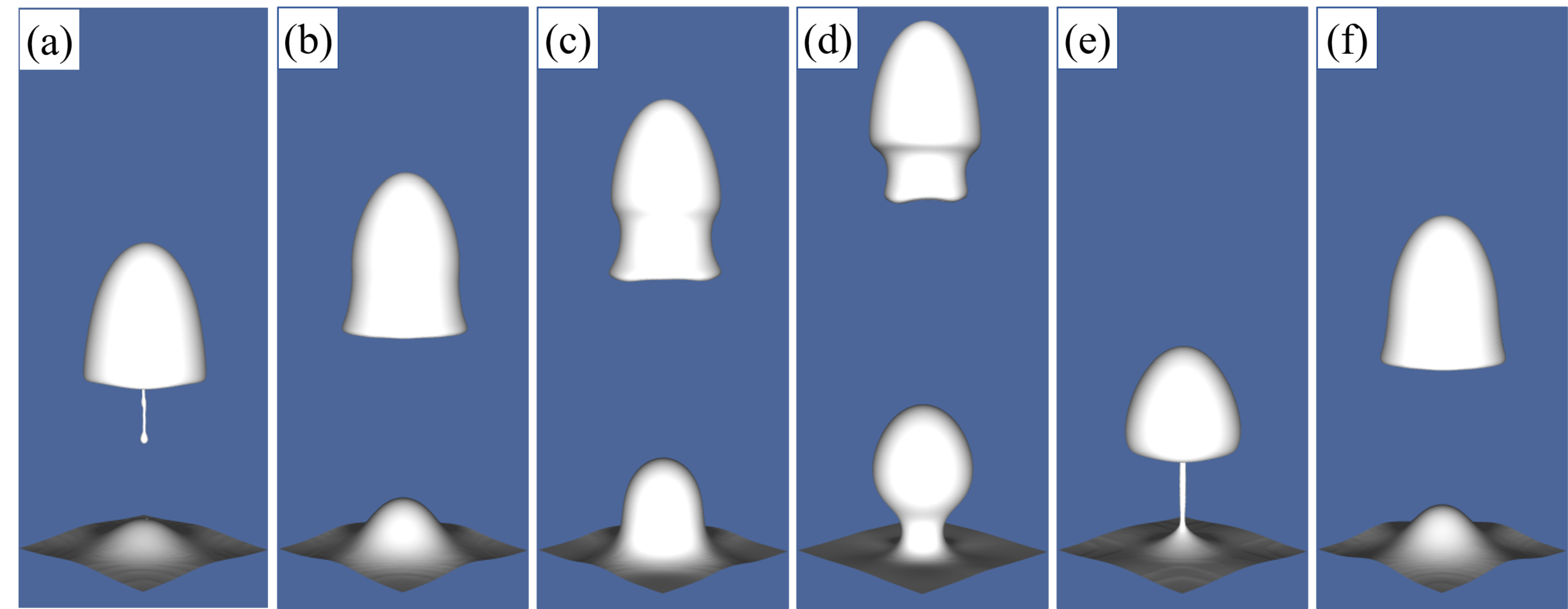}
	\caption{Vapor bubble generation during the film boiling flow without a MF at different dimensionless times, which correspond to the blue region in Fig. \ref{nu-b0}. (a) $t=51.9$, (b) $t=53.7$, (c) $t= 55.6$, (d) $t=57.4$, (e) $t=59.3$ and (f) $t=63.0$.}
	\label{shape-b0}
\end{figure}

\begin{figure}
    \centering
    \includegraphics[height=5.5cm]{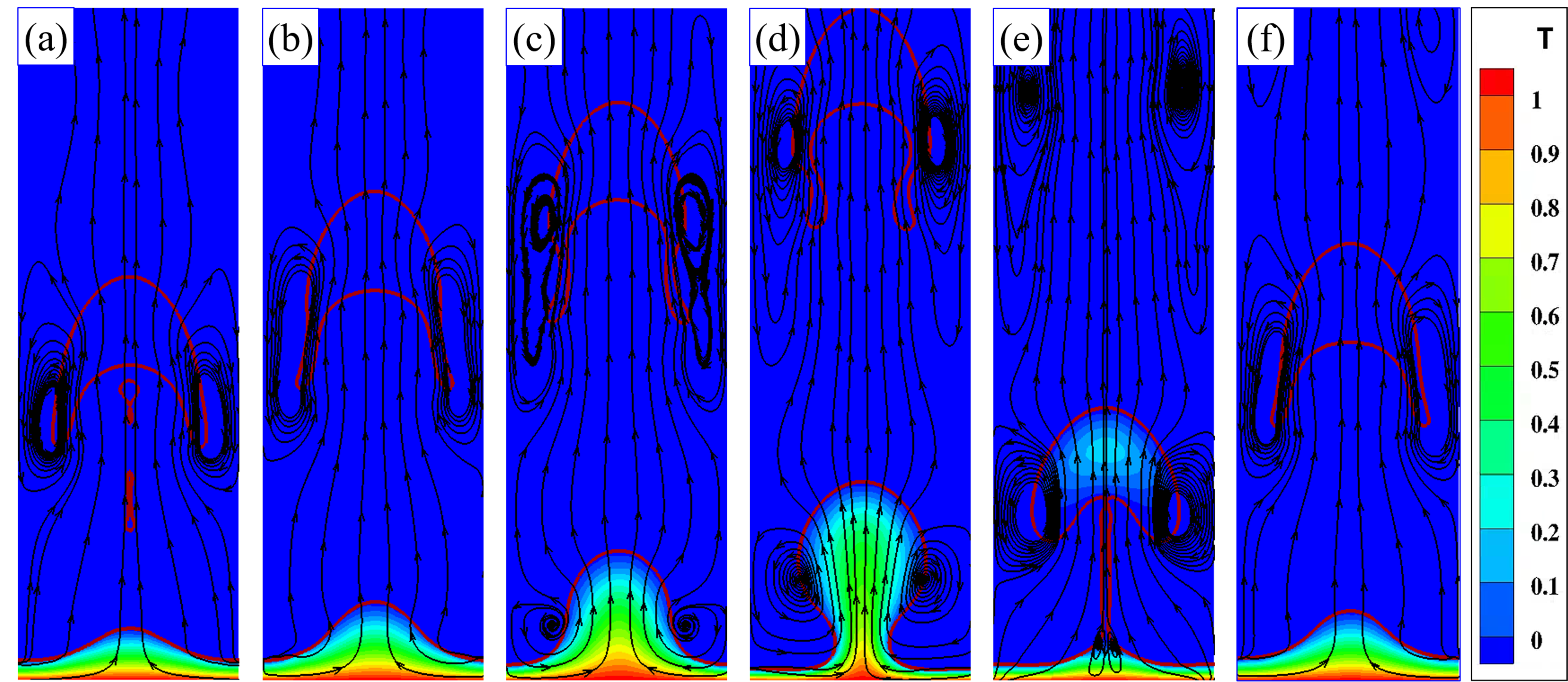}
    \caption{Temperature profile (contours) and velocity streamlines (black arrow lines) in the vertical cross-section ($XOY$) of the computational domain. The time instants from panels (a-f) correspond to those displayed in Fig. \ref{shape-b0}(a-f).}
\label{tem-b0}
\end{figure}

We begin the presentation of our results by analyzing the film boiling phenomenon without MF and compare with existing boiling mechanisms and empirical models. Fig. \ref{nu-b0} shows the temporal evolution of the space-averaged $Nu$. It can be seen that $Nu$ oscillates periodically around the correlation proposed by \citet{Klimenko1}, given that $Nu_k=0.19(GrPr)^{\frac{1}{3}}\times (0.89Ja^{-\frac{1}{3}})$, this agreement indicates the credibility of the numerical simulation results. This periodic oscillation corresponds to the shedding of bubbles. The evolution of the liquid/vapor interface for the fifth bubble cycle (from $t=51.9$ to $t=63.0$), corresponding to the blue region in Fig. \ref{nu-b0}, is depicted in Fig. \ref{shape-b0}, while Fig. \ref{tem-b0} presents the corresponding streamline and temperature field distributions within the vertical cross-section ($XOY$). It is noteworthy that during the fifth bubble cycle, the bubble generated in the fourth bubble cycle has just detached, resulting in the presence of a rising bubble in Fig. \ref{shape-b0}(a-d). The film boiling process within one bubble cycle can be divided into four stages: Stage one: From $t=51.9$ to $t=53.7$, the phase change causes the expansion of the vapor phase volume, and the bulge grows rapidly. Stage two: From $t=53.7$ to $t=55.6$, the Rayleigh-Taylor instability occurs, leading to the formation of a significant bulge and vortices. Stage three: From $t=55.6$ to $t=59.3$, the neck connecting the thin film and bubble starts to contract, leading to a mushroom-shaped bubble and the vortex structure. Stage four: From $t=59.3$ to $t=63.0$, the bubble pinches off, and a new bubble cycle initiates. Furthermore, within one bubble cycle, the vapor film thickness initially increases due to the expansion of the vapor phase volume and the growth of the bulge. Later, it decreases when the neck phenomenon occurs. Conversely, the Nusselt number decreases first and then increases within one bubble cycle, as explained by Eq. (\ref{nu}). The temperature profile reveals a linear distribution from the overheated wall to the liquid/vapor interface throughout the film boiling process. A smaller distance between the liquid/vapor interface and the overheated wall corresponds to a more significant temperature gradient, resulting in a higher $Nu$. In contrast, a larger distance leads to a smaller $Nu$. These findings align with previous studies \citep{Juric1,zhang1}. Next, we investigate the influence of a vertical MF on the film boiling dynamics.

%Fig 4

%Fig 5

\subsubsection{Effect of the vertical MF} \label{single-vertical MF}

%Fig 6
\begin{figure}
    \centering
    \begin{overpic}
		[width=10.5cm,tics=1]{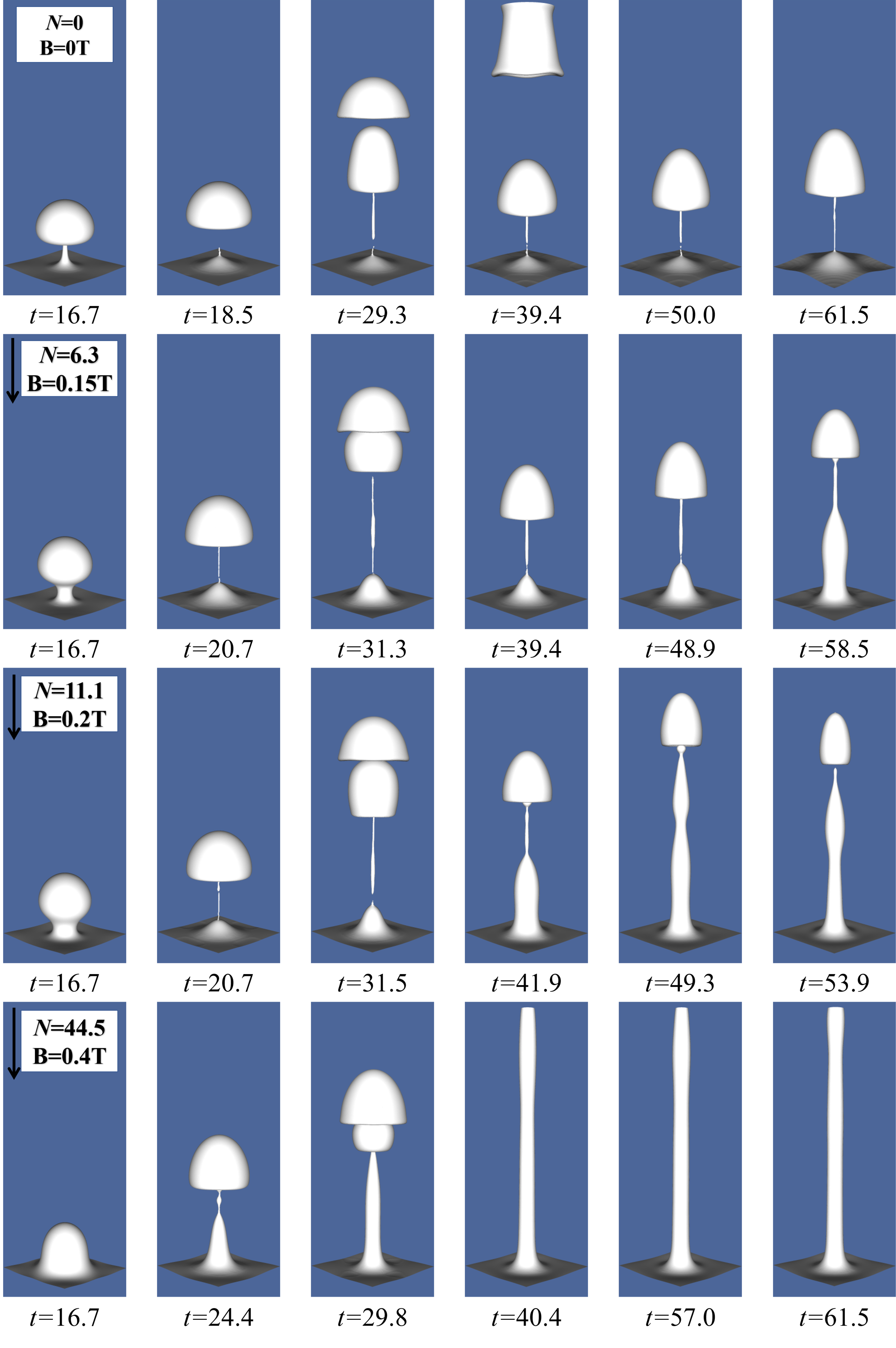}
		\put(-3.5,98.1){\large (a)}
		\put(-3.5,73.3){\large (b)}
		\put(-3.5,48.6){\large (c)}
		\put(-3.5,23.8){\large (d)}
	\end{overpic}    
    \caption{Temporal evolution of the bubble shapes during the film boiling flow without and with the vertical magnetic fields. (a) $N=0$ (B = 0 T), (b) $N=6.3$ (B = 0.15 T), (c) $N=11.1$ (B = 0.2 T), and  (d) $N=44.5$ (B = 0.4 T).}
\label{compare-ver-sin}
\end{figure}

We examine the influence of the vertical MF applied along the $x$ direction on film boiling flows. The horizontal directions perpendicular to the applied MF are represented by $y$ and $z$. Thus, it is essential to highlight that under the vertical MF, the MHD effect is expected to be isotropic in any horizontal plane parallel to $YOZ$. Consequently, the results in two vertical cross-sections, namely the $XOY$ and $XOZ$ planes, are the same. Therefore, we only illustrate findings for the $XOY$ cross-section. We vary the MF intensities with $\textrm{B =  0.05 T}$, $\textrm{B = 0.1 T}$, $\textrm{B = 0.15 T}$, $\textrm{B = 0.2 T}$ and $\textrm{B = 0.4 T}$. \ra{According to Eq. (\ref{nn}), the corresponding interaction numbers (Hartmann numbers) are $N=0.7$ ($Ha=1.9$), $N=2.8$ ($Ha=3.8$), $N=6.3$ ($Ha=5.7$), $N=11.1$ ($Ha=7.6$), and $N=44.5$ ($Ha=15.3$), respectively.} The MF intensity choice considers a spectrum ranging from inertia being predominant to the Lorentz force being dominant.

%Fig 7
\begin{figure}
	\centering
	\includegraphics[height=5.5cm]{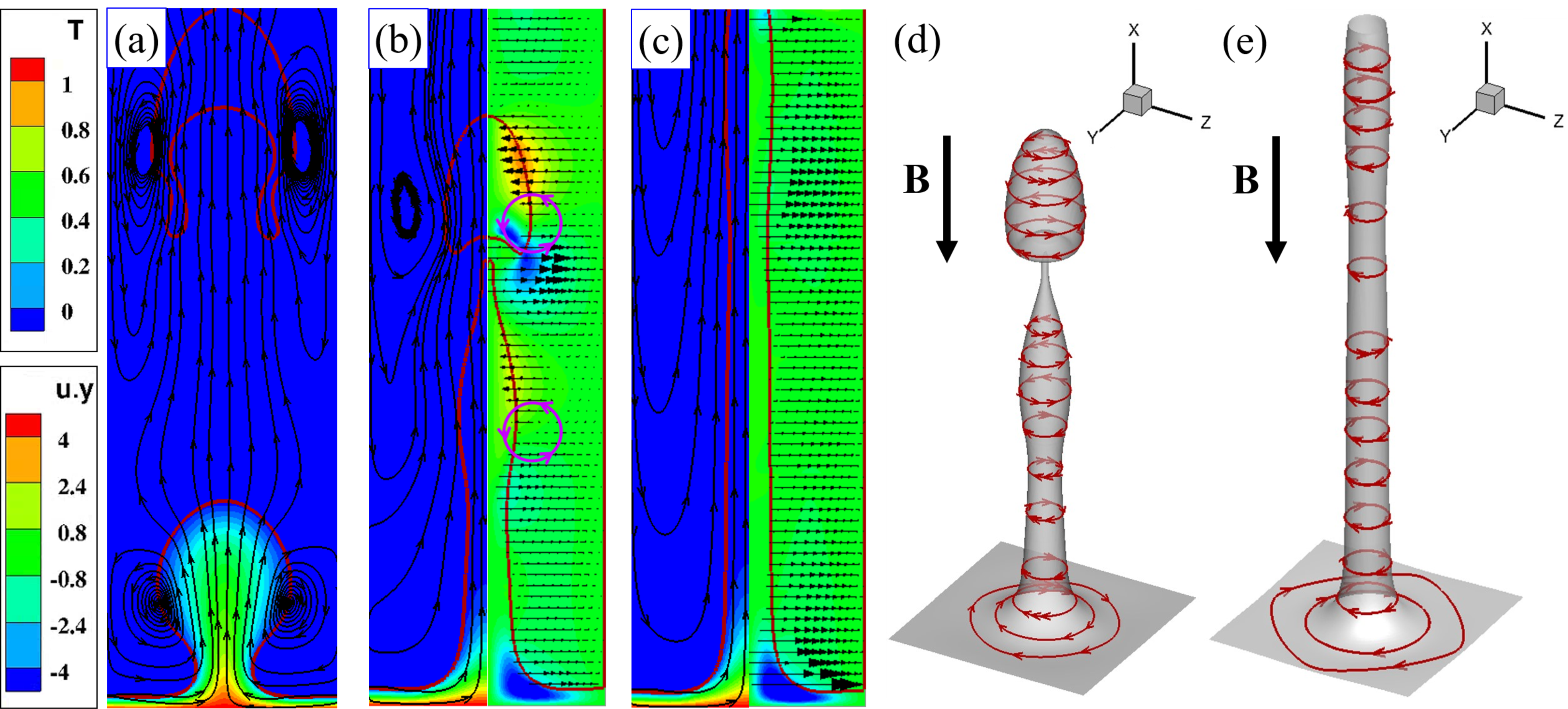}
    \caption{(a-c) Comparison of the temperature field and streamline patterns in the vertical cross-section ($XOY$) under different vertical magnetic fields. The right-hand side of panels (b) and (c) shows the comparison of the Lorentz force (black arrows) and horizontal velocity distribution ($u_y$, \add{the positive direction: from left to right}) (contours).  (d-e) The distribution of the current line. (a) $N=0$, $t=57.4$. (b) and (d) $N=11.1$, $t=53.7$. (c) and (e) $N=44.5$, $t=40.7$.}
\label{compare-ver-sin-streamline}
\end{figure}

To illustrate the MHD effect, we compare the liquid/vapor interface patterns within four bubble generation cycles, and the results are presented in Fig. \ref{compare-ver-sin}. From the first and second columns, it can be seen that the vertical MF inhibits the necking phenomenon and delays the detachment of the bubble. For instance, at $t=16.7$, the bubble is about to fall off when $N=0$, while the necking phenomenon does not appear when $N = 44.5$. Besides, as the MF strength increases, the periodic detachment of the bubble is gradually suppressed, and the vapor jet emerges. At $N=0$, the periodic detachment of the bubble caused by Rayleigh-Taylor instability can be observed, and the bubble detachment locations for each bubble cycle are almost the same. However, when $N=6.3$ and $N=11.1$, this phenomenon can only be observed in the previous short period. After the boiling flow completely develops, the vapor jet arises, and a small bubble is released from the top of the jet, corresponding to the onset of vapor jet instability.  Also, the bubble detachment location increases, implying that the MF suppresses the onset of Rayleigh-Taylor instability. When $N$ increases to $44.5$, the detachment of the bubble is barely observable, and boiling takes place in the form of a steady vapor column after the flow completely develops. In this stage, small amplitude surface waves can still be observed at the interface, indicating that the high-intensity MF further inhibits the occurrence of jet instability. 

To investigate the impact of a vertical MF on film boiling flow and its consequent influence on the boiling pattern, we initially compare the streamlines for cases $N=0$, $N=11.1$ and $N=44.5$, as depicted in Fig. \ref{compare-ver-sin-streamline}(a-c). It can be seen that with an increase in MF intensity, the suppression of the vortex structure, owing to the MHD effect, becomes more pronounced. In Fig. \ref{compare-ver-sin-streamline}(a), where $N=0$, vortices are observed both at locations where the bubble has detached and where the bubble is still connected to the thin film. However, in the left-half panel of Fig. \ref{compare-ver-sin-streamline}(b), where $N=11.1$, vortices are only present at the location where the bubble has detached. Vortices no longer appear when $N$ increases to 44.5, as depicted in the left-half panel of Fig. \ref{compare-ver-sin-streamline}(c). Additionally, close inspection of the bubble shapes reveals that the interface curvature at the bottom of the bulge decreases, and the necking phenomenon is thus weakened.

Now, we aim to elucidate the mechanisms underlying the aforementioned phenomenon. First, the distribution of the current line surrounding the liquid/vapor interface is plotted in Fig. \ref{compare-ver-sin-streamline}(d) and \ref{compare-ver-sin-streamline}(e). It can be seen that the current lines form closed loops in two cases, indicating that the computation of the induced current is conserved. The right-hand panels of Fig. \ref{compare-ver-sin-streamline}(b) and \ref{compare-ver-sin-streamline}(c) present the distribution of the Lorentz force (black arrows) and the horizontal component of velocity (contours, $u_y$ in the $XOY$ cross-section). It is essential to clarify that, due to a significantly large ratio of electric conductivities ($\sigma_{e, l}/\sigma_{e, v}$), the Lorentz forces in the vapor phase can be neglected so that they only manifest in the liquid phase. It is evident that the Lorentz force induced by the vertical MF acts in the opposite direction to $u_y$. Notably, under moderate MF intensity (Fig. \ref{compare-ver-sin-streamline}(b)), the Lorentz force near the upper part of the bubble points towards the inside of the bubble, while at the lower part, it points towards the outside. This results in a torque opposite to the direction of the local vortex, as indicated by the purple circle with arrows in the diagram. Consequently, the horizontal velocity is weakened, consistent with the color scale results in Fig. \ref{compare-ver-sin-streamline}(a-c). This explains the suppression of vortex formation, and such decay in vortices counteracts the pressure difference across the vapor-liquid interface. Thus, the surface tension, which balances this pressure difference in the form of $\sigma \kappa = p_{v}-p_{l}$, should be reduced correspondingly. The magnitude of surface tension is positively correlated with the curvature of the interface, and thus, the presence of the Lorentz force reduces interface curvature, making the bubble more challenging to pinch off. Simultaneously, due to the rapid evaporation of the liquid, the continuous generation of vapor converges towards the bulge, ultimately leading to the formation of a vapor jet under the influence of buoyancy. For vapor jets, bubble detachment is triggered by the amplification of instabilities and surface waves \citep{lin1}. Then, in the scenario with a large MF intensity ($N=44.5$, Fig. \ref{compare-ver-sin-streamline}(c)), the Lorentz forces entirely suppress the vortices and lead to uniformly distributed $u_y$, with the flow directing toward the center of the vapor column and upward flow. In this context, the Lorentz force also plays a role in suppressing the growth of surface waves while concurrently sustaining boiling in the form of a vapor column. In \S\ref{sec:remove MF}, an additional case based on the large MF intensity case $(N=44.5)$ will be further introduced to validate this finding.

%Fig 8
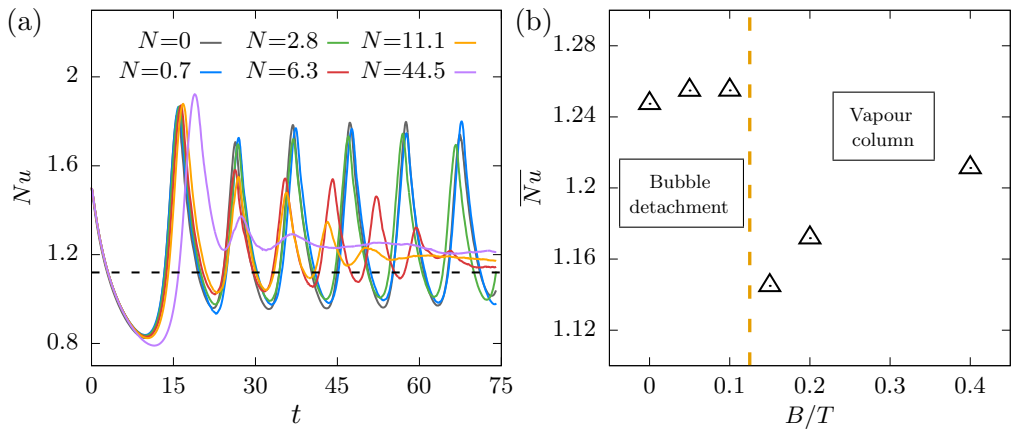
\begin{figure}
	\scalebox{1}{\input{figure/sin-ver-nu.tex}}
	\caption{(a) Temporal evolution of the space-averaged $Nu$ in the single-mode film boiling under different vertical MF intensities. The dashed line represents the theoretical results of \citet{Klimenko1}. (b) Variation of the space- and time-averaged $Nu$ number ($\overline{Nu}$) with vertical MF intensity in the case of the single-mode film boiling. The dimensional parameter $B$ is used on the horizontal axis to enhance clarity in presenting the value distribution.}
	\label{nu-ver-sin}	
\end{figure}

The temporal evolution of the space-averaged Nusselt number ($Nu$) under different vertical MFs is shown in Fig. \ref{nu-ver-sin}(a). It can be observed that as the vertical MF intensity increases, the $Nu$ transitions from approximately constant amplitude oscillation ($N \leq 2.8$) to damped oscillation ($N \geq 6.3$). This observation is consistent with the variations in the bubble detachment pattern under the influence of a vertical MF, suggesting that the vertical MF serves as a damping force, restraining bubble pinch-off behavior. Specifically, for cases with $N=6.3$, $N=11.1$, and $N=44.5$, the initially oscillated $Nu$ eventually stabilizes at a constant value. This stabilization occurs because, at these MF intensities, film boiling tends to maintain a stable jet form after the full development of boiling flow, as depicted in Fig. \ref{compare-ver-sin}. Bubble detachment occurs at locations far from the overheated wall or even without bubble detachment, keeping the interface near the overheated wall relatively stable. Given that, during saturated film boiling, the temperature field in the vapor phase exhibits a nearly linear distribution, the temperature distribution remains unchanged, as observed in Figs. \ref{compare-ver-sin} and \ref{compare-ver-sin-streamline}.

Fig. \ref{nu-ver-sin}(b) illustrates the space- and time-averaged Nusselt number ($\overline{Nu}$) calculated using Eq. (\ref{time-nu}) for different MF intensities. In cases of constant amplitude oscillation ($N=0$, $N=0.7$, and $N=2.8$), the $Nu$ is averaged over time from the second trough to the seventh trough in Fig. \ref{nu-ver-sin}(a), in order to eliminate the influence of the first bubble cycle. The final constant value is utilized in the case of damped oscillation, where $N=6.3$, $N=11.1$, and $N=44.5$. The dimensional parameter $B$ is used on the horizontal axis to enhance clarity in presenting the distribution of $\overline{Nu}$ under different MF intensities. The chart is demarcated with an orange dashed line based on the pattern of bubble detachment and $Nu$ variation. The left side of the dashed line represents the periodic bubble detachment area, where the $Nu$ exhibits constant amplitude oscillation. In this region, the vertical MF results in a slightly higher $\overline{Nu}$ than the case without a MF. On the right side of the dashed line is the vapor column boiling area, where the $Nu$ exhibits damped oscillation. The $\overline{Nu}$ sharply declines in the transition zone. After that, the $\overline{Nu}$ gradually increases with the increase in MF intensity.

%Fig 9
\begin{figure}
	\centering
    \includegraphics[width=13.5cm]{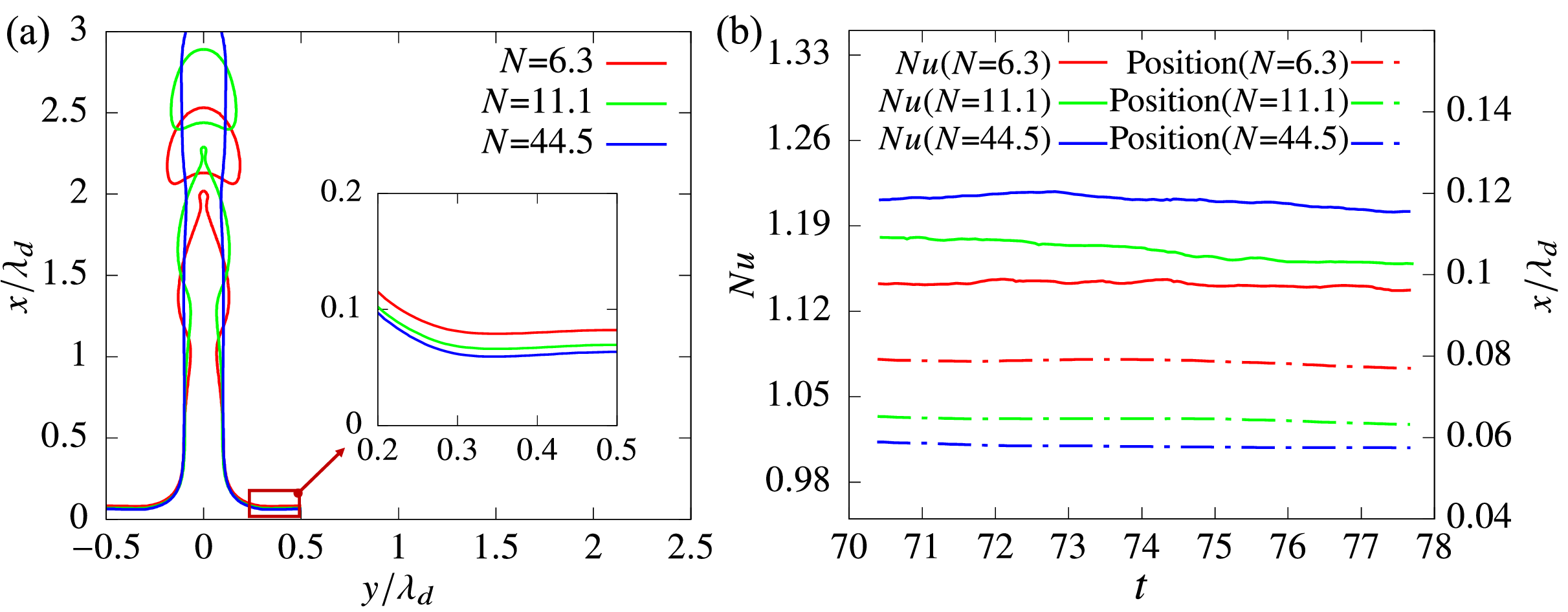}
	\caption{(a) Comparison of the bubble shapes at $t=70.4$ under three MF intensities ($N=6.3$, $N=11.1$, and $N=44.5$). (b) Temporal evolution of the space-averaged Nusselt number ($Nu$) and the minimum distance between the liquid/vapor interface and the overheated wall over time after boiling reaches a steady state under these three MF intensities.}
\label{ver-single-position}	
\end{figure}

%Fig 10
\begin{figure}
	\centering
    \includegraphics[width=13.5cm]{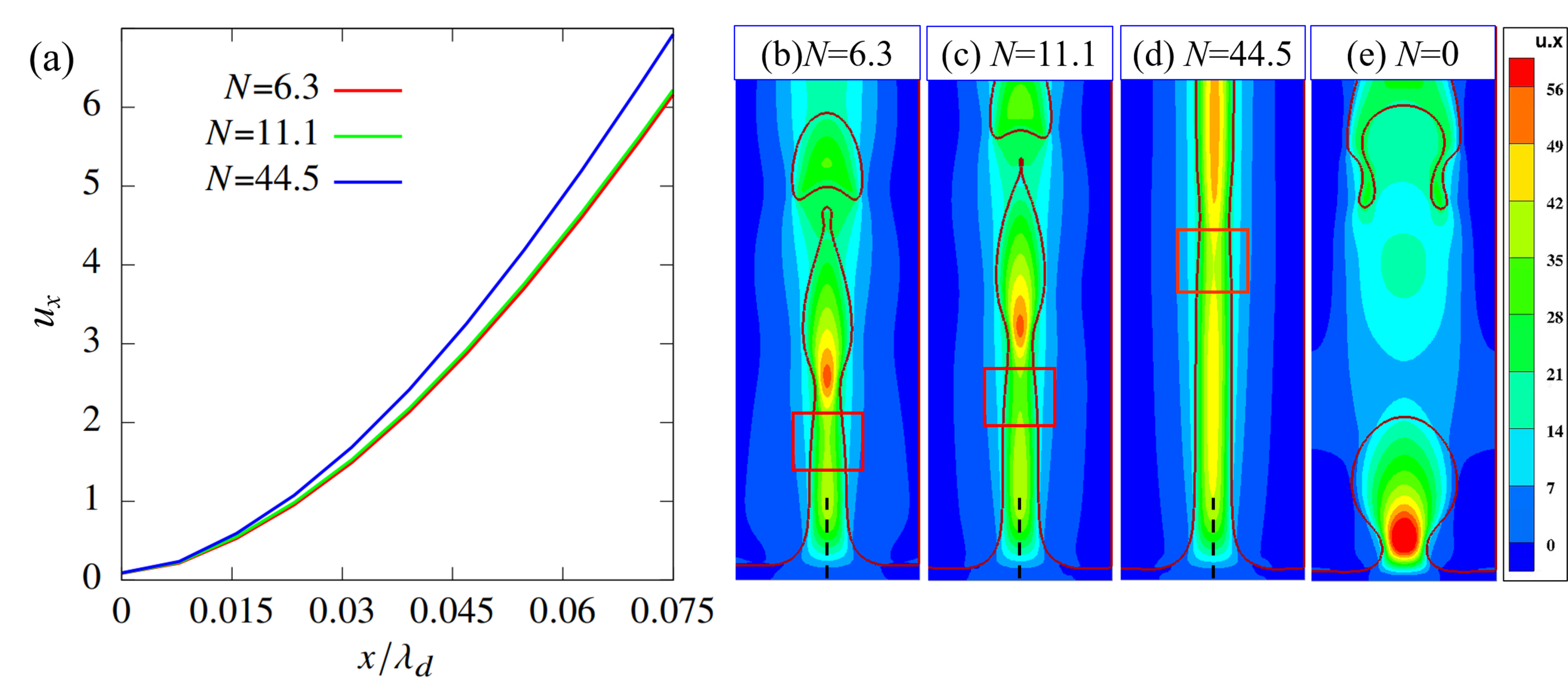}
    \caption{(a) Variation of the vertical component of velocity ($u_x$, \add{the positive direction: from bottom to top}) with dimensionless height ($x/\lambda_d$) on the axis line near the overheated wall (see black dashed line in panel b) under three different magnetic field intensities ($N=6.3$, $N=11.1$, and $N=44.5$) at $t=70.4$. Panels (b)-(d) show the distribution of $u_x$ for $N=6.3$, $N=11.1$, and $N=44.5$ at $t=70.4$, respectively. (e) The distribution of $u_x$ for $N=0$ at $t=57.4$.}
    \label{ver-single-ux}	
\end{figure}

To understand the reason behind the increase in the $\overline{Nu}$ with the rise in MF intensity in this region ($N \geq 6.3$), we compare the bubble shapes at $t=70.4$ under three different MF intensities, as shown in Fig. \ref{ver-single-position}(a). Note that in the same picture, the embedded panel illustrates the vapor interface in the vapor film region ($y > 0.2$),  and Fig. \ref{ver-single-position}(b) presents the temporal evolution of the minimum distance between the interface and the overheated wall, and the development of the corresponding $Nu$ is also depicted. It can be observed that the vapor thickness decreases with the increase in MF intensity, leading to an increase in the $Nu$ and $\overline{Nu}$. Now, the question arises: why does the vapor thickness decrease by generating a vapor jet as the MF intensity rises? Fig. \ref{ver-single-ux}(a) explains the reasons by plotting the variation of the vertical component of velocity $(u_x)$ with height on the axisymmetric line near the overheated wall (as shown by the black dashed line in Fig. \ref{ver-single-ux}(b-d)) under three different MF intensities ($N=6.3$, $N=11.1$, and $N=44.5$) at $t=70.4$. This figure shows that the greater the MF intensity, the higher the vertical velocity component near the overheated wall on the axisymmetric line. This allows vapor generated through evaporation to flow quickly from the near-wall region into the vapor jet, resulting in a smaller minimum distance between the vapor interface and the overheated wall. Analyzing the vertical component of the velocity contour ($u_x$) plot (Fig. \ref{ver-single-ux}(b-d)), it is observed that the vertical velocity is highest at the necking point of the jet, leading to a blocking effect that there is a sharp decrease in vertical velocity at its lower end (marked in the red box in the pictures). This decrease suppresses the process of upstream vapor entering the jet interior. Previous analyses have indicated that an increase in MF intensity suppresses jet instability, causing the necking point of the jet to move farther from the overheated wall. Therefore, the necking phenomenon has a weaker inhibitory effect on the velocity near the wall on the symmetry axis. In summary, when boiling occurs in the form of a jet, the location of bubble detachment influences the velocity at which vapor enters the center of the jet. The higher the bubble detachment position, the greater the vapor velocity in the jet, making vapor overflow more quickly. Thus, a higher MF intensity decreases the vapor film thickness near the overheated wall, promoting the $Nu$ and $\overline{Nu}$.

The abrupt cliff-like decrease in the $\overline{Nu}$ during the transition region in Fig. \ref{nu-ver-sin}(b) can be explained through the same analytical method. Fig. \ref{ver-single-ux}(e) presents the distribution of $u_x$ for $N=0$ at $t=57.4$, when the necking phenomenon caused by Rayleigh-Taylor instability is distinctly visible. When film boiling occurs in the form of a periodic bubble detachment, the location of bubble necking is near the vapor film and close to the overheated wall. Thus, the necking phenomenon brings about a sharp increase in vertical velocity, directly translating into an abrupt rise in vapor overflow velocity, as shown by the color scale in Fig. \ref{ver-single-ux}(e). This facilitates the rapid influx of vapor generated by evaporation into the bulge, significantly reducing the vapor film thickness during the bubble necking stage. As a result, the $Nu$ increases, enhancing the $\overline{Nu}$. %This also explains the temperature field distribution observed in Fig. \ref{tem-b0} and Fig. \ref{compare-ver-sin-streamline}(a). During the bubble-necking stage, the rapid influx of vapor from the bottom leads to comparatively higher temperatures within the bulge.

\subsubsection{Effect of the horizontal MF}\label{single-horizontal MF}

%Fig 11
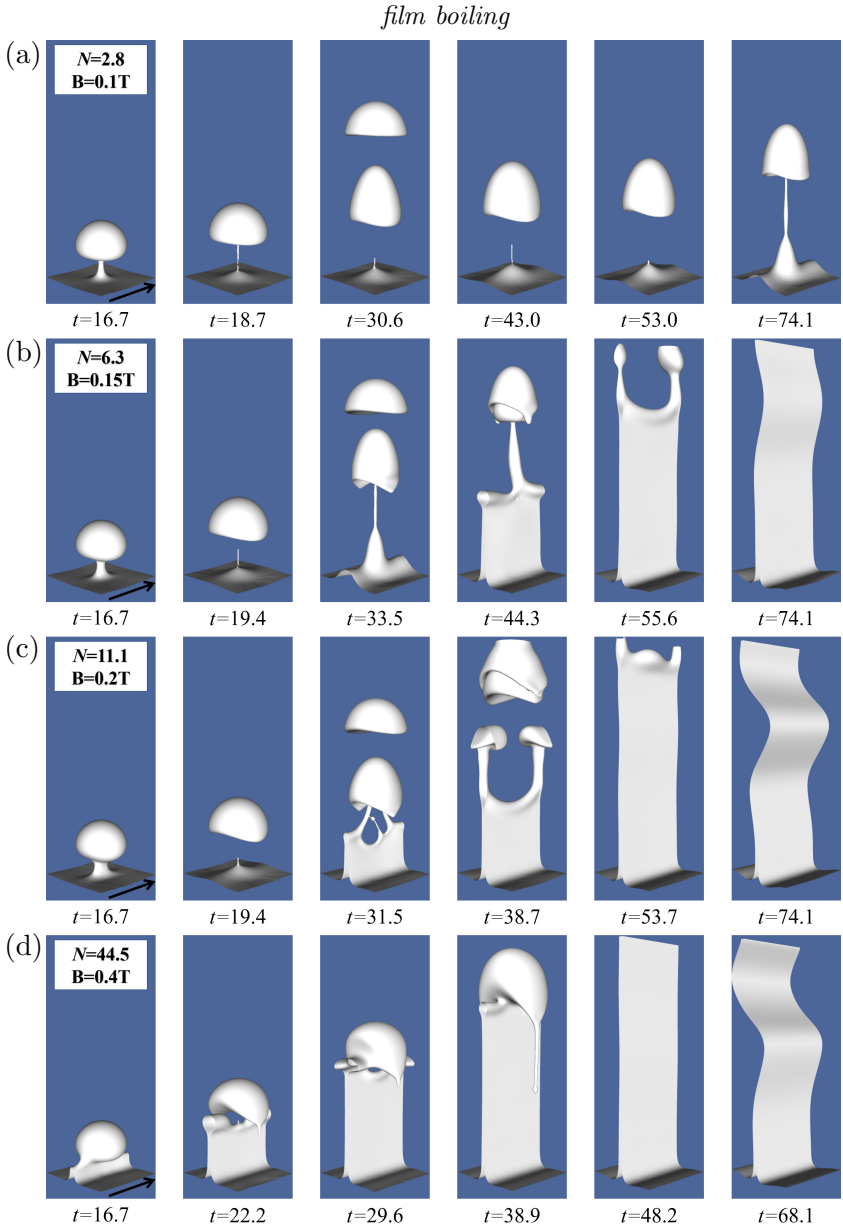
\begin{figure}
    \centering
    \begin{overpic}
       [width=10.5cm,tics=1]{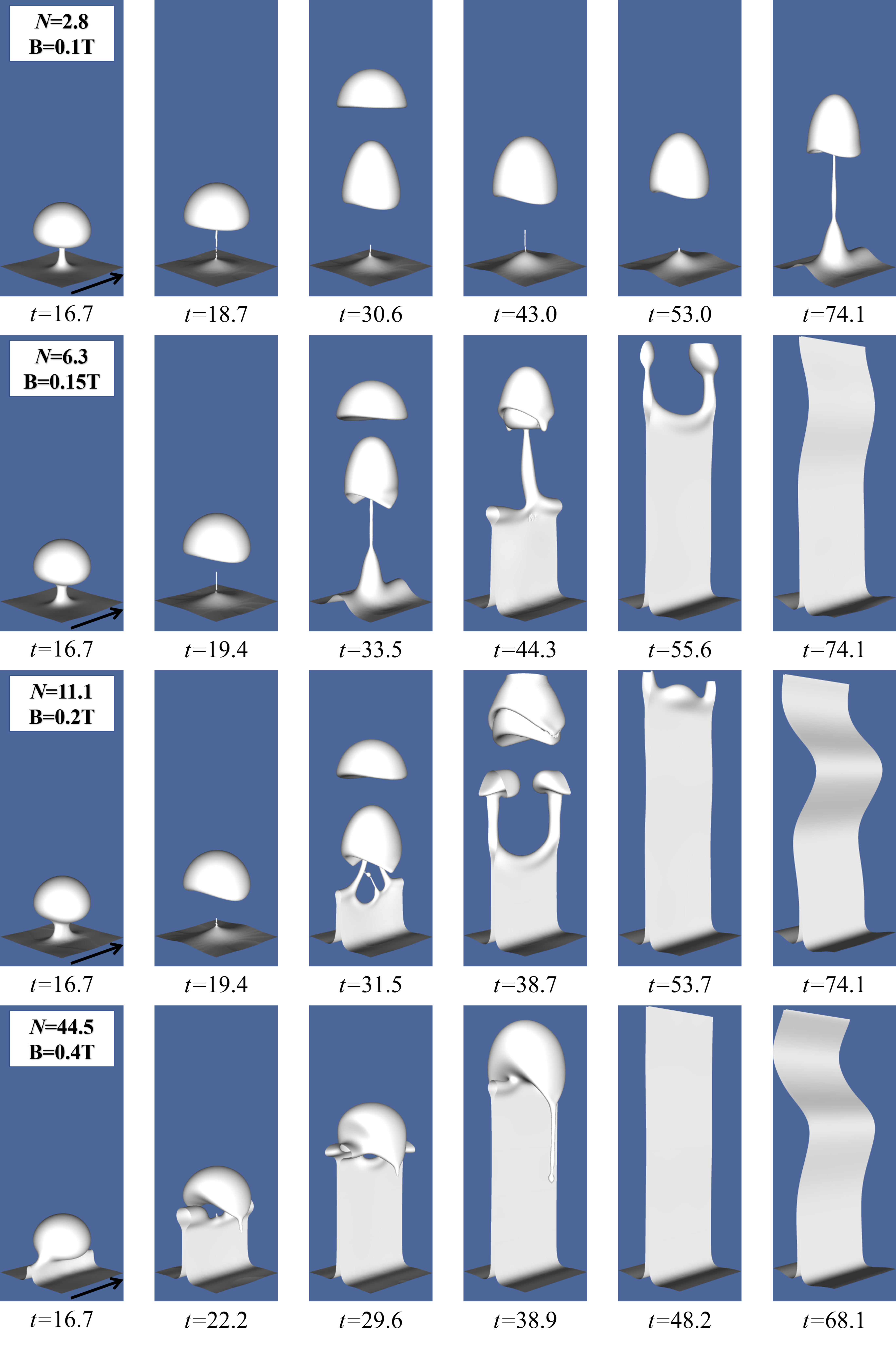}
		\put(-3.5,98.1){\large (a)}
		\put(-3.5,73.3){\large (b)}
		\put(-3.5,48.6){\large (c)}
		\put(-3.5,23.8){\large (d)}
	\end{overpic}  
    \caption{Temporal evolution of the vapor bubble shapes during the film boiling flow without and with the horizontal MFs. (a) $N=2.8$ $\textrm{(B = 0.1 T)}$, (b) $N=6.3$ $\textrm{(B = 0.15 T)}$, (c) $N=11.1$ $\textrm{(B = 0.2 T)}$ and (d) $N=44.5$ $\textrm{(B = 0.4 T)}$.}
\label{compare-hor-sin}
\end{figure}

Next, we investigate the effect of a horizontal MF on the single-mode boiling configuration. Here, the MF is applied in the $y$ direction. Thus, the vertical $XOY$ plane is parallel to the MF, while the vertical $XOZ$ plane is perpendicular to the MF. The choice of MF strengths is the same as those used in the case of vertical MF configuration, namely $N=0.7$ $\textrm{(B = 0.05 T)}$, $N=2.8$ $\textrm{(B = 0.1 T)}$, $N=6.3$ $\textrm{(B = 0.15 T)}$, $N=11.1$ $\textrm{(B = 0.2 T)}$ and $N=44.5$ $\textrm{(B = 0.4 T)}$.

Fig. \ref{compare-hor-sin} depicts the evolution of the liquid/vapor interface patterns under different horizontal MF intensities. Anisotropy is observed in the boiling flow when a horizontal MF is applied. For instance, in the latter part of the process ($t > 30.6$) for $N=2.8$, columnar bulges out at the thin film grow along the MF direction, while the detaching bubble is also non-axisymmetric. With further increases in the MF intensity ($N=6.3$, $N=11.1$, and $N=44.5$), this characteristic manifests much earlier that a two-dimensional boiling flow forms and the vapor features as a thin sheet along the direction of the MF after complete development. Furthermore, after the vapor sheet grows to a certain height (see those scenarios at $t = 74.1$), the amplification of minor disturbances can lead to the instability of the vapor sheet. In the study of sheet instability \citep{hagerty1}, two modes of instability for sheets have been identified: the sinuous mode and the varicose mode. From Fig. \ref{compare-hor-sin}, it is evident that the flow corresponds to the sinuous mode.

%Fig 12
\begin{figure}
	\centering
    \includegraphics[height=5.3cm]{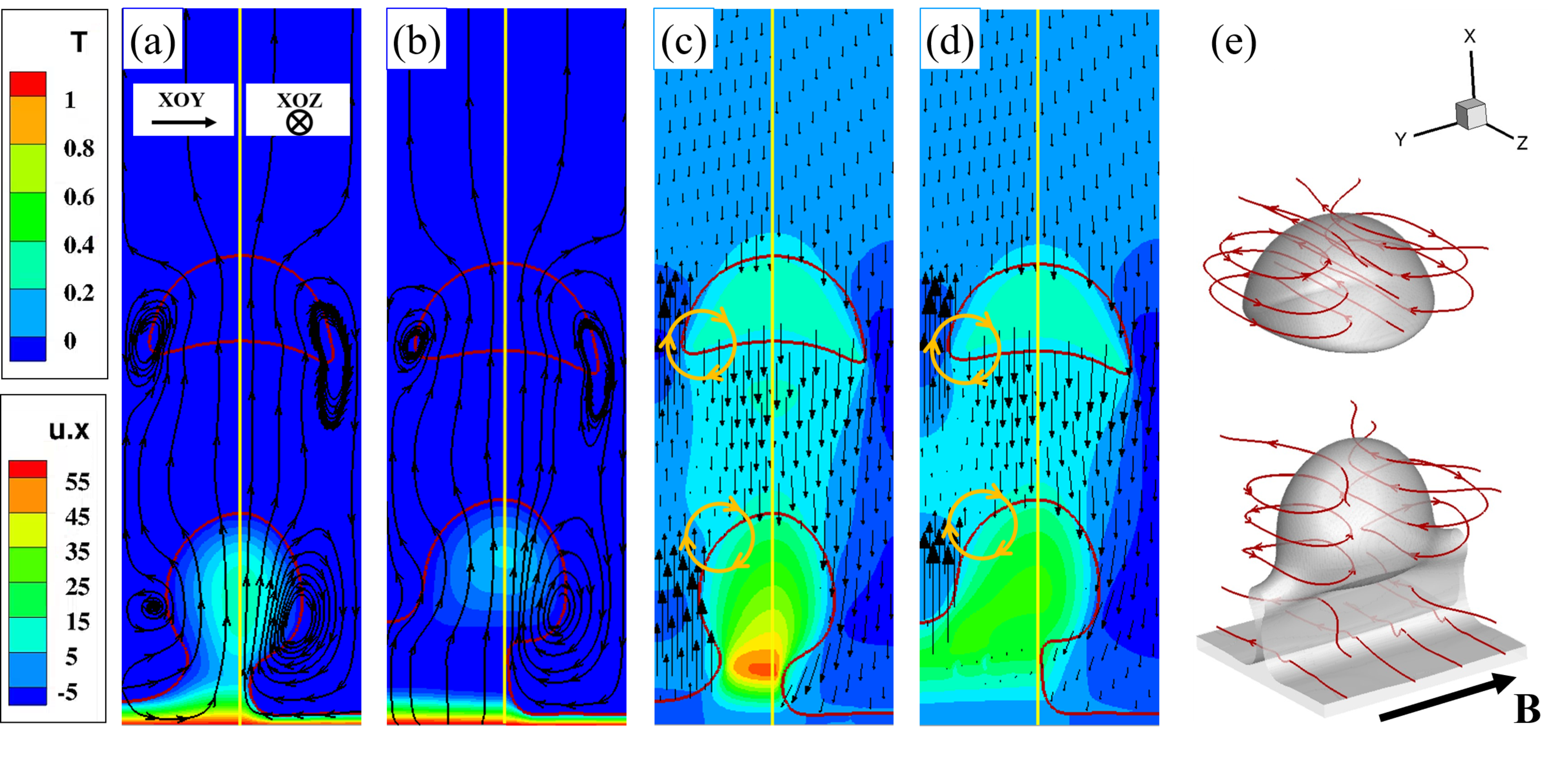}
    \caption{(a-b) Comparison of temperature field and streamline patterns in two vertical cross-sections ($XOY$: parallel to the magnetic field, on the left-hand side; $XOZ$: perpendicular to the magnetic field, on the right-hand side) under different horizontal magnetic field intensities. (c-d) Comparison of the Lorentz forces (black arrows) and the distribution of the vertical component of velocity ($u_x$, contours) corresponding to panels (a-b). (e) Distribution of the current lines. (a) and (c) $N=6.3$, $t=27.8$. (b), (d), and (e) $N=11.1$, $t=27.8$.}
\label{compare-sin-hor}
\end{figure}

To understand those scenarios caused by the horizontal MFs, we examine the temperature and streamline distribution in two vertical cross-sections: the $XOY$ plane parallel to the MF and the $XOZ$ plane perpendicular to the MF. These distributions are plotted for two different MF intensities ($N=6.3$ and $N=11.1$) at $t=27.8$, as illustrated in Fig. \ref{compare-sin-hor}(a) and \ref{compare-sin-hor}(b), respectively. The corresponding distribution of Lorentz forces and vertical velocity component ($u_x$) are presented in Fig. \ref{compare-sin-hor}(c-d). Fig. \ref{compare-sin-hor}(e) shows the distribution of current lines under the horizontal MF. It can be observed that the current lines are not closed. This occurs because we set periodic boundary conditions around the computational domain, causing the current lines to pass through the boundaries. By comparing the velocity streamline distribution (see Fig. \ref{compare-sin-hor}(a) and \ref{compare-sin-hor}(b)), it becomes evident that the horizontal MF primarily influences the flow vortices in the direction perpendicular to the MF, which suppresses the vortex on the side of the bubble near the wall. When $N$ increases to 11.1, the vortex completely disappears, and no downward flow is observed on the $XOY$ plane. Then, the distribution of Lorentz force, as displayed in Fig. \ref{compare-sin-hor}(c) and \ref{compare-sin-hor}(d), explains the disappearance of the vortex perpendicular to the MF. From left-half panels of Fig. \ref{compare-sin-hor}(c) and \ref{compare-sin-hor}(d), it is observed that in the $XOY$ plane, the direction of the Lorentz force is upward on the side of the bubble near the wall, while the Lorentz force is directed vertically downward at the top of the bubble. This results in a torque opposite to the direction of the local vortex, as indicated by the orange circle with arrows in the diagrams. So, the horizontal MF tends to suppress vortices perpendicular to the MF. As a result, vortices gradually disappear, and the necking phenomenon weakens, causing the interface to move away from the overheated wall under the influence of continuously generated vapor. Eventually, a columnar bulge along the MF direction is formed. In contrast, for the $XOZ$ plane perpendicular to the MF (see right-half panels of Fig. \ref{compare-sin-hor}(c) and \ref{compare-sin-hor}(d)), the direction of the Lorentz force is uniformly vertical downward, and the flow field structure does not exhibit significant changes with variations in MF intensity. Thus, the necking phenomenon still occurs in this plane, forming a vapor sheet. These conclusions align with the theoretical prediction of MHD effects\citep{sommeria1, davidson1}, demonstrating that the Lorentz force tends to decrease the velocity gradient parallel to the MF to reduce Joule dissipation. Furthermore, the temperature distribution depicted in Fig. \ref{compare-sin-hor}(a) and \ref{compare-sin-hor}(b) exhibits a pattern similar to that of the vertical MF cases. It can be seen that as the MF intensity increases, the temperature inside the bulge becomes lower. This trend is in line with the distribution of the vertical component of velocity $(u_x)$ shown in Fig. \ref{compare-sin-hor}(c) and \ref{compare-sin-hor}(d), and the underlying reason is analogous to the vertical MF situation. Specifically, with the increase in MF intensity, the weakening of the necking phenomenon leads to a reduction in the velocity of vapor entering the bulge, making heat transfer more challenging. Consequently, the internal temperature becomes lower.

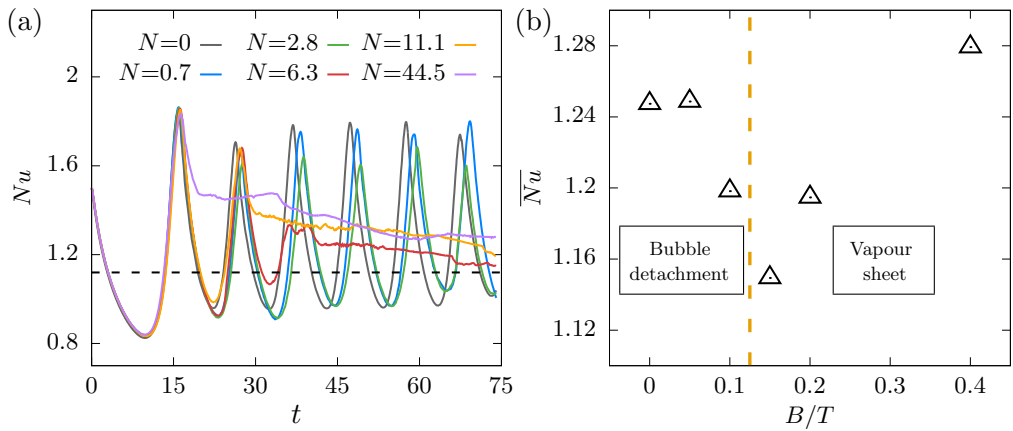
\begin{figure}
	\centering
	\scalebox{1}{\input{figure/sin-hor-nu.tex}}
	\caption{(a) Temporal evolution of the space-averaged $Nu$ under different horizontal MF intensities in single-mode film boiling. The dashed line represents the theoretical results of \citet{Klimenko1}. (b) Variation of the space and time averaged $Nu$ number ($\overline{Nu}$) with horizontal MF intensity in the case of the single-mode film boiling. The dimensional parameter B is used on the horizontal axis to enhance clarity.}
	\label{nu-hor-sin}	
\end{figure}

Fig. \ref{nu-hor-sin}(a) presents the temporal evolution of the $Nu$ under varying horizontal MF intensities. At lower MF intensities $(N<6.3)$, the $Nu$ variation demonstrates approximate periodic oscillations, aligning with the periodic detachment of bubbles. Notably, when $N=2.8$, a considerable reduction in amplitude occurs, attributed to the initiation of the horizontal MF effect. This effect induces the formation of a columnar bulge along the MF direction at the vapor film's bottom, leading to an increased vapor layer thickness and subsequently causing a decline in $Nu$. As the MF intensity increases $(N \geq 6.3)$, $Nu$ undergoes a brief period of periodic oscillations in the initial stages due to bubble detachment. However, once a stable vapor sheet is established, a transition to a more stable state occurs, and higher MF intensities accelerate this transition.

%Fig 13

%Fig 14
\begin{figure}
	\centering
    \includegraphics[width=13.5cm]{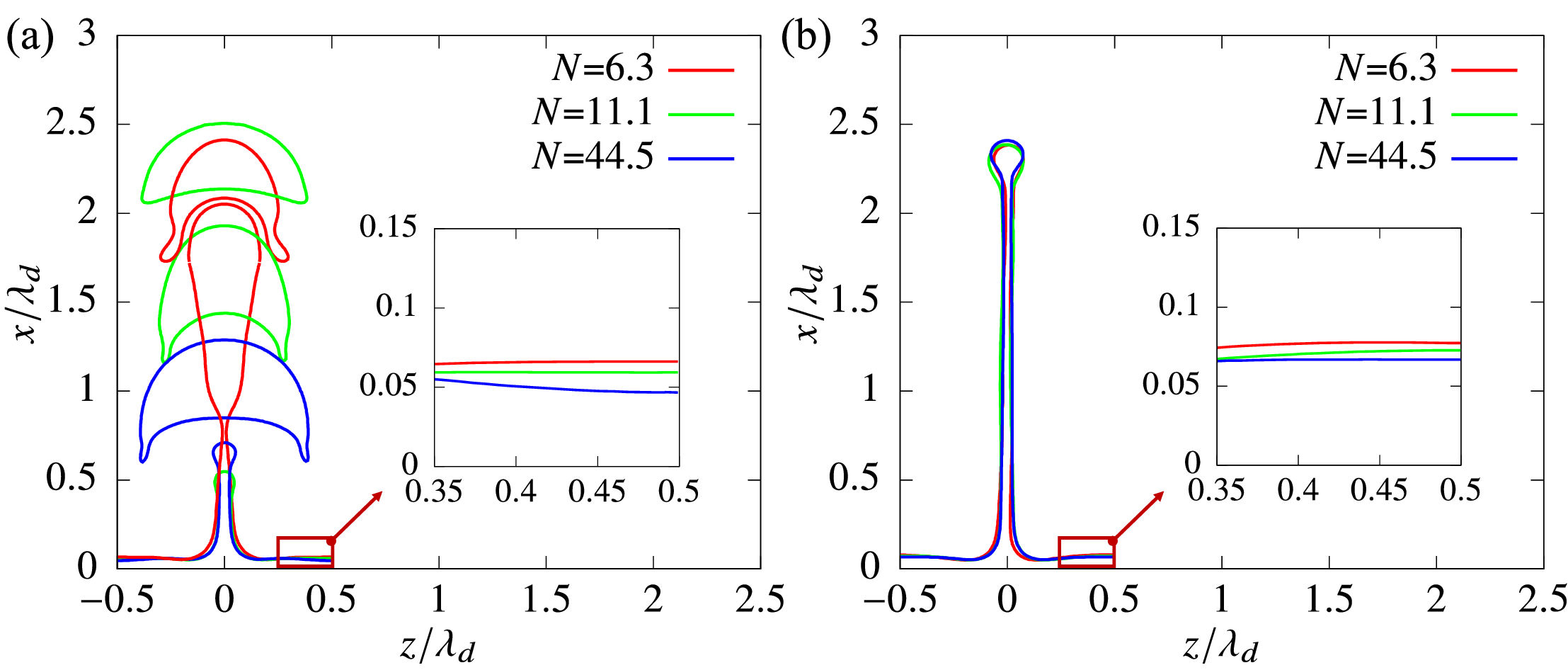}
    \caption{Comparison of the bubble shapes in the vertical cross-section ($XOZ$ plane) perpendicular to the magnetic field under three different MF intensities ($N=6.3$, $N=11.1$, and $N=44.5$). In panel (a), the red, green, and blue lines depict the results for $N=6.3$ at $t=42.6$, $N=11.1$ at $t=33.3$, and $N=44.5$ at $t=22.2$, respectively. In panel (b), the red, green, and blue lines depict the results for $N=6.3$ at $t=57.4$, $N=11.1$ at $t=50.0$, and $N=44.5$ at $t=40.7$, respectively.}
\label{hor-single-position}	
\end{figure}

%Fig 15 
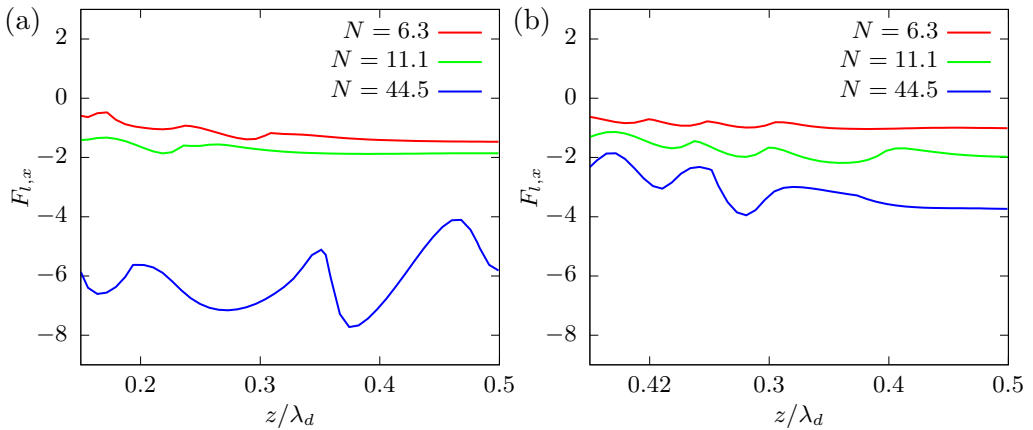
\begin{figure}
	\centering
    \scalebox{1}{\input{figure/single-hor-lf-compare.tex}}
    \caption{Comparison of the vertical component of the Lorentz force ($F_{l,x}$) on the liquid/vapor interface under different MF intensities corresponding to Fig. \ref{hor-single-position}. In panel (a), the red, green, and blue lines depict the results for $N=6.3$ at $t=42.6$, $N=11.1$ at $t=33.3$, and $N=44.5$ at $t=22.2$, respectively. In panel (b), the red, green, and blue lines depict the results for $N=6.3$ at $t=57.4$, $N=11.1$ at $t=50.0$, and $N=44.5$ at $t=40.7$, respectively.}
\label{hor-single-flx1}	
\end{figure}

The $\overline{Nu}$ for the aforementioned cases is plotted in Fig. \ref{nu-hor-sin}(b). Similar to the vertical MF, for small MF intensity cases $(N<6.3)$, the average of the $Nu$ from the second trough to the seventh trough in Fig. \ref{nu-hor-sin}(a) is selected, while for large MF intensity cases, the final value is employed. Based on the final character of the film boiling, the chart is divided into two regions. The left side represents the periodic detachment region of the bubble. In this region, the $\overline{Nu}$ experiences a sharp decrease after the MF reaches a certain intensity. This is consistent with the previously mentioned reduction in the amplitude of the $Nu$, both attributed to the appearance of the elongated bulge in the vapor film. The right side is the vapor sheet boiling region, where the $\overline{Nu}$ increases as the MF intensity increases. To investigate the reason, we compare the shapes of the interface in the $XOZ$ plane at the initial formation of the vapor sheet (Fig. \ref{hor-single-position}(a)) and when the vapor sheet stably exists (Fig. \ref{hor-single-position}(b)) under these three different MF intensities. It is clear that there is a significant difference in the thickness of the thin film near the overheated wall, particularly away from the vapor sheet side, as indicated by the red square in the diagram. The higher the MF strength, the smaller the thickness, resulting in a larger $Nu$ and $\overline{Nu}$. From Fig. \ref{compare-sin-hor}(c) and \ref{compare-sin-hor}(d), we know that in the $XOZ$ plane, the Lorentz force is consistently directed downward. Therefore, we hypothesize that the Lorentz force might suppress the expansion of the vapor film near the overheated wall. To verify this hypothesis, we output the vertical component of the Lorentz force ($F_{l,x}$) on the near wall interface corresponding to Fig. \ref{hor-single-position}, as depicted in Fig. \ref{hor-single-flx1}. It can be observed that as the MF intensity increases, the downward effect of the Lorentz force becomes more pronounced. In other words, the suppression of the expansion of the vapor film near the overheated wall is stronger, resulting in a smaller thickness of the vapor film and a larger $\overline{Nu}$.

\rb{After examining the MHD effect on the growth of vapor bubbles, it is useful to compare our findings with those of previous studies. The most relevant work is by \cite{chandrasekhar1}, who used linear stability analysis to investigate the effects of vertical and horizontal magnetic fields on Rayleigh-Taylor instability in inviscid interfacial flows. Specifically, the exponential growth rate $n$ of disturbances for a flat interface between inviscid fluids in ideal MHD, with a vertical MF applied to the interface, is given by:
$n^{3}+n^{2} k B_{0}\left(\sqrt{\rho_l/\rho_0}+\sqrt{\rho_v/\rho_0}\right)+n\left(k^{2} B_{0}^{2}-(\rho_l-\rho_v)/(\rho_l+\rho_v) g k\right)=2 (\rho_l-\rho_v)/(\rho_l+\rho_v) g k^{2} B_{0}/(\sqrt{\rho_l/\rho_0}+\sqrt{\rho_v/\rho_0})$, where $g$ is the gravitational acceleration, $k$ is the wavenumber, $B_0 = B / \sqrt{\rho_0 \mu_0}$, $\rho_0 = (\rho_l + \rho_v) / 2$, and $\mu_0$ is the vacuum magnetic permeability. Notably, there is no critical wavelength, so all modes are unstable, but a larger $B_0$ enhances the stability of the interface. This aligns with our results, which show that the vapor bubble grows more slowly under a stronger vertical MF. However, the RT instability in boiling flow cannot be completely eliminated, even with a very strong MF. For the horizontal MF parallel to the interface, linear stability analysis also reveals anisotropic effects induced by the external MF. Horizontal disturbances transverse to the $\mathbf{B}$-lines are unaffected by the horizontal MF, while disturbances parallel to the $\mathbf{B}$-lines have an exponential growth rate $n$ given by:
$n^2=gk(\rho_l-\rho_v)/(\rho_l+\rho_v)-k^2{B_0}^2$,
with $k$ denoting the wavenumber of the disturbances in that direction. This shows that a sufficiently strong MF can suppress Rayleigh-Taylor instability in the direction parallel to the $\mathbf{B}$-lines but has little impact in the direction transverse to the $\mathbf{B}$-lines, consistent with our numerical results. We also acknowledge the importance of future studies addressing these challenges by incorporating a phase change model into the framework of linear stability analysis for a more comprehensive understanding of boiling coupled RT instability under MHD conditions.}

\subsubsection{What happens after MF removal?}\label{sec:remove MF}

To further understand the MHD effect on film boiling, in this section, we investigate what happens when the MF is suddenly removed after film boiling reaches a steady state under MF (vapor column boiling for vertical MF scenario and vapor sheet boiling for horizontal MF scenario).

%Fig 16
\begin{figure}
	\centering
    \includegraphics[height=5.3cm]{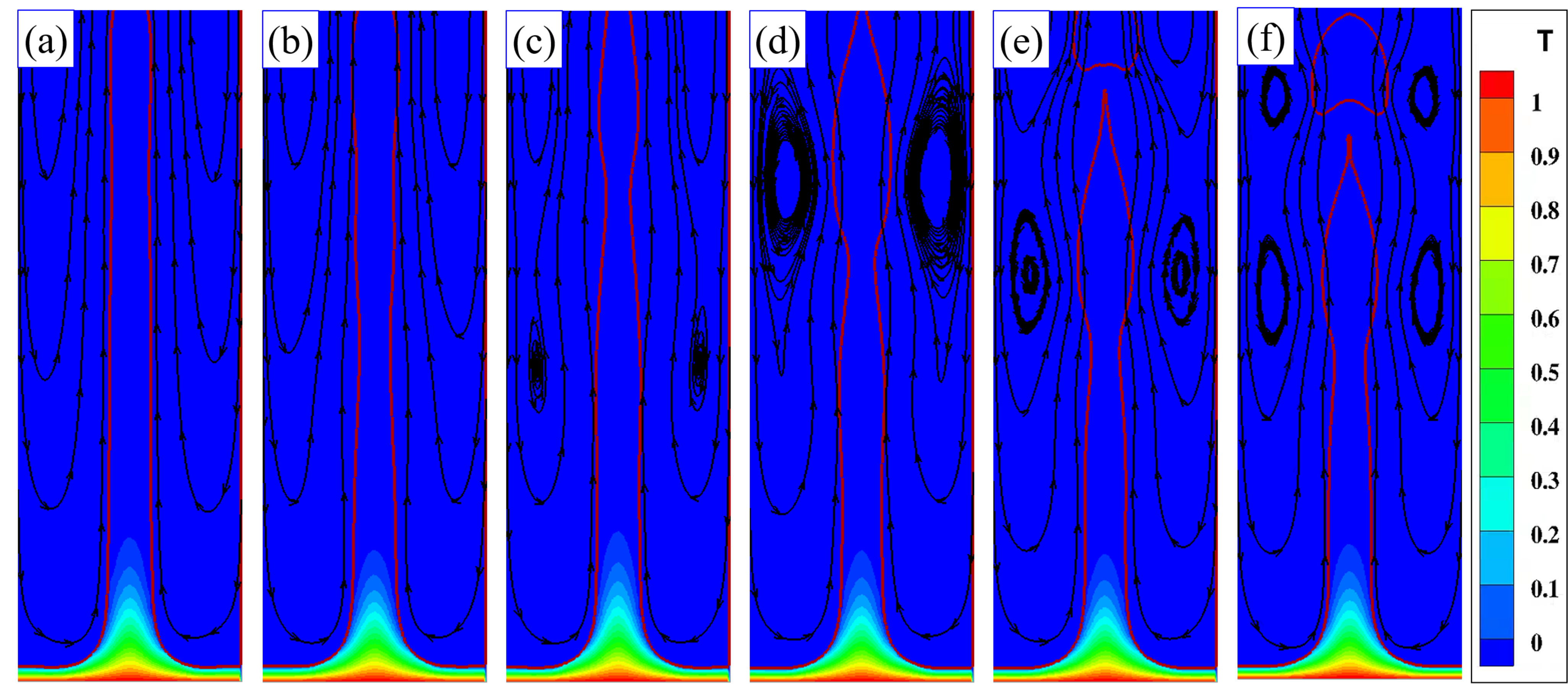}
	\caption{Temporal evolution of the temperature profile (color contours) and velocity streamline pattern (black arrow lines) in the vertical cross-section ($XOY$) after removing the magnetic field at $t=44.4$ for $N=44.5$. (a) $t=44.4$, (b) $t=50.0$, (c) $t=51.9$, (d) $t=54.6$, and (e) $t=56.5$ and $t=59.3$. It can be seen that the removal of the magnetic field leads to vapor jet instability, resulting in the reappearance of bubble detachment.}
	\label{compare-ver-b04-0}
\end{figure}

%Fig 17
\begin{figure}
    \centering
    \scalebox{1}{\input{figure/single-ver-remove-MF.tex}}
	\caption{Comparison of the temporal evolution of (a) the space-averaged Nusselt number ($Nu$) and (b) the evaporation rate ($\dot{m} S_{\Gamma}$) with the vertical MF and after the removal of the MF at $t=44.4$ for $N=44.5$. Note that the oscillation of the curves, corresponding to the magnetic field removal, is due to the reappearance of vapor bubble detachment.}
	\label{ver-remove-nu-rate}	
\end{figure}
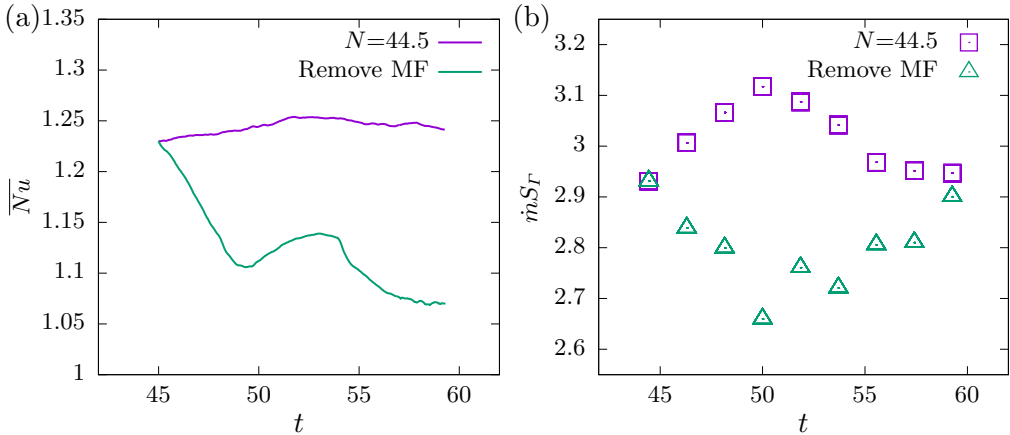

Initially, we examine the scenario based on the vertical MF with large intensity $(N=44.5)$. The MF is suddenly removed when the film boiling reaches a relatively stable state $(t=44.4)$. The results are depicted in Fig. \ref{compare-ver-b04-0}. It is observed that, after the removal of the MF, the amplitude of surface waves progressively increases, and vortices appear, ultimately leading to the detachment of small bubbles. This observation supports the findings discussed in \S\ref{single-vertical MF} that the Lorentz force plays a vital role in suppressing the growth of surface waves and the subsequent detachment of the vapor bubble. The comparison of the temporal evolution of the $Nu$ with the MF and upon its removal is illustrated in Fig. \ref{ver-remove-nu-rate}(a). After removing the vertical MF, a noticeable decrease in the $Nu$ can be observed. This is attributed to the fact that, upon the removal of the vertical MF, the stable columnar boiling pattern can no longer be sustained, leading to the detachment of bubbles at the top of the vapor column (see Fig. \ref{compare-ver-b04-0}). As analyzed in \S\ref{single-vertical MF}, the bubble detachment behavior inhibits the vapor influx into the columnar bulge, causing the near wall vapor film to thicken and reducing $Nu$. Fig. \ref{ver-remove-nu-rate}(b) compares the evaporation rate (here, we measure the evaporation rate using the mass flow at the interface ($\dot{m} S_{\Gamma}$)) between the cases, revealing a significant decrease in the evaporation rate after the removal of the MF, which is also caused by the increase in thickness of the near wall vapor film. Note that the oscillation of $\dot{m} S_{\Gamma}$ after removing the MF is caused by the reappeared detachment of the vapor bubble. The change in evaporation rate provides another perspective on the reason for bubble detachment. After removing the MF, the decrease in evaporation rate prevents film boiling from supplying sufficient vapor for stable columnar bulge.

%Fig 18
\begin{figure}
    \centering
    \includegraphics[height=5.3cm]{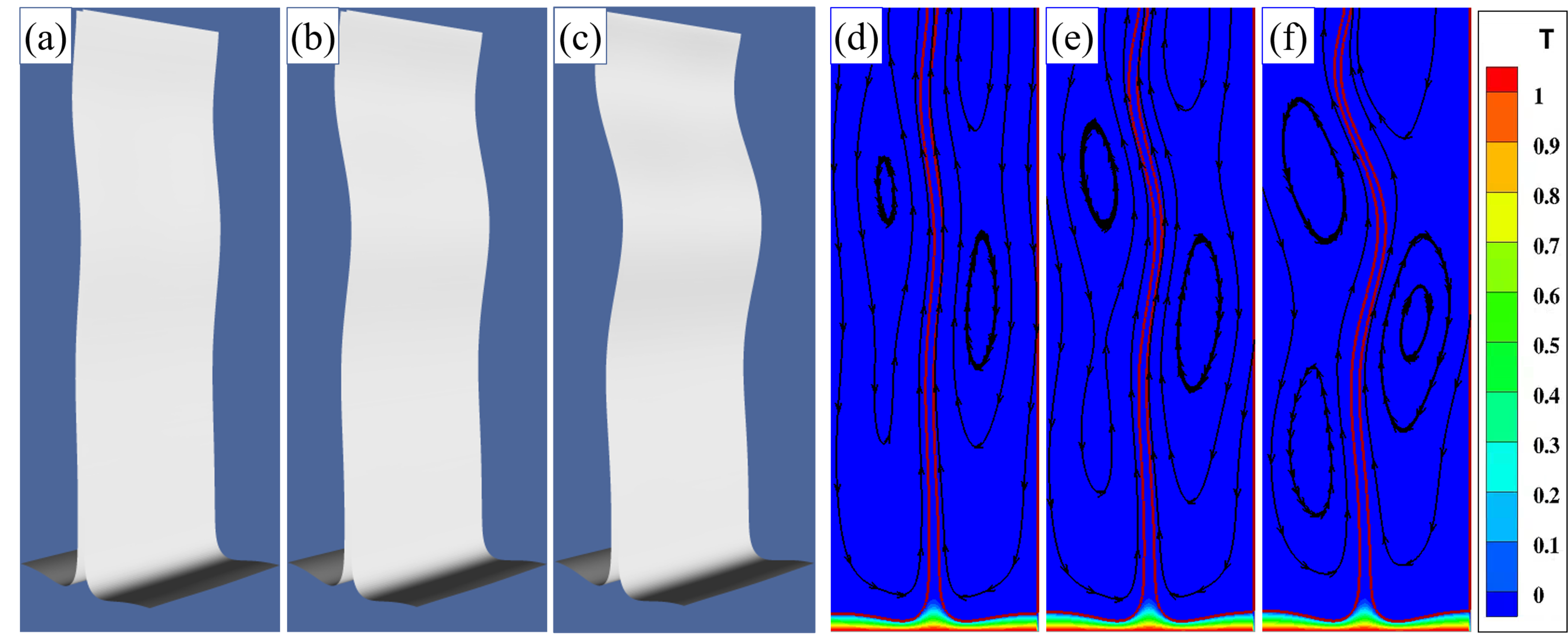}
    \caption{(a-c) The temporal evolution of the liquid/vapor interface after removing the horizontal MF at $t=44.4$ for $N=44.5$. (d-f) Temperature profile (color contours) and velocity streamline pattern (black arrow lines) in the vertical cross-section ($XOZ$) corresponding to panels (a-c). Panels (a,d), (b,e), and (c,f) correspond to $t=59.3$, $t=63.0$, and $t=66.7$, respectively.
    }
    \label{remove-hor-MF-shape}
\end{figure}

%Fig 19
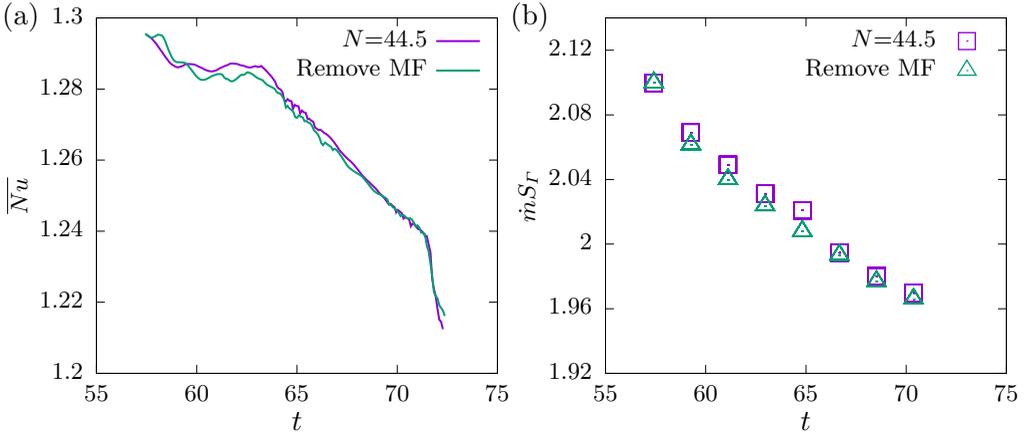
\begin{figure}
	\centering
   \scalebox{1}{\input{figure/single-hor-remove-MF.tex}}
	\caption{Comparison of the temporal evolution of (a) the space-averaged Nusselt number ($Nu$) and (b) the evaporation rate ($\dot{m} S_{\Gamma}$) with the horizontal MF and after the removal of the magnetic field at $t=57.4$ for $N=22.2$. We observe that in both plots, the values are almost identical before and after removing the horizontal magnetic field.}
	\label{hor-remove-nu-rate}	
\end{figure}

Now, we examine the scenario based on a horizontal MF with the intensity of $N=22.2$. When film boiling reaches a relatively stable vapor sheet boiling at $t=57.4$, the MF is suddenly removed. However, intriguingly, we observe a very interesting result: film boiling in the vapor sheet state does not appear to be affected after removing the horizontal MF, as depicted in Fig. \ref{remove-hor-MF-shape}. Furthermore, by comparing the $Nu$ and evaporation rate (Fig. \ref{hor-remove-nu-rate}), it is observed that with and without the presence of a horizontal MF, both $Nu$ and evaporation rate are almost identical. This indicates that the vapor sheet formed under horizontal MF is initially stable, in contrast to the vapor column generated under vertical MF. To explain this, we should remember that the vertical MF eliminates the vortices around the vapor bubble in an axisymmetrical manner, and such a controlling effect leads to the generation of an isotropic vapor column. Therefore, the axisymmetrical Lorentz forces, directing radial-wards, maintain the force balance around the vapor column. Consequently, if the Lorentz forces are removed to break such force balance, the downward flow must develop due to any disturbance that emerges on the vapor column, resulting in an expanding vortex region and, hence, a detached bubble. Nevertheless, the horizontal MF suppresses the vortices in the parallel plane ($XOY$) but has little influence on the vortices in the perpendicular plane ($XOZ$). Besides, Fig. \ref{compare-sin-hor}(c) and \ref{compare-sin-hor}(d) reveals that corresponding to the stable vapor sheet, the Lorentz forces direct almost downwards in terms of $F_{l,x} \propto u_x\cdot B_y^2$, because any $z-$ component of $\boldsymbol{F}_{l}$ requires the participation of $u_z$, which almost disappears for a stable and two-dimensional vapor sheet. It indicates that for the vapor sheet scenario under horizontal MF, the Lorentz forces play little role in the force balance in radial direction, and thus, removing it causes little influence on the vapor pattern.

%This is consistent with the conclusions drawn in \S\ref{single-horizontal MF} that, with a horizontal MF, the downward Lorentz force in the cross-section ($XOZ$ plane) perpendicular to the MF inhibits the expansion of the vapor film near the overheated wall, resulting in a smaller vapor film thickness, larger $Nu$, and higher evaporation rate ($\dot{m} S_{\Gamma}$). However, this subtle difference does not lead to significant variations in the overall boiling system. In other words, after the sudden removal of the horizontal MF, a sufficient amount of vapor can still be generated to maintain a stable vapor sheet.

\subsection{Multi-mode film boiling} \label{sec:multi-mode}

For multi-mode film boiling scenarios with and without MFs, the computational domain size is increased to $4\lambda_d \times 4\lambda_d \times 4\lambda_d$. To eliminate the influence of artificial disturbances, a randomly initial perturbation is introduced and defined by the following equation:
\begin{equation}
	x=\frac{\lambda_d}{16}+0.05 / n \sum_{i=1}^{n} A(i)\left(\cos \left( \frac{2 \pi i y}{3\lambda_d} \right)+\sin \left( \frac{2 \pi i z}{3\lambda_d} \right)+\cos \left( \frac{2 \pi i y}{3\lambda_d} \right)+\sin \left( \frac{2 \pi i z}{3\lambda_d} \right) \right),
\end{equation}
where $n=10$ denotes the number of random waves, and $0 \leq A(i) \leq 1$ represents a random number. We vary the MF intensities to $\textrm{B = 0 T}$, $\textrm{B = 0.1 T}$, $\textrm{B = 0.2 T}$, $\textrm{B = 0.4 T}$ and $\textrm{B = 0.6 T}$. \ra{The corresponding interaction numbers (Hartmann numbers) for both vertical and horizontal magnetic fields are as follows: $N=0$ ($Ha=0$), $N=2.8$ ($Ha=3.8$), $N=11.1$ ($Ha=7.6$), $N=44.5$ ($Ha=15.3$), and $N=100.1$ ($Ha=22.9$), respectively.}
 
\subsubsection{Effect of the vertical MF}

%Fig 20
\begin{figure}
    \centering
	\begin{overpic}
		[width=12.3cm,tics=1]{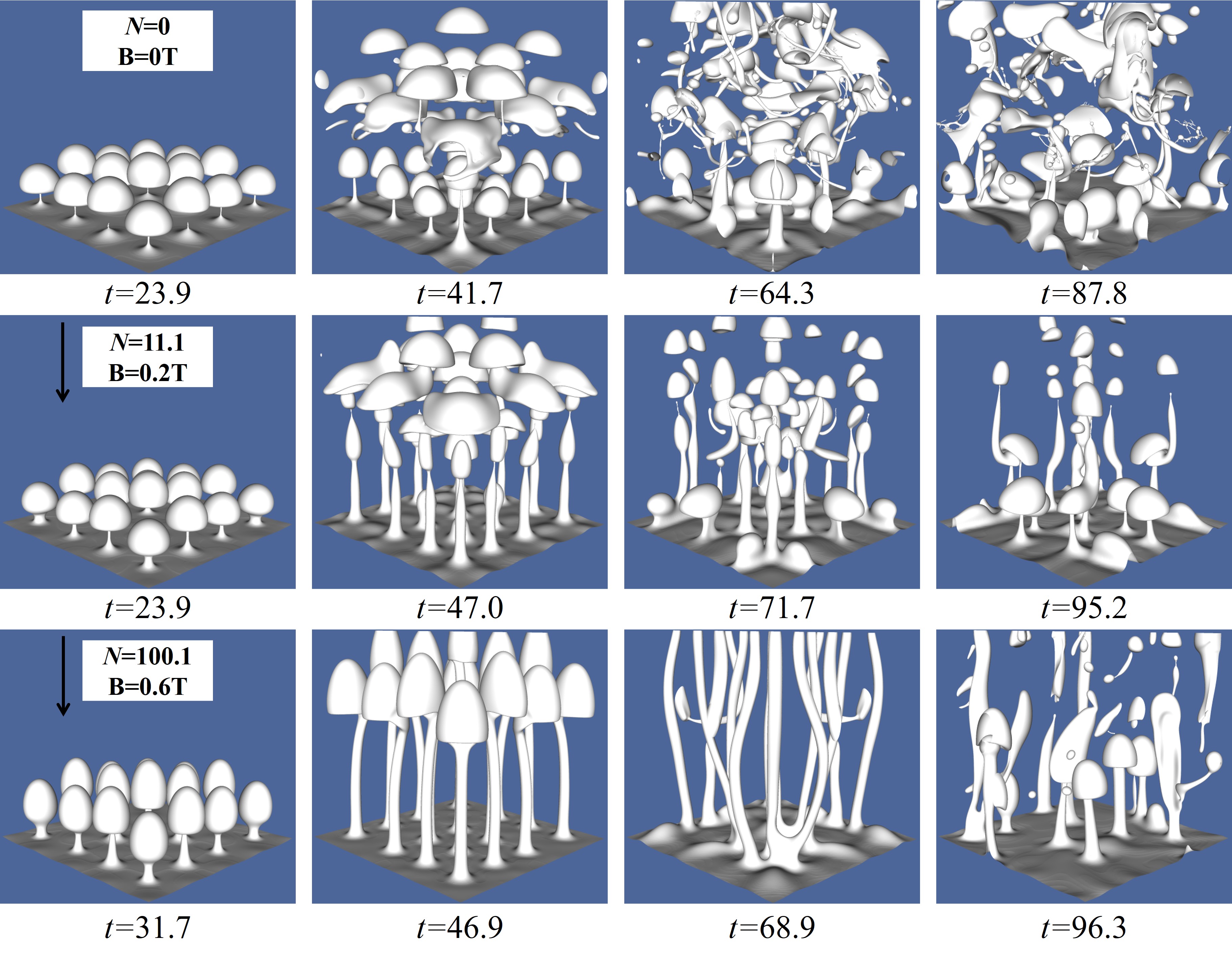}
		\put(-4,76){\large (a)}
		\put(-4,50.5){\large (b)}
		\put(-4,24.8){\large (c)}
	\end{overpic} 
    \caption{Temporal evolution of the bubble shapes during multi-mode film boiling without and with the vertical MFs. (a) $N=0$ (B=0.0 T), (b) $N=11.1$ (B=0.2 T), and (c) $N=100.1$ (B=0.6 T).}
\label{compare-ver-multi}
\end{figure}

\begin{figure}
	\centering
    \includegraphics[width=12cm]{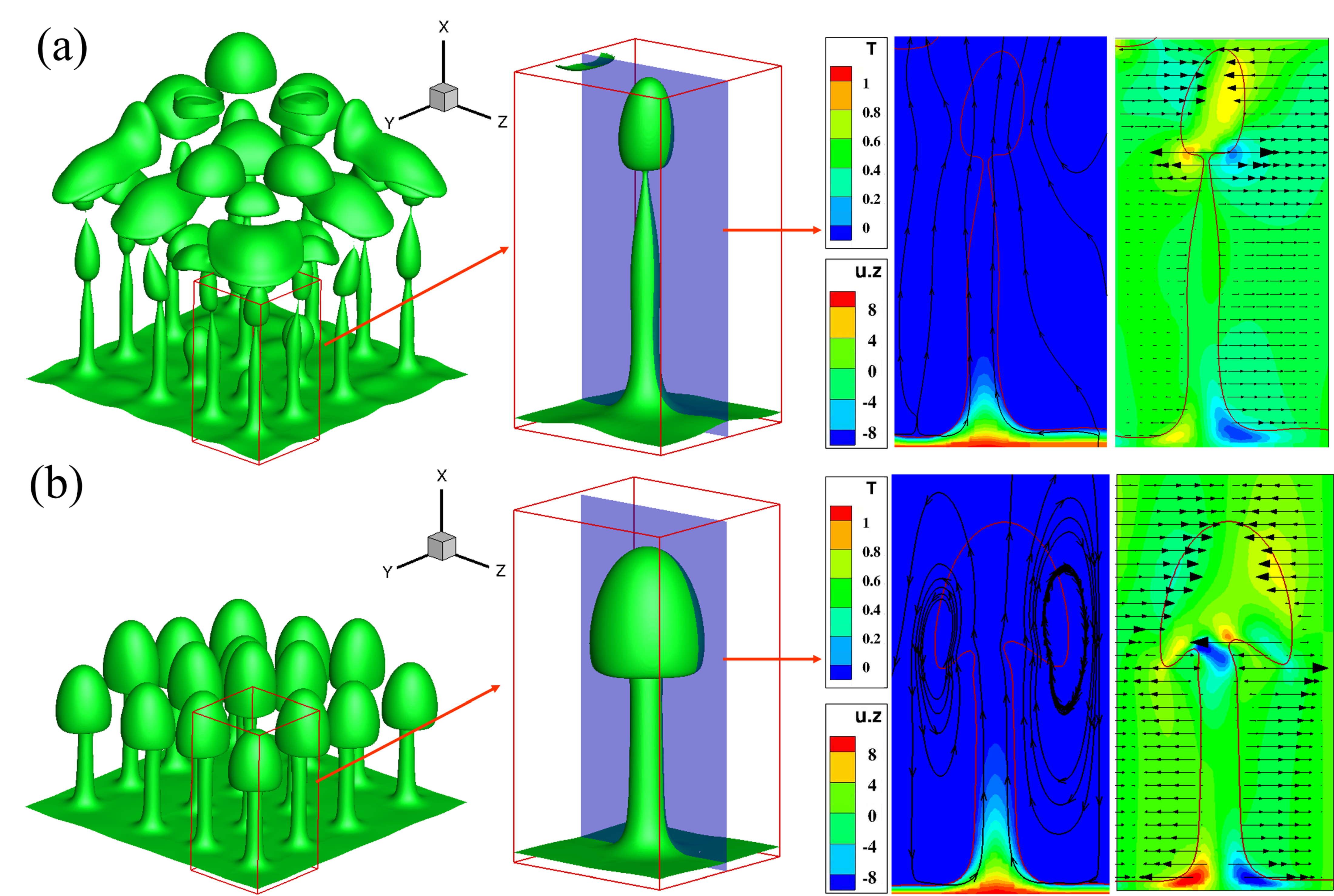}	  
	\caption{Distribution of the temperature field, the streamline patterns, the Lorentz forces, and the horizontal component of velocity ($u_z$, \add{the positive direction: from left to right}), in the vertical cross-section of a single bubble in multi-mode film boiling under different vertical MFs. (a) $N=11.1$, $t=47.2$ and (b) $N=100.1$, $t=37.0$.}
	\label{compare-ver-multi-fl}
\end{figure}

\begin{figure}
	\hspace{-17mm}
	\scalebox{0.94}{\input{figure/mut-ver-nu.tex}}
	\caption{Temporal evolution of the space-averaged $Nu$ under different vertical MF intensities in the case of the multi-mode film boiling. The dashed line represents the theoretical results of \citet{Klimenko1}.}
	\label{ver-nu-mut}	
\end{figure}
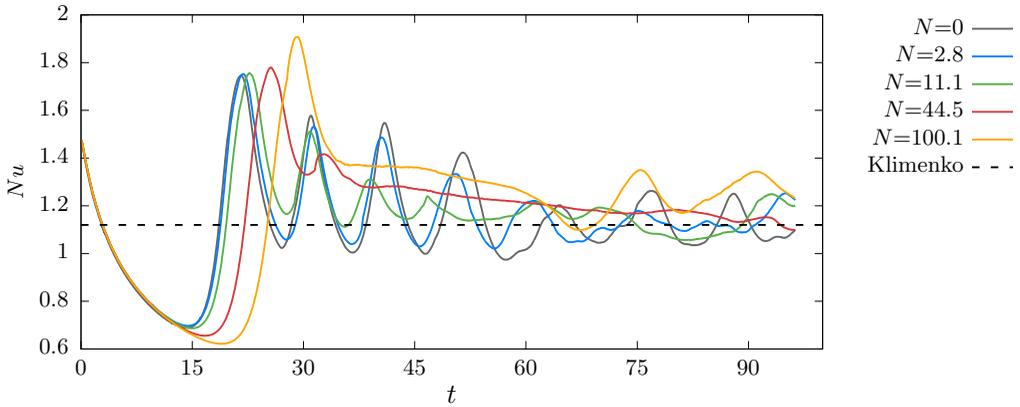

Fig. \ref{compare-ver-multi} depicts the evolution of the liquid/vapor interface under three different vertical MFs. In the initial stage $(t<60.0)$ of film boiling, as observed in the first two columns of figures, the distribution of bubbles aligns with the theory proposed by \citet{Zuber1} for the most dangerous wavelength ($\lambda_d$). The distance between adjacent bubbles correlates with $\lambda_d$ even if the initial disturbances are randomly imposed. Additionally, in the early stage, the interaction between bubbles is not pronounced, allowing for a clear observation that under the multi-bubble model, the impact of a vertical MF on bubble shapes follows the conclusions drawn from the single-mode film boiling: as the MF strength increases, the film boiling transitions from the periodic detachment of bubbles from the bottom to the vapor jet releasing small bubbles from the top. This evolution ultimately leads to vapor columnar boiling without any bubble detachment, a pattern confirmed by Fig. \ref{compare-ver-multi-fl}, which presents the vertical cross-section of a single bubble in the multi-mode, illustrating the distribution of the Lorentz force, the $z-$ component of velocity $(u_z)$, the temperature field, and the velocity streamlines. A comparison with Fig. \ref{compare-ver-sin-streamline} reveals that in the multi-mode boiling, the effect of the vertical MF aligns with the observations in the single-mode boiling. The evolution of $Nu$ also mirrors the single-mode cases in the early stages. \rb{As the magnetic field strength increases, the Nusselt number shifts from approximate periodic oscillations to damped oscillations. A slight increase in the averaged value of $Nu$ for large $N$ is also observed, as shown in Fig. \ref{ver-nu-mut}.} \\

\begin{figure}
	\centering
    \includegraphics[width=13cm]{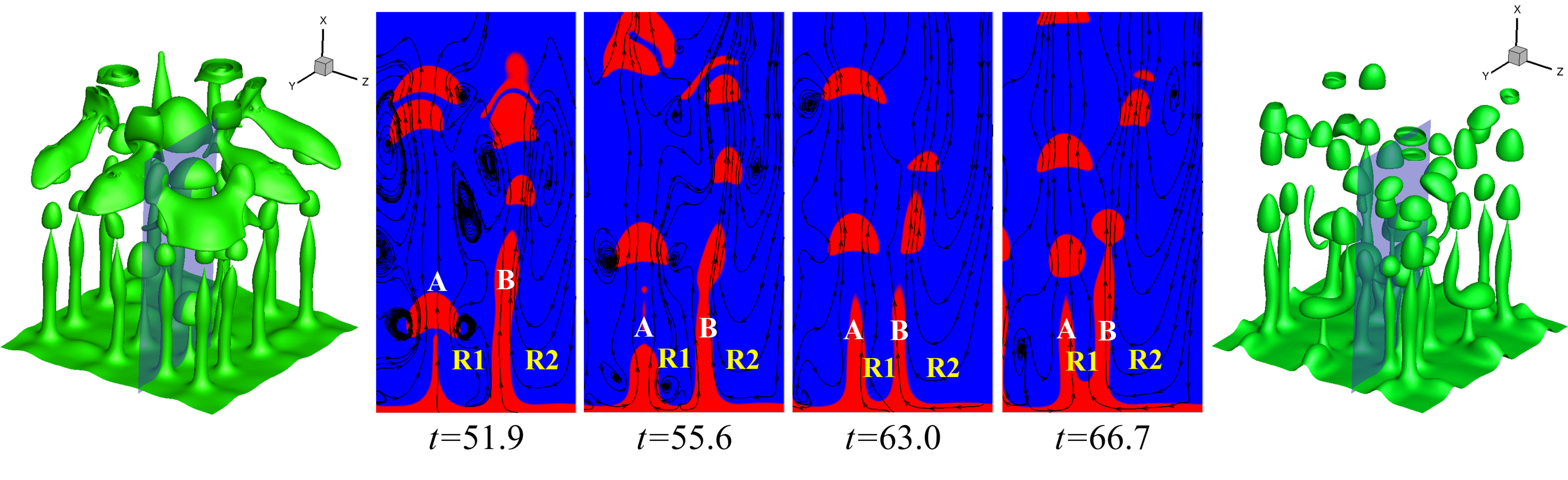}	
\caption{Interaction of two neighboring bubble sites and the evolution of the streamline patterns at different time instants from $t=51.9$ to $t=66.7$ under the vertical MF with $N=11.1$.}
\label{mut-ver-two bubble}
\end{figure}

Next, let us focus on the later stages of the film boiling process under vertical MFs. In the last two columns ($t > 64$) of Fig. \ref{compare-ver-multi}, it becomes evident that the positions of bubbles during boiling undergo displacement in both scenarios—with and without MF. Particularly, such boiling pattern comprising vapor columns under strong vertical MF becomes unstable too. This displacement is primarily attributed to the non-uniform growth of bubbles generated by the initial random perturbation, leading to amplified interactions between bubbles over time. This phenomenon is depicted in Fig. \ref{mut-ver-two bubble}, chosen with $N=11.1$ as an example since this behavior is observed consistently regardless of the presence of the MF. In Fig. \ref{mut-ver-two bubble}, two adjacent bubbles (A and B) are selected for observation, and the evolution of their shapes and velocity streamlines in the vertical plane ($XOY$) at different time instants is presented. At $t=51.9$, the detachment behaviors of bubbles A and B differ, resulting in asymmetry in the flow field on either side of bubble B, as indicated by regions R1 and R2 in the figure. Vortices are present in region R1, while region R2 lacks vortices. As is well-known, regions with vortices exhibit lower pressure. Consequently, bubble B tends to approach bubble A. At $t=63.0$, the two bubbles are very close together, and there is almost no horizontal flow in region R1, while horizontal flow persists in region R2. This causes bubble B to approach bubble A further, eventually leading to their fusion and resulting in a chaotic flow state. Additionally, when bubbles merge, vapor overflows into new positions to form bubbles. The entire boiling system is no longer in a state where bubbles simultaneously detach and generate, but rather, the formation, detachment, and merging of bubbles occur simultaneously. Consequently, at this stage, the height fluctuation of the near-wall thin film is no longer significant. Therefore, the $Nu$ number variation curve shown in Fig. \ref{ver-nu-mut} exhibits small amplitude oscillations around the correlation proposed by \citet{Klimenko1} during the latter stage.

%Fig 21

%Fig 22

%Fig 23

\subsubsection{Effect of the horizontal MF}

%Fig 24
\begin{figure}
    \centering
	\begin{overpic}
		[width=12.3cm,tics=1]{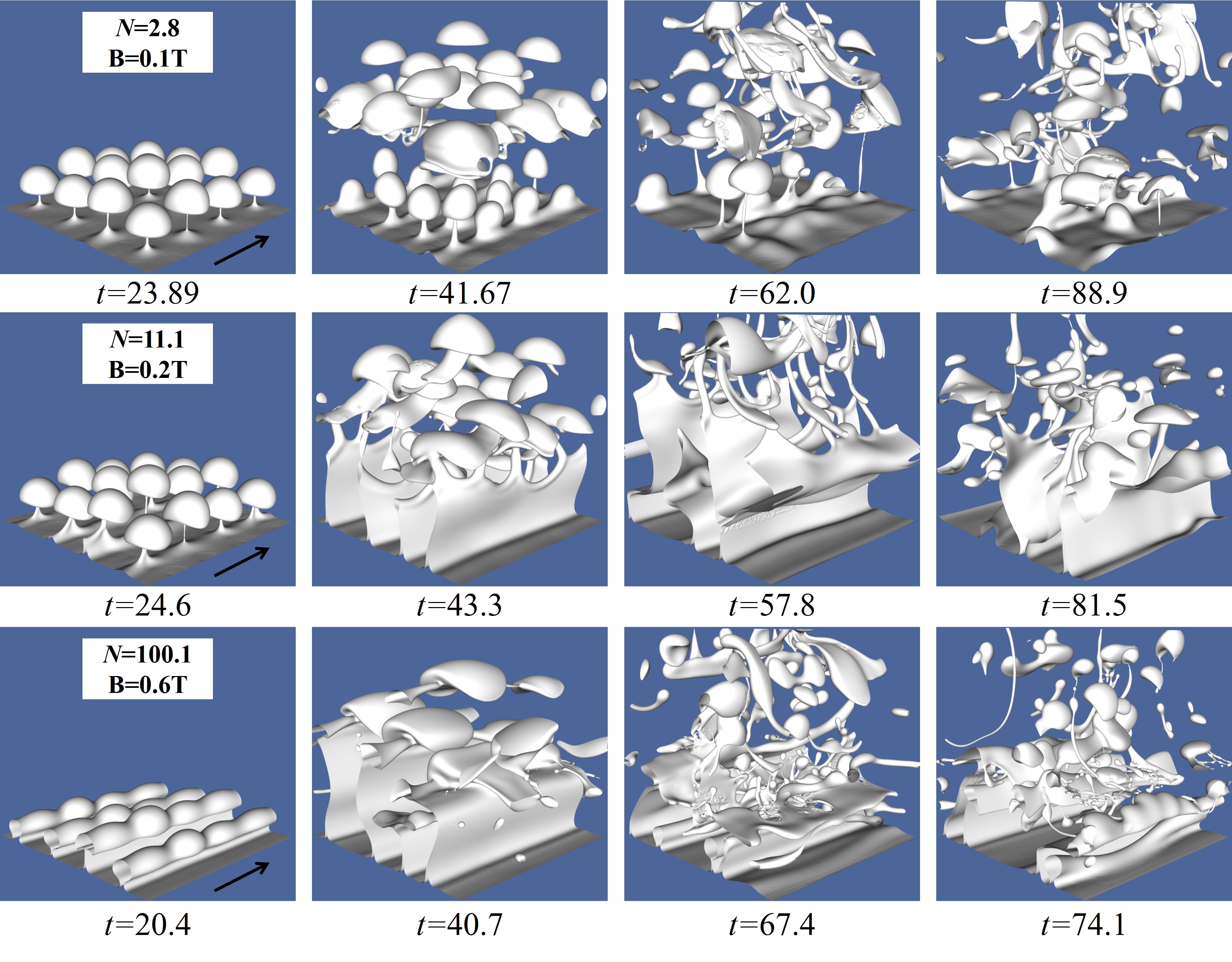}
		\put(-4,76){\large (a)}
		\put(-4,50.5){\large (b)}
		\put(-4,24.8){\large (c)}
	\end{overpic} 
    \caption{Temporal evolution of the bubble shapes during the multi-mode film boiling flow without and with the horizontal MFs. From the top to the bottom row ((a)-(c)): $N=2.8$ $\textrm{(B=0.1~T)}$, $N=11.1$ $\textrm{(B=0.2~T)}$, $N=100.1$ $\textrm{(B=0.6~T)}$.}
\label{compare-hor-multi}
\end{figure}

\begin{figure}
	\centering
    \includegraphics[width=12cm]{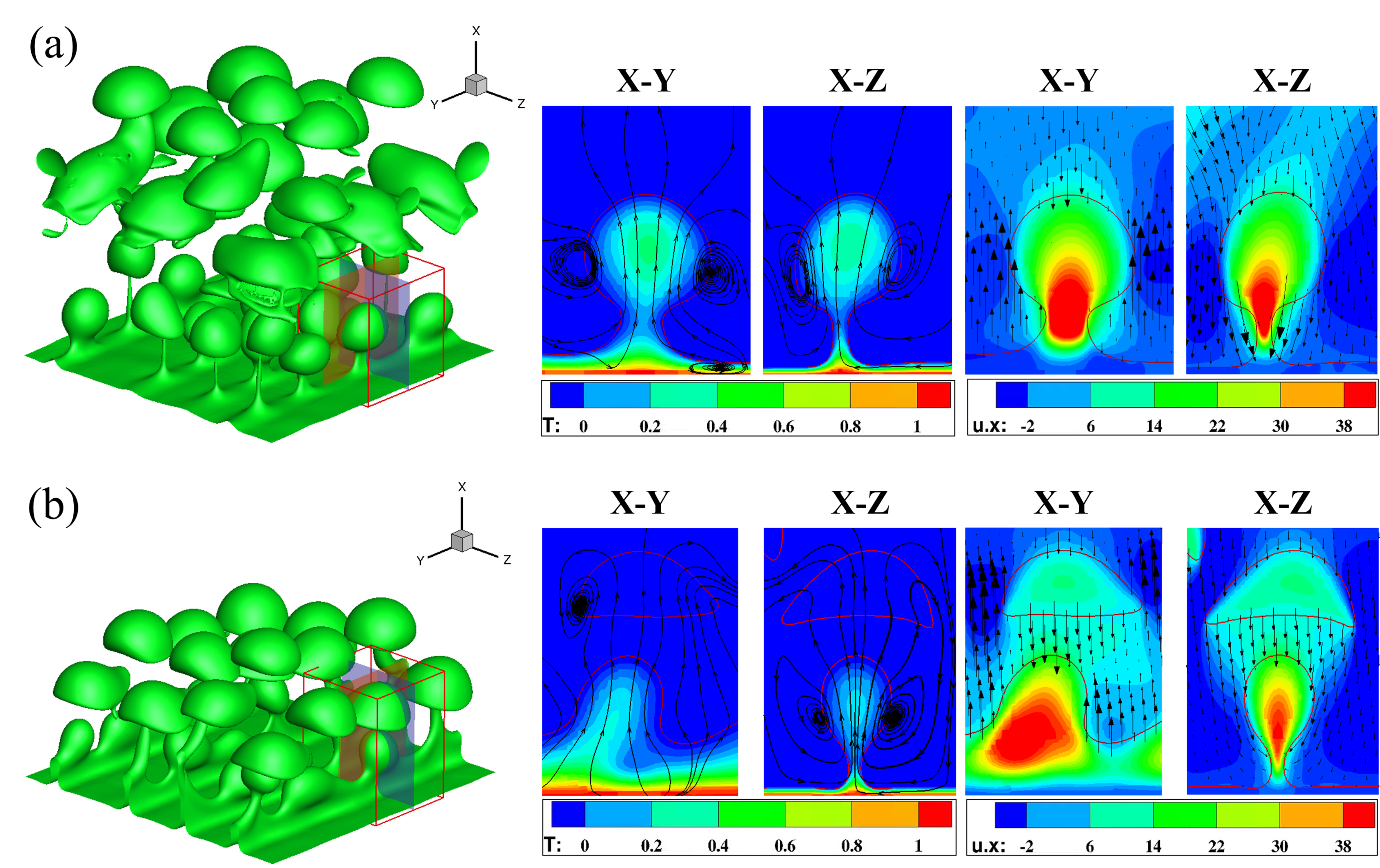}	    
    \caption{Distribution of the temperature field, the streamline patterns, the Lorentz forces, and the vertical component of velocity $(u_x)$, in two vertical cross-sections ($XOY$: parallel to the MF and $XOZ$: perpendicular to the MF) of a single bubble in the multi-mode film boiling under the horizontal MFs. (a) $N=2.8$, $t=42.6$. (b) $N=11.1$, $t=31.5$.}
\label{compare-hor-multi-fl}
\end{figure}

\begin{figure}
	\hspace{-17mm}
	\scalebox{0.94}{\input{figure/mut-hor-nu.tex}}
	\caption{Temporal evolution of the space-averaged $Nu$ under different horizontal MF intensities in the case of the multi-mode film boiling. The dashed line represents the theoretical results of \citet{Klimenko1}.}
	\label{hor-nu-mut}	
\end{figure}
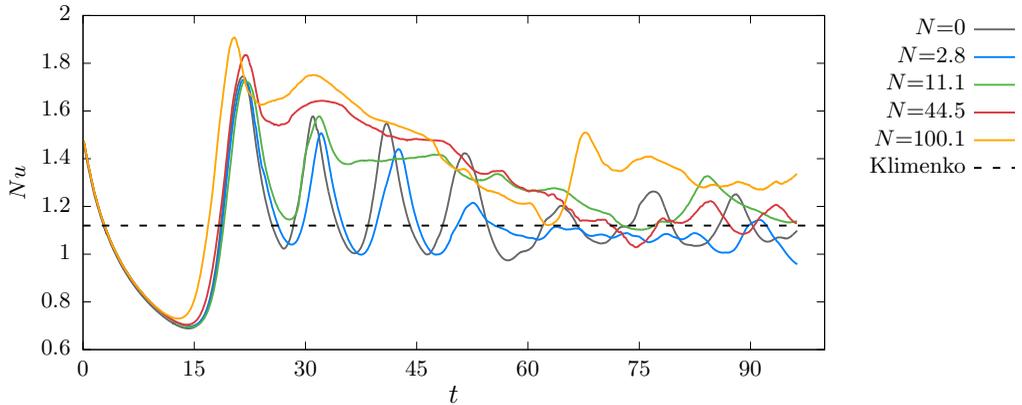

Fig. \ref{compare-hor-multi} presents the evolution of the liquid/vapor interface under three different horizontal MFs. In the initial stage of boiling flow, as shown in the first column of Fig. \ref{compare-hor-multi}, the relatively weak interaction between bubbles results in the effect of the horizontal MF on each bubble in the multi-mode cases being consistent with the conclusion drawn for the single-mode situation. With an increase in MF strength, the boiling flow tends to become two-dimensional along the MF direction. For small MF strength ($N=2.8$ in Fig. \ref{compare-hor-multi}), boiling still occurs in the form of bubble detachment. For large MF strength ($N=11.1$ and 100.1 in Fig. \ref{compare-hor-multi}), boiling takes place in the form of a vapor sheet. Fig. \ref{compare-hor-multi-fl}, using $N=2.8$ and $N=11.1$ as examples, illustrates the distribution of the flow field, temperature, Lorentz forces, and vertical velocity component $(u_x)$ in two cross-sections of a single bubble in the multi-mode case during the initial stage of film boiling. The results align with the single-mode situation (Fig. \ref{compare-sin-hor}). In other words, the mechanism of the horizontal MF cases is consistent with that of single-mode situations. Additionally, the variation of $Nu$, as shown in Fig. \ref{hor-nu-mut}, is also similar to the single-mode cases before $t=60.0$.

%Fig 25

%Fig 26

\begin{figure}
	\centering
	\includegraphics[width=13cm]{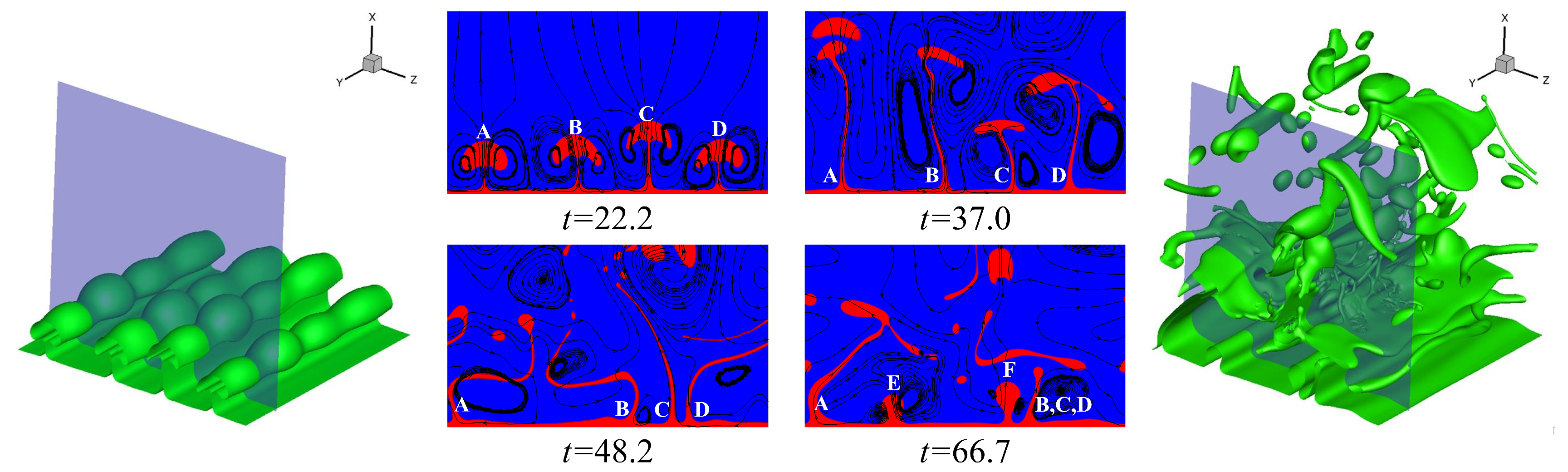}	
	\caption{Dynamics of vapor sheets and the evolution of the streamline pattern for different time instants range from $t=22.2$ to $t=66.7$, with the horizontal MFs $(N=100.1)$.}
    \label{mut-hor-four bubbles}
\end{figure}

Similar to the case with a vertical MF, as boiling progresses to a certain extent, the flow becomes chaotic due to the influence of bubble interactions, as shown in the last two columns ($t > 57.0$) of Fig. \ref{compare-hor-multi}. Moreover, the flow exhibits more chaotic behavior in horizontal MF cases. This is attributed to the fact that, under the horizontal MF, each single column of bubbles along the direction of the MF forms a vapor sheet. As boiling proceeds, the interaction between vapor sheets intensifies compared to bubbles and vapor columns, ultimately resulting in larger-scale rupture or fusion of vapor sheets, exerting a more significant influence on the entire flow system. Fig. \ref{mut-hor-four bubbles} illustrates the evolution of streamline patterns in the vertical cross-section perpendicular to the MF under the condition of $N=100.1$. It can be observed that, as boiling progresses, the asymmetry of vortices on both sides of the four vapor sheets (A, B, C, D) becomes more pronounced. The vapor sheets are then twisted, eventually leading to the fusion of vapor sheets B, C, and D at the base and fragmentation at the top, as shown in Fig. \ref{mut-hor-four bubbles} at $t=48.2$. The fragments produce small bubbles, making the flow system more chaotic. After fusion, new vapor sheets E and F are generated, as shown in Fig. \ref{mut-hor-four bubbles} at $t=66.7$, corresponding to the sudden increase in $Nu$ number in Fig. \ref{hor-nu-mut}. For other cases where $N<100.1$, the variation in $Nu$ exhibits smaller fluctuations in the later stages. This is because, in these situations, the merging and generation of bubbles or vapor sheets occur almost simultaneously, resulting in a smaller fluctuation in the distance between the interface and the overheated wall.

%Fig 27

\rb{Moreover, it is essential to note that bubbly flow in multi-mode film boiling eventually transitions into turbulence or pseudo-turbulence, making the accurate resolution of bubble coalescence and collisions a significant area of ongoing research. The complexity of these interactions is beyond the scope of the present work. For example, Volume of Fluid (VOF) methods often simplify the coalescence process \citep{scardovelli1999direct}. Additionally, recent work by \cite{innocenti2021direct} suggests that a spatial resolution of $\Delta = 2d/Ar$ (where $Ar = \rho_l g^{1/2} d^{3/2}/\mu_l$, and $d$ is the typical size of the bubble) is sufficient for convergence in the spectra of liquid velocity fluctuations in bubbly flows. In our study, the multi-mode film boiling flows are characterized by an Archimedes number of $Ar \approx 120$. The spatial resolution at grid level 7, which corresponds to $\Delta = d/96$, meets the convergence criteria suggested by Innocenti et al. While this does not constitute a full grid convergence verification, it provides confidence in the resolution used to capture the essential physics of heat and mass transfer during multi-mode film boiling. Future research will focus on a more detailed study of vapor bubbly flows under external magnetic fields, including a comprehensive statistical analysis and examination of spectra of kinetic energy and velocity fluctuations on various grids, to rigorously address these complex interactions in the presence of external MFs.}
 
\section{Conclusions} \label{sec:conclusion}

Direct numerical simulations of the 3D film boiling process under magnetic fields (MFs) have been conducted using a numerical method based on the combined magnetohydrodynamics (MHD) and phase change model developed in our previous studies \cite{zhao1,zhang1}. The present study analyzes the effects of vertical and horizontal MFs on the morphology and heat transfer characteristics of liquid film boiling. The results demonstrate that the MF primarily suppresses vortices within the cross-section parallel to the MF direction. Consequently, the boiling pattern and heat transfer are altered depending on the MF direction.

In the presence of a vertical magnetic field, we observe that the flow structure remains isotropic in single-mode scenarios. As the strength of the MF increases, film boiling gradually transitions from periodic bubble detachment to a columnar jet formation. This transition occurs because the induced Lorentz forces counteract the horizontal velocity component, thereby suppressing the generation of vortices near the bubbles and inhibiting their detachment. Regarding the heat transfer characteristics corresponding to the change in boiling form, the time variation of the Nusselt number $(Nu)$ changes from approximately periodic oscillations to damped oscillations. This transition occurs because, during boiling in the form of a columnar jet, bubble detachment happens away from the wall, resulting in a relatively stable bottom vapor film. In terms of overall heat transfer efficiency, the impact of small MF strength in the bubble periodic detachment area on the space- and time-average Nusselt number ($\overline{Nu}$) is not substantial. However, under high MF strength in the vapor column boiling area, $\overline{Nu}$ increases with the rising MF strength. Additionally, a cliff-like sharp decrease in $\overline{Nu}$ occurs during the transition from the region of periodic bubble detachment to the columnar boiling region. This drastic change is attributed to the different boiling forms resulting in varying bubble detachment positions, consequently affecting the thickness of the bottom vapor film. Furthermore, it is observed that after removing the MF in the case of complete columnar jetting boiling, the boiling flow tended to revert to the unstable state observed without a magnetic field.

In the presence of a horizontal magnetic field, film boiling gradually transitions from periodic bubble detachment to a two-dimensional planar vapor film sheet as the MF strength increases. This shift occurs because the induced Lorentz force opposes the vertical velocity component within the cross-section parallel to the MF, suppressing vortex generation on both sides of the bubbles and preventing necking phenomena. Meanwhile, the flow field within the cross-section perpendicular to the MF remains largely unaffected, prompting the boiling flow to adopt a two-dimensional flow along the MF direction. Regarding heat transfer characteristics, the time variation of $Nu$ shifts from approximately periodic oscillations to nearly linear changes. 
This transition is attributed to the formation of a relatively stable vapor film during boiling in the form of a planar vapor film sheet. Concerning overall heat transfer efficiency, in the region of bubble periodic detachment, the initial effects of the horizontal MF lead to a decrease in $\overline{Nu}$ with increasing MF strength. However, $\overline{Nu}$ increases with rising MF strength in the planar vapor sheet boiling area. This is because, during boiling in the form of a planar vapor sheet, the Lorentz force acting vertically downward on the bottom vapor film becomes more pronounced with increasing MF strength, resulting in a thinner bottom vapor film thickness. Unlike the vertical MF, it was observed that under complete planar vapor film jetting boiling conditions, removing the MF still maintained the original state of the boiling flow.

In multi-mode film boiling, irrespective of whether it is under vertical or horizontal MF conditions, the early stages of boiling exhibit similar patterns observed in single-mode film boiling, with consistent effects of the MF on boiling morphology and heat transfer characteristics. However, once the boiling flow reaches complete development, the interaction between bubbles increases, leading to the complexity of the boiling flow and the emergence of numerous scattered small bubbles. Furthermore, under horizontal MF conditions, the flow structure tends to be more chaotic compared to vertical MF conditions. During this stage, bubble formation, detachment, and merging occur simultaneously, causing the $Nu$ to deviate from its original variation pattern with a significantly reduced amplitude. 
\\
\section{Appendix A.}

\begin{figure}
			\centering
			\scalebox{0.75}{\input{figure/test-31.tex}}	
			\caption{ Comparison of the vapor-liquid interface without (red line) and with (blue line) a vertical MF for (a) $Gr=0.392$, (b) $Gr=1.745$ and (c) $Gr=24.265$, while 1 and 2 are the time moments just after bubble detachment in the absence of a MF. (a-1) $t=18.9$ and  (a-2) $t=37.6$; (b-1) $t=17.5$ and  (b-2) $t=29.8$; (c-1) $t=17.9$ and  (c-2) $t=35.8$.}
			\label{case1}
		\end{figure}
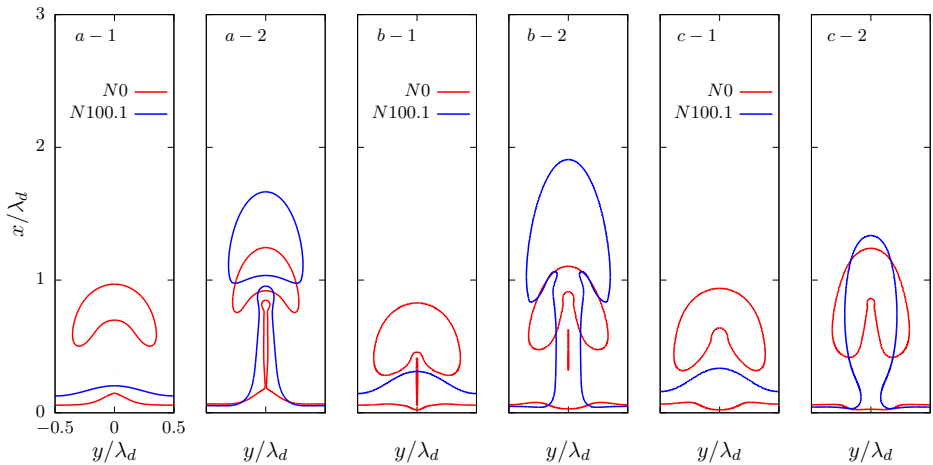

\rb{Here, we demonstrate that even when using more realistic physical properties for the liquid metal, the MHD effects on the rising characteristics of the vapor bubbles remain qualitatively consistent. As detailed in \S 2.4, the value of $Gr$ in the present study is set at $0.392$. For the additional test cases, we have simulated scenarios with $Gr$ values of 1.745 and 24.265, which are one and two orders of magnitude larger than the original value, respectively. Correspondingly, the grid resolutions were increased to 1536 $\times$ 512 $\times$ 512 and 3072 $\times$ 1024 $\times$ 1024, representing 8 and 64 times the grid numbers of the $Gr = 0.392$ case. Figure~\ref{case1} shows snapshots of the liquid/vapor interface in the $XOY$ vertical plane: panel (a) represents $Gr = 0.392$, panel (b) represents $Gr = 1.745$, and panel (c) represents $Gr = 24.265$. In each case, the two sub-figures depict the first two bubble detachment cycles; the blue line represents the non-MHD case, while the red line shows the MHD case at $N = 100.08$. The results indicate that even with higher $Gr$ numbers, the influence of the vertical magnetic field on film boiling remains consistent: the vertical magnetic field suppresses the Rayleigh-Taylor instability, causing a delay or even complete suppression of vapor bubble detachment. These additional simulations support the conclusion that the adjustments in physical properties do not significantly alter the fundamental physical processes observed in the original simulations.}\\

\noindent{\bf Acknowledgement:} The authors gratefully acknowledge the supports from  National Key R $\&$ D Program of China (2023YFA1011000, 2022YFE03130000) and from the NSFC (12222208, 51927812). K.C.S. thanks IIT Hyderabad for the financial support through grant IITH/CHE/F011/SOCH1.\\

\noindent{\bf Declaration of interests:} The authors report no conflict of interest.

\bibliographystyle{jfm}
\bibliography{jfm}
		
\end{document}

%% file: figure/grid-test.tex
% GNUPLOT: LaTeX picture with Postscript
\begingroup
  \makeatletter
  \providecommand\color[2][]{%
    \GenericError{(gnuplot) \space\space\space\@spaces}{%
      Package color not loaded in conjunction with
      terminal option `colourtext'%
    }{See the gnuplot documentation for explanation.%
    }{Either use 'blacktext' in gnuplot or load the package
      color.sty in LaTeX.}%
    \renewcommand\color[2][]{}%
  }%
  \providecommand\includegraphics[2][]{%
    \GenericError{(gnuplot) \space\space\space\@spaces}{%
      Package graphicx or graphics not loaded%
    }{See the gnuplot documentation for explanation.%
    }{The gnuplot epslatex terminal needs graphicx.sty or graphics.sty.}%
    \renewcommand\includegraphics[2][]{}%
  }%
  \providecommand\rotatebox[2]{#2}%
  \@ifundefined{ifGPcolor}{%
    \newif\ifGPcolor
    \GPcolorfalse
  }{}%
  \@ifundefined{ifGPblacktext}{%
    \newif\ifGPblacktext
    \GPblacktexttrue
  }{}%
  % define a \g@addto@macro without @ in the name:
  \let\gplgaddtomacro\g@addto@macro
  % define empty templates for all commands taking text:
  \gdef\gplbacktext{}%
  \gdef\gplfronttext{}%
  \makeatother
  \ifGPblacktext
    % no textcolor at all
    \def\colorrgb#1{}%
    \def\colorgray#1{}%
  \else
    % gray or color?
    \ifGPcolor
      \def\colorrgb#1{\color[rgb]{#1}}%
      \def\colorgray#1{\color[gray]{#1}}%
      \expandafter\def\csname LTw\endcsname{\color{white}}%
      \expandafter\def\csname LTb\endcsname{\color{black}}%
      \expandafter\def\csname LTa\endcsname{\color{black}}%
      \expandafter\def\csname LT0\endcsname{\color[rgb]{1,0,0}}%
      \expandafter\def\csname LT1\endcsname{\color[rgb]{0,1,0}}%
      \expandafter\def\csname LT2\endcsname{\color[rgb]{0,0,1}}%
      \expandafter\def\csname LT3\endcsname{\color[rgb]{1,0,1}}%
      \expandafter\def\csname LT4\endcsname{\color[rgb]{0,1,1}}%
      \expandafter\def\csname LT5\endcsname{\color[rgb]{1,1,0}}%
      \expandafter\def\csname LT6\endcsname{\color[rgb]{0,0,0}}%
      \expandafter\def\csname LT7\endcsname{\color[rgb]{1,0.3,0}}%
      \expandafter\def\csname LT8\endcsname{\color[rgb]{0.5,0.5,0.5}}%
    \else
      % gray
      \def\colorrgb#1{\color{black}}%
      \def\colorgray#1{\color[gray]{#1}}%
      \expandafter\def\csname LTw\endcsname{\color{white}}%
      \expandafter\def\csname LTb\endcsname{\color{black}}%
      \expandafter\def\csname LTa\endcsname{\color{black}}%
      \expandafter\def\csname LT0\endcsname{\color{black}}%
      \expandafter\def\csname LT1\endcsname{\color{black}}%
      \expandafter\def\csname LT2\endcsname{\color{black}}%
      \expandafter\def\csname LT3\endcsname{\color{black}}%
      \expandafter\def\csname LT4\endcsname{\color{black}}%
      \expandafter\def\csname LT5\endcsname{\color{black}}%
      \expandafter\def\csname LT6\endcsname{\color{black}}%
      \expandafter\def\csname LT7\endcsname{\color{black}}%
      \expandafter\def\csname LT8\endcsname{\color{black}}%
    \fi
  \fi
    \setlength{\unitlength}{0.0500bp}%
    \ifx\gptboxheight\undefined%
      \newlength{\gptboxheight}%
      \newlength{\gptboxwidth}%
      \newsavebox{\gptboxtext}%
    \fi%
    \setlength{\fboxrule}{0.5pt}%
    \setlength{\fboxsep}{1pt}%
\begin{picture}(8640.00,3024.00)%
    \gplgaddtomacro\gplbacktext{%
      \csname LTb\endcsname%%
      \put(323,462){\makebox(0,0)[r]{\strut{}$0$}}%
      \put(323,1012){\makebox(0,0)[r]{\strut{}$0.25$}}%
      \put(323,1562){\makebox(0,0)[r]{\strut{}$0.5$}}%
      \put(323,2111){\makebox(0,0)[r]{\strut{}$0.75$}}%
      \put(323,2661){\makebox(0,0)[r]{\strut{}$1$}}%
      \put(422,297){\makebox(0,0){\strut{}$-0.5$}}%
      \put(972,297){\makebox(0,0){\strut{}$-0.25$}}%
      \put(1522,297){\makebox(0,0){\strut{}$0$}}%
      \put(2072,297){\makebox(0,0){\strut{}$0.25$}}%
      \put(2622,297){\makebox(0,0){\strut{}$0.5$}}%
      \put(160,2661){\makebox(0,0)[r]{\strut{}\large{(a)}}}%
    }%
    \gplgaddtomacro\gplfronttext{%
      \csname LTb\endcsname%%
      \put(-40,1561){\rotatebox{-270}{\makebox(0,0){\strut{}\large $y/ \lambda_d $}}}%
      \put(1522,77){\makebox(0,0){\strut{}\large $x/ \lambda_d $}}%
      \csname LTb\endcsname%%
      \put(2163,2468){\makebox(0,0)[r]{\strut{}Level6}}%
      \csname LTb\endcsname%%
      \put(2163,2303){\makebox(0,0)[r]{\strut{}Level7}}%
      \csname LTb\endcsname%%
      \put(2163,2138){\makebox(0,0)[r]{\strut{}Level8}}%
    }%
    \gplgaddtomacro\gplbacktext{%
      \csname LTb\endcsname%%
      \put(3150,462){\makebox(0,0)[r]{\strut{}$0$}}%
      \put(3150,1012){\makebox(0,0)[r]{\strut{}$0.25$}}%
      \put(3150,1562){\makebox(0,0)[r]{\strut{}$0.5$}}%
      \put(3150,2111){\makebox(0,0)[r]{\strut{}$0.75$}}%
      \put(3150,2661){\makebox(0,0)[r]{\strut{}$1$}}%
      \put(3249,297){\makebox(0,0){\strut{}$-0.5$}}%
      \put(3799,297){\makebox(0,0){\strut{}$-0.25$}}%
      \put(4349,297){\makebox(0,0){\strut{}$0$}}%
      \put(4899,297){\makebox(0,0){\strut{}$0.25$}}%
      \put(5449,297){\makebox(0,0){\strut{}$0.5$}}%
      \put(2987,2661){\makebox(0,0)[r]{\strut{}\large{(b)}}}
    }%
    \gplgaddtomacro\gplfronttext{%
      \csname LTb\endcsname%%
      \put(2787,1561){\rotatebox{-270}{\makebox(0,0){\strut{}\large $y/ \lambda_d $}}}%
      \put(4349,77){\makebox(0,0){\strut{}\large $x/ \lambda_d $}}%
    }%
    \gplgaddtomacro\gplbacktext{%
      \csname LTb\endcsname%%
      \put(6030,462){\makebox(0,0)[r]{\strut{}$0.6$}}%
      \put(6030,776){\makebox(0,0)[r]{\strut{}$0.8$}}%
      \put(6030,1090){\makebox(0,0)[r]{\strut{}$1$}}%
      \put(6030,1404){\makebox(0,0)[r]{\strut{}$1.2$}}%
      \put(6030,1719){\makebox(0,0)[r]{\strut{}$1.4$}}%
      \put(6030,2033){\makebox(0,0)[r]{\strut{}$1.6$}}%
      \put(6030,2347){\makebox(0,0)[r]{\strut{}$1.8$}}%
      \put(6030,2661){\makebox(0,0)[r]{\strut{}$2$}}%
      \put(6129,297){\makebox(0,0){\strut{}$0$}}%
      \put(6679,297){\makebox(0,0){\strut{}$5$}}%
      \put(7229,297){\makebox(0,0){\strut{}$10$}}%
      \put(7778,297){\makebox(0,0){\strut{}$15$}}%
      \put(8328,297){\makebox(0,0){\strut{}$20$}}%
      \put(5867,2661){\makebox(0,0)[r]{\strut{}\large{(c)}}}%
    }%
    \gplgaddtomacro\gplfronttext{%
      \csname LTb\endcsname%%
      \put(5687,1561){\rotatebox{-270}{\makebox(0,0){\strut{}$Nu$}}}%
      \put(7228,77){\makebox(0,0){\strut{}\large $t$}}%
    }%
    \gplbacktext
    \put(0,0){\includegraphics{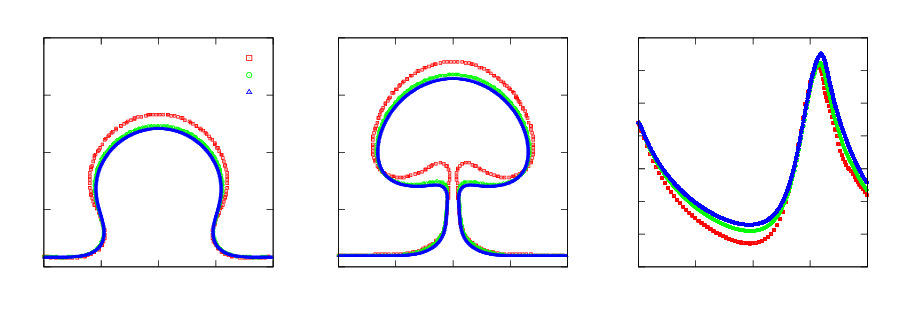}}%
    \gplfronttext
  \end{picture}%
\endgroup

%% file: figure/nu-b0.tex
% GNUPLOT: LaTeX picture with Postscript
\begingroup
  \makeatletter
  \providecommand\color[2][]{%
    \GenericError{(gnuplot) \space\space\space\@spaces}{%
      Package color not loaded in conjunction with
      terminal option `colourtext'%
    }{See the gnuplot documentation for explanation.%
    }{Either use 'blacktext' in gnuplot or load the package
      color.sty in LaTeX.}%
    \renewcommand\color[2][]{}%
  }%
  \providecommand\includegraphics[2][]{%
    \GenericError{(gnuplot) \space\space\space\@spaces}{%
      Package graphicx or graphics not loaded%
    }{See the gnuplot documentation for explanation.%
    }{The gnuplot epslatex terminal needs graphicx.sty or graphics.sty.}%
    \renewcommand\includegraphics[2][]{}%
  }%
  \providecommand\rotatebox[2]{#2}%
  \@ifundefined{ifGPcolor}{%
    \newif\ifGPcolor
    \GPcolorfalse
  }{}%
  \@ifundefined{ifGPblacktext}{%
    \newif\ifGPblacktext
    \GPblacktexttrue
  }{}%
  % define a \g@addto@macro without @ in the name:
  \let\gplgaddtomacro\g@addto@macro
  % define empty templates for all commands taking text:
  \gdef\gplbacktext{}%
  \gdef\gplfronttext{}%
  \makeatother
  \ifGPblacktext
    % no textcolor at all
    \def\colorrgb#1{}%
    \def\colorgray#1{}%
  \else
    % gray or color?
    \ifGPcolor
      \def\colorrgb#1{\color[rgb]{#1}}%
      \def\colorgray#1{\color[gray]{#1}}%
      \expandafter\def\csname LTw\endcsname{\color{white}}%
      \expandafter\def\csname LTb\endcsname{\color{black}}%
      \expandafter\def\csname LTa\endcsname{\color{black}}%
      \expandafter\def\csname LT0\endcsname{\color[rgb]{1,0,0}}%
      \expandafter\def\csname LT1\endcsname{\color[rgb]{0,1,0}}%
      \expandafter\def\csname LT2\endcsname{\color[rgb]{0,0,1}}%
      \expandafter\def\csname LT3\endcsname{\color[rgb]{1,0,1}}%
      \expandafter\def\csname LT4\endcsname{\color[rgb]{0,1,1}}%
      \expandafter\def\csname LT5\endcsname{\color[rgb]{1,1,0}}%
      \expandafter\def\csname LT6\endcsname{\color[rgb]{0,0,0}}%
      \expandafter\def\csname LT7\endcsname{\color[rgb]{1,0.3,0}}%
      \expandafter\def\csname LT8\endcsname{\color[rgb]{0.5,0.5,0.5}}%
    \else
      % gray
      \def\colorrgb#1{\color{black}}%
      \def\colorgray#1{\color[gray]{#1}}%
      \expandafter\def\csname LTw\endcsname{\color{white}}%
      \expandafter\def\csname LTb\endcsname{\color{black}}%
      \expandafter\def\csname LTa\endcsname{\color{black}}%
      \expandafter\def\csname LT0\endcsname{\color{black}}%
      \expandafter\def\csname LT1\endcsname{\color{black}}%
      \expandafter\def\csname LT2\endcsname{\color{black}}%
      \expandafter\def\csname LT3\endcsname{\color{black}}%
      \expandafter\def\csname LT4\endcsname{\color{black}}%
      \expandafter\def\csname LT5\endcsname{\color{black}}%
      \expandafter\def\csname LT6\endcsname{\color{black}}%
      \expandafter\def\csname LT7\endcsname{\color{black}}%
      \expandafter\def\csname LT8\endcsname{\color{black}}%
    \fi
  \fi
    \setlength{\unitlength}{0.0500bp}%
    \ifx\gptboxheight\undefined%
      \newlength{\gptboxheight}%
      \newlength{\gptboxwidth}%
      \newsavebox{\gptboxtext}%
    \fi%
    \setlength{\fboxrule}{0.5pt}%
    \setlength{\fboxsep}{1pt}%
\begin{picture}(6912.00,3225.60)%
    \gplgaddtomacro\gplbacktext{%
      \csname LTb\endcsname%%
      \put(1551,462){\makebox(0,0)[r]{\strut{}$0.6$}}%
      \put(1551,841){\makebox(0,0)[r]{\strut{}$0.8$}}%
      \put(1551,1220){\makebox(0,0)[r]{\strut{}$1$}}%
      \put(1551,1599){\makebox(0,0)[r]{\strut{}$1.2$}}%
      \put(1551,1978){\makebox(0,0)[r]{\strut{}$1.4$}}%
      \put(1551,2357){\makebox(0,0)[r]{\strut{}$1.6$}}%
      \put(1551,2736){\makebox(0,0)[r]{\strut{}$1.8$}}%
      \put(1551,3115){\makebox(0,0)[r]{\strut{}$2$}}%
      \put(1650,297){\makebox(0,0){\strut{}$0$}}%
      \put(2372,297){\makebox(0,0){\strut{}$15$}}%
      \put(3094,297){\makebox(0,0){\strut{}$30$}}%
      \put(3817,297){\makebox(0,0){\strut{}$45$}}%
      \put(4539,297){\makebox(0,0){\strut{}$60$}}%
      \put(5261,297){\makebox(0,0){\strut{}$75$}}%
    }%
    \gplgaddtomacro\gplfronttext{%
      \csname LTb\endcsname%%
      \put(1149,1788){\rotatebox{-270}{\makebox(0,0){\strut{}$ Nu $}}}%
      \put(3455,77){\makebox(0,0){\strut{}\large$ t$}}%
      \csname LTb\endcsname%%
      \put(6386,3005){\makebox(0,0)[r]{\strut{} $ N=0 $}}%
      \csname LTb\endcsname%%
      \put(6386,2785){\makebox(0,0)[r]{\strut{}$Klimenko$}}%
    }%
    \gplbacktext
    \put(0,0){\includegraphics{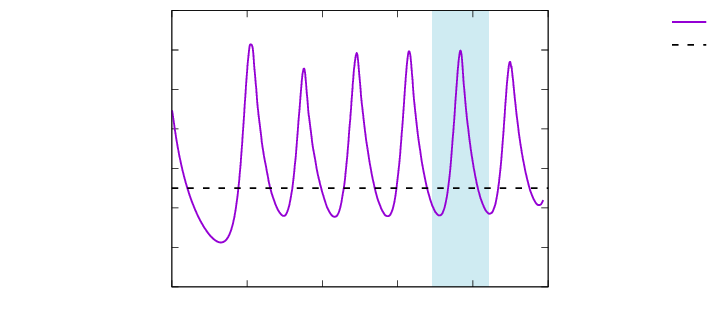}}%
    \gplfronttext
  \end{picture}%
\endgroup

%% file: figure/sin-ver-nu.tex
% GNUPLOT: LaTeX picture with Postscript
\begingroup
  \makeatletter
  \providecommand\color[2][]{%
    \GenericError{(gnuplot) \space\space\space\@spaces}{%
      Package color not loaded in conjunction with
      terminal option `colourtext'%
    }{See the gnuplot documentation for explanation.%
    }{Either use 'blacktext' in gnuplot or load the package
      color.sty in LaTeX.}%
    \renewcommand\color[2][]{}%
  }%
  \providecommand\includegraphics[2][]{%
    \GenericError{(gnuplot) \space\space\space\@spaces}{%
      Package graphicx or graphics not loaded%
    }{See the gnuplot documentation for explanation.%
    }{The gnuplot epslatex terminal needs graphicx.sty or graphics.sty.}%
    \renewcommand\includegraphics[2][]{}%
  }%
  \providecommand\rotatebox[2]{#2}%
  \@ifundefined{ifGPcolor}{%
    \newif\ifGPcolor
    \GPcolorfalse
  }{}%
  \@ifundefined{ifGPblacktext}{%
    \newif\ifGPblacktext
    \GPblacktexttrue
  }{}%
  % define a \g@addto@macro without @ in the name:
  \let\gplgaddtomacro\g@addto@macro
  % define empty templates for all commands taking text:
  \gdef\gplbacktext{}%
  \gdef\gplfronttext{}%
  \makeatother
  \ifGPblacktext
    % no textcolor at all
    \def\colorrgb#1{}%
    \def\colorgray#1{}%
  \else
    % gray or color?
    \ifGPcolor
      \def\colorrgb#1{\color[rgb]{#1}}%
      \def\colorgray#1{\color[gray]{#1}}%
      \expandafter\def\csname LTw\endcsname{\color{white}}%
      \expandafter\def\csname LTb\endcsname{\color{black}}%
      \expandafter\def\csname LTa\endcsname{\color{black}}%
      \expandafter\def\csname LT0\endcsname{\color[rgb]{1,0,0}}%
      \expandafter\def\csname LT1\endcsname{\color[rgb]{0,1,0}}%
      \expandafter\def\csname LT2\endcsname{\color[rgb]{0,0,1}}%
      \expandafter\def\csname LT3\endcsname{\color[rgb]{1,0,1}}%
      \expandafter\def\csname LT4\endcsname{\color[rgb]{0,1,1}}%
      \expandafter\def\csname LT5\endcsname{\color[rgb]{1,1,0}}%
      \expandafter\def\csname LT6\endcsname{\color[rgb]{0,0,0}}%
      \expandafter\def\csname LT7\endcsname{\color[rgb]{1,0.3,0}}%
      \expandafter\def\csname LT8\endcsname{\color[rgb]{0.5,0.5,0.5}}%
    \else
      % gray
      \def\colorrgb#1{\color{black}}%
      \def\colorgray#1{\color[gray]{#1}}%
      \expandafter\def\csname LTw\endcsname{\color{white}}%
      \expandafter\def\csname LTb\endcsname{\color{black}}%
      \expandafter\def\csname LTa\endcsname{\color{black}}%
      \expandafter\def\csname LT0\endcsname{\color{black}}%
      \expandafter\def\csname LT1\endcsname{\color{black}}%
      \expandafter\def\csname LT2\endcsname{\color{black}}%
      \expandafter\def\csname LT3\endcsname{\color{black}}%
      \expandafter\def\csname LT4\endcsname{\color{black}}%
      \expandafter\def\csname LT5\endcsname{\color{black}}%
      \expandafter\def\csname LT6\endcsname{\color{black}}%
      \expandafter\def\csname LT7\endcsname{\color{black}}%
      \expandafter\def\csname LT8\endcsname{\color{black}}%
    \fi
  \fi
    \setlength{\unitlength}{0.0500bp}%
    \ifx\gptboxheight\undefined%
      \newlength{\gptboxheight}%
      \newlength{\gptboxwidth}%
      \newsavebox{\gptboxtext}%
    \fi%
    \setlength{\fboxrule}{0.5pt}%
    \setlength{\fboxsep}{1pt}%
\begin{picture}(7560.00,3225.60)%
    \gplgaddtomacro\gplbacktext{%
      \csname LTb\endcsname%%
      \put(518,628){\makebox(0,0)[r]{\strut{}$0.8$}}%
      \put(518,1291){\makebox(0,0)[r]{\strut{}$1.2$}}%
      \put(518,1954){\makebox(0,0)[r]{\strut{}$1.6$}}%
      \put(518,2618){\makebox(0,0)[r]{\strut{}$2$}}%
      \put(617,297){\makebox(0,0){\strut{}$0$}}%
      \put(1231,297){\makebox(0,0){\strut{}$15$}}%
      \put(1846,297){\makebox(0,0){\strut{}$30$}}%
      \put(2460,297){\makebox(0,0){\strut{}$45$}}%
      \put(3075,297){\makebox(0,0){\strut{}$60$}}%
      \put(3689,297){\makebox(0,0){\strut{}$75$}}%
      \put(116,3020){{\makebox(0,0){\strut{}{\large (a)}}}}%
    }%
    \gplgaddtomacro\gplfronttext{%
      \csname LTb\endcsname%%
      \put(116,1788){\rotatebox{-270}{\makebox(0,0){\strut{}$Nu$}}}%
      \put(2153,77){\makebox(0,0){\strut{}\large $t$}}%
      \csname LTb\endcsname%%
      \put(1364,2872){\makebox(0,0)[r]{\strut{}$N$=0}}%
      \csname LTb\endcsname%%
      \put(1364,2652){\makebox(0,0)[r]{\strut{}$N$=0.7}}%
      \csname LTb\endcsname%%
      \put(2318,2872){\makebox(0,0)[r]{\strut{}$N$=2.8}}%
      \csname LTb\endcsname%%
      \put(2318,2652){\makebox(0,0)[r]{\strut{}$N$=6.3}}%
      \csname LTb\endcsname%%
      \put(3272,2872){\makebox(0,0)[r]{\strut{}$N$=11.1}}%
      \csname LTb\endcsname%%
      \put(3272,2652){\makebox(0,0)[r]{\strut{}$N$=44.5}}%
    }%
    \gplgaddtomacro\gplbacktext{%
      \csname LTb\endcsname%%
      \put(4400,727){\makebox(0,0)[r]{\strut{}$1.12$}}%
      \put(4400,1258){\makebox(0,0)[r]{\strut{}$1.16$}}%
      \put(4400,1789){\makebox(0,0)[r]{\strut{}$1.2$}}%
      \put(4400,2319){\makebox(0,0)[r]{\strut{}$1.24$}}%
      \put(4400,2850){\makebox(0,0)[r]{\strut{}$1.28$}}%
      \put(4800,297){\makebox(0,0){\strut{}$0$}}%
      \put(5401,297){\makebox(0,0){\strut{}$0.1$}}%
      \put(6002,297){\makebox(0,0){\strut{}$0.2$}}%
      \put(6603,297){\makebox(0,0){\strut{}$0.3$}}%
      \put(7204,297){\makebox(0,0){\strut{}$0.4$}}%
        \put(3932,3020){{\makebox(0,0){\strut{}{\large (b)}}}}%
    }%
    \gplgaddtomacro\gplfronttext{%
      \csname LTb\endcsname%%
      \put(3932,1788){\rotatebox{-270}{\makebox(0,0){\strut{}$\overline{Nu}$}}}%
      \put(6001,77){\makebox(0,0){\strut{}$B/T$}}%
       \put(4800,1789){\strut{}\scalebox{0.8}{Bubble}}	
      \put(4650,1600){\strut{}\scalebox{0.8}{detachment}}	
      \put(4580,1500){\framebox(920,500){}}
      \put(6300,2289){\strut{}\scalebox{0.8}{Vapour}}	
      \put(6320,2100){\strut{}\scalebox{0.8}{column}}	
      \put(6180,2000){\framebox(750,500){}}
    }%
    \gplbacktext
    \put(0,0){\includegraphics{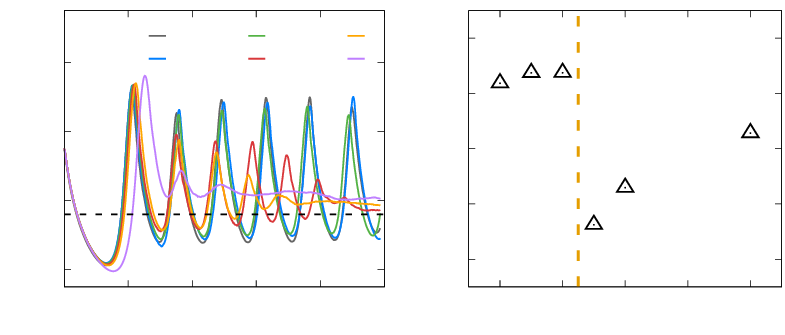}}%
    \gplfronttext
  \end{picture}%
\endgroup

%% file: figure/sin-hor-nu.tex
% GNUPLOT: LaTeX picture with Postscript
\begingroup
  \makeatletter
  \providecommand\color[2][]{%
    \GenericError{(gnuplot) \space\space\space\@spaces}{%
      Package color not loaded in conjunction with
      terminal option `colourtext'%
    }{See the gnuplot documentation for explanation.%
    }{Either use 'blacktext' in gnuplot or load the package
      color.sty in LaTeX.}%
    \renewcommand\color[2][]{}%
  }%
  \providecommand\includegraphics[2][]{%
    \GenericError{(gnuplot) \space\space\space\@spaces}{%
      Package graphicx or graphics not loaded%
    }{See the gnuplot documentation for explanation.%
    }{The gnuplot epslatex terminal needs graphicx.sty or graphics.sty.}%
    \renewcommand\includegraphics[2][]{}%
  }%
  \providecommand\rotatebox[2]{#2}%
  \@ifundefined{ifGPcolor}{%
    \newif\ifGPcolor
    \GPcolorfalse
  }{}%
  \@ifundefined{ifGPblacktext}{%
    \newif\ifGPblacktext
    \GPblacktexttrue
  }{}%
  % define a \g@addto@macro without @ in the name:
  \let\gplgaddtomacro\g@addto@macro
  % define empty templates for all commands taking text:
  \gdef\gplbacktext{}%
  \gdef\gplfronttext{}%
  \makeatother
  \ifGPblacktext
    % no textcolor at all
    \def\colorrgb#1{}%
    \def\colorgray#1{}%
  \else
    % gray or color?
    \ifGPcolor
      \def\colorrgb#1{\color[rgb]{#1}}%
      \def\colorgray#1{\color[gray]{#1}}%
      \expandafter\def\csname LTw\endcsname{\color{white}}%
      \expandafter\def\csname LTb\endcsname{\color{black}}%
      \expandafter\def\csname LTa\endcsname{\color{black}}%
      \expandafter\def\csname LT0\endcsname{\color[rgb]{1,0,0}}%
      \expandafter\def\csname LT1\endcsname{\color[rgb]{0,1,0}}%
      \expandafter\def\csname LT2\endcsname{\color[rgb]{0,0,1}}%
      \expandafter\def\csname LT3\endcsname{\color[rgb]{1,0,1}}%
      \expandafter\def\csname LT4\endcsname{\color[rgb]{0,1,1}}%
      \expandafter\def\csname LT5\endcsname{\color[rgb]{1,1,0}}%
      \expandafter\def\csname LT6\endcsname{\color[rgb]{0,0,0}}%
      \expandafter\def\csname LT7\endcsname{\color[rgb]{1,0.3,0}}%
      \expandafter\def\csname LT8\endcsname{\color[rgb]{0.5,0.5,0.5}}%
    \else
      % gray
      \def\colorrgb#1{\color{black}}%
      \def\colorgray#1{\color[gray]{#1}}%
      \expandafter\def\csname LTw\endcsname{\color{white}}%
      \expandafter\def\csname LTb\endcsname{\color{black}}%
      \expandafter\def\csname LTa\endcsname{\color{black}}%
      \expandafter\def\csname LT0\endcsname{\color{black}}%
      \expandafter\def\csname LT1\endcsname{\color{black}}%
      \expandafter\def\csname LT2\endcsname{\color{black}}%
      \expandafter\def\csname LT3\endcsname{\color{black}}%
      \expandafter\def\csname LT4\endcsname{\color{black}}%
      \expandafter\def\csname LT5\endcsname{\color{black}}%
      \expandafter\def\csname LT6\endcsname{\color{black}}%
      \expandafter\def\csname LT7\endcsname{\color{black}}%
      \expandafter\def\csname LT8\endcsname{\color{black}}%
    \fi
  \fi
    \setlength{\unitlength}{0.0500bp}%
    \ifx\gptboxheight\undefined%
      \newlength{\gptboxheight}%
      \newlength{\gptboxwidth}%
      \newsavebox{\gptboxtext}%
    \fi%
    \setlength{\fboxrule}{0.5pt}%
    \setlength{\fboxsep}{1pt}%
\begin{picture}(7560.00,3225.60)%
    \gplgaddtomacro\gplbacktext{%
      \csname LTb\endcsname%%
      \put(518,628){\makebox(0,0)[r]{\strut{}$0.8$}}%
      \put(518,1291){\makebox(0,0)[r]{\strut{}$1.2$}}%
      \put(518,1954){\makebox(0,0)[r]{\strut{}$1.6$}}%
      \put(518,2618){\makebox(0,0)[r]{\strut{}$2$}}%
      \put(617,297){\makebox(0,0){\strut{}$0$}}%
      \put(1231,297){\makebox(0,0){\strut{}$15$}}%
      \put(1846,297){\makebox(0,0){\strut{}$30$}}%
      \put(2460,297){\makebox(0,0){\strut{}$45$}}%
      \put(3075,297){\makebox(0,0){\strut{}$60$}}%
      \put(3689,297){\makebox(0,0){\strut{}$75$}}%
      \put(116,3020){{\makebox(0,0){\strut{}{\large (a)}}}}%
    }%
    \gplgaddtomacro\gplfronttext{%
      \csname LTb\endcsname%%
      \put(116,1788){\rotatebox{-270}{\makebox(0,0){\strut{}$Nu$}}}%
      \put(2153,77){\makebox(0,0){\strut{}\large $t$}}%
      \csname LTb\endcsname%%
      \put(1364,2856){\makebox(0,0)[r]{\strut{}$N$=0}}%
      \csname LTb\endcsname%%
      \put(1364,2636){\makebox(0,0)[r]{\strut{}$N$=0.7}}%
      \csname LTb\endcsname%%
      \put(2318,2856){\makebox(0,0)[r]{\strut{}$N$=2.8}}%
      \csname LTb\endcsname%%
      \put(2318,2636){\makebox(0,0)[r]{\strut{}$N$=6.3}}%
      \csname LTb\endcsname%%
      \put(3272,2856){\makebox(0,0)[r]{\strut{}$N$=11.1}}%
      \csname LTb\endcsname%%
      \put(3272,2636){\makebox(0,0)[r]{\strut{}$N$=44.5}}%
    }%
    \gplgaddtomacro\gplbacktext{%
      \csname LTb\endcsname%%
      \put(4400,727){\makebox(0,0)[r]{\strut{}$1.12$}}%
      \put(4400,1258){\makebox(0,0)[r]{\strut{}$1.16$}}%
      \put(4400,1789){\makebox(0,0)[r]{\strut{}$1.2$}}%
      \put(4400,2319){\makebox(0,0)[r]{\strut{}$1.24$}}%
      \put(4400,2850){\makebox(0,0)[r]{\strut{}$1.28$}}%
      \put(4800,297){\makebox(0,0){\strut{}$0$}}%
      \put(5401,297){\makebox(0,0){\strut{}$0.1$}}%
      \put(6002,297){\makebox(0,0){\strut{}$0.2$}}%
      \put(6603,297){\makebox(0,0){\strut{}$0.3$}}%
      \put(7204,297){\makebox(0,0){\strut{}$0.4$}}%
      \put(3932,3020){{\makebox(0,0){\strut{}{\large (b)}}}}%
    }%
    \gplgaddtomacro\gplfronttext{%
      \csname LTb\endcsname%%
      \put(3932,1788){\rotatebox{-270}{\makebox(0,0){\strut{}$\overline{Nu}$}}}%
      \put(6001,77){\makebox(0,0){\strut{}$B/T$}}%
      \put(4800,1289){\strut{}\scalebox{0.8}{Bubble}}	
      \put(4650,1100){\strut{}\scalebox{0.8}{detachment}}	
      \put(4580,1000){\framebox(920,500){}}
      \put(6300,1289){\strut{}\scalebox{0.8}{Vapour}}	
      \put(6400,1100){\strut{}\scalebox{0.8}{sheet}}	
      \put(6180,1000){\framebox(750,500){}}
    }%
    \gplbacktext
    \put(0,0){\includegraphics{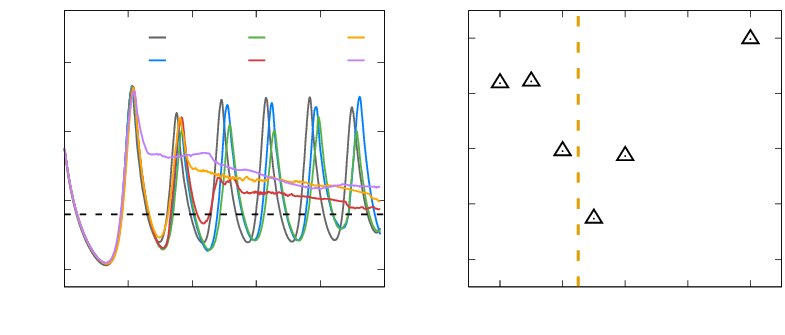}}%
    \gplfronttext
  \end{picture}%
\endgroup

%% file: figure/single-hor-lf-compare.tex
% GNUPLOT: LaTeX picture with Postscript
\begingroup
  \makeatletter
  \providecommand\color[2][]{%
    \GenericError{(gnuplot) \space\space\space\@spaces}{%
      Package color not loaded in conjunction with
      terminal option `colourtext'%
    }{See the gnuplot documentation for explanation.%
    }{Either use 'blacktext' in gnuplot or load the package
      color.sty in LaTeX.}%
    \renewcommand\color[2][]{}%
  }%
  \providecommand\includegraphics[2][]{%
    \GenericError{(gnuplot) \space\space\space\@spaces}{%
      Package graphicx or graphics not loaded%
    }{See the gnuplot documentation for explanation.%
    }{The gnuplot epslatex terminal needs graphicx.sty or graphics.sty.}%
    \renewcommand\includegraphics[2][]{}%
  }%
  \providecommand\rotatebox[2]{#2}%
  \@ifundefined{ifGPcolor}{%
    \newif\ifGPcolor
    \GPcolorfalse
  }{}%
  \@ifundefined{ifGPblacktext}{%
    \newif\ifGPblacktext
    \GPblacktexttrue
  }{}%
  % define a \g@addto@macro without @ in the name:
  \let\gplgaddtomacro\g@addto@macro
  % define empty templates for all commands taking text:
  \gdef\gplbacktext{}%
  \gdef\gplfronttext{}%
  \makeatother
  \ifGPblacktext
    % no textcolor at all
    \def\colorrgb#1{}%
    \def\colorgray#1{}%
  \else
    % gray or color?
    \ifGPcolor
      \def\colorrgb#1{\color[rgb]{#1}}%
      \def\colorgray#1{\color[gray]{#1}}%
      \expandafter\def\csname LTw\endcsname{\color{white}}%
      \expandafter\def\csname LTb\endcsname{\color{black}}%
      \expandafter\def\csname LTa\endcsname{\color{black}}%
      \expandafter\def\csname LT0\endcsname{\color[rgb]{1,0,0}}%
      \expandafter\def\csname LT1\endcsname{\color[rgb]{0,1,0}}%
      \expandafter\def\csname LT2\endcsname{\color[rgb]{0,0,1}}%
      \expandafter\def\csname LT3\endcsname{\color[rgb]{1,0,1}}%
      \expandafter\def\csname LT4\endcsname{\color[rgb]{0,1,1}}%
      \expandafter\def\csname LT5\endcsname{\color[rgb]{1,1,0}}%
      \expandafter\def\csname LT6\endcsname{\color[rgb]{0,0,0}}%
      \expandafter\def\csname LT7\endcsname{\color[rgb]{1,0.3,0}}%
      \expandafter\def\csname LT8\endcsname{\color[rgb]{0.5,0.5,0.5}}%
    \else
      % gray
      \def\colorrgb#1{\color{black}}%
      \def\colorgray#1{\color[gray]{#1}}%
      \expandafter\def\csname LTw\endcsname{\color{white}}%
      \expandafter\def\csname LTb\endcsname{\color{black}}%
      \expandafter\def\csname LTa\endcsname{\color{black}}%
      \expandafter\def\csname LT0\endcsname{\color{black}}%
      \expandafter\def\csname LT1\endcsname{\color{black}}%
      \expandafter\def\csname LT2\endcsname{\color{black}}%
      \expandafter\def\csname LT3\endcsname{\color{black}}%
      \expandafter\def\csname LT4\endcsname{\color{black}}%
      \expandafter\def\csname LT5\endcsname{\color{black}}%
      \expandafter\def\csname LT6\endcsname{\color{black}}%
      \expandafter\def\csname LT7\endcsname{\color{black}}%
      \expandafter\def\csname LT8\endcsname{\color{black}}%
    \fi
  \fi
    \setlength{\unitlength}{0.0500bp}%
    \ifx\gptboxheight\undefined%
      \newlength{\gptboxheight}%
      \newlength{\gptboxwidth}%
      \newsavebox{\gptboxtext}%
    \fi%
    \setlength{\fboxrule}{0.5pt}%
    \setlength{\fboxsep}{1pt}%
\begin{picture}(7560.00,3225.60)%
    \gplgaddtomacro\gplbacktext{%
      \csname LTb\endcsname%%
      \put(452,683){\makebox(0,0)[r]{\strut{}$-8$}}%
      \put(452,1125){\makebox(0,0)[r]{\strut{}$-6$}}%
      \put(452,1567){\makebox(0,0)[r]{\strut{}$-4$}}%
      \put(452,2010){\makebox(0,0)[r]{\strut{}$-2$}}%
      \put(452,2452){\makebox(0,0)[r]{\strut{}$0$}}%
      \put(452,2894){\makebox(0,0)[r]{\strut{}$2$}}%
      \put(999,297){\makebox(0,0){\strut{}$0.2$}}%
      \put(1896,297){\makebox(0,0){\strut{}$0.3$}}%
      \put(2792,297){\makebox(0,0){\strut{}$0.4$}}%
      \put(3689,297){\makebox(0,0){\strut{}$0.5$}}%
       \put(116,3020){{\makebox(0,0){\strut{}{\large (a)}}}}%
    }%
    \gplgaddtomacro\gplfronttext{%
      \csname LTb\endcsname%%
      \put(116,1788){\rotatebox{-270}{\makebox(0,0){\strut{}$F_{l,x}$}}}%
      \put(2120,77){\makebox(0,0){\strut{}$z/ \lambda_d $}}%
      \csname LTb\endcsname%%
      \put(3164,2942){\makebox(0,0)[r]{\strut{}$N=6.3$}}%
      \csname LTb\endcsname%%
      \put(3164,2722){\makebox(0,0)[r]{\strut{}$N=11.1$}}%
      \csname LTb\endcsname%%
      \put(3164,2502){\makebox(0,0)[r]{\strut{}$N=44.5$}}%
    }%
    \gplgaddtomacro\gplbacktext{%
      \csname LTb\endcsname%%
      \put(4268,683){\makebox(0,0)[r]{\strut{}$-8$}}%
      \put(4268,1125){\makebox(0,0)[r]{\strut{}$-6$}}%
      \put(4268,1567){\makebox(0,0)[r]{\strut{}$-4$}}%
      \put(4268,2010){\makebox(0,0)[r]{\strut{}$-2$}}%
      \put(4268,2452){\makebox(0,0)[r]{\strut{}$0$}}%
      \put(4268,2894){\makebox(0,0)[r]{\strut{}$2$}}%
      \put(4815,297){\makebox(0,0){\strut{}$0.42$}}%
      \put(5711,297){\makebox(0,0){\strut{}$0.3$}}%
      \put(6608,297){\makebox(0,0){\strut{}$0.4$}}%
      \put(7504,297){\makebox(0,0){\strut{}$0.5$}}%
      \put(3932,3020){{\makebox(0,0){\strut{}{\large (b)}}}}%
    }%
    \gplgaddtomacro\gplfronttext{%
      \csname LTb\endcsname%%
      \put(3932,1788){\rotatebox{-270}{\makebox(0,0){\strut{}$F_{l,x}$}}}%
      \put(5935,77){\makebox(0,0){\strut{}$z/ \lambda_d $}}%
      \csname LTb\endcsname%%
      \put(6979,2942){\makebox(0,0)[r]{\strut{}$N=6.3$}}%
      \csname LTb\endcsname%%
      \put(6979,2722){\makebox(0,0)[r]{\strut{}$N=11.1$}}%
      \csname LTb\endcsname%%
      \put(6979,2502){\makebox(0,0)[r]{\strut{}$N=44.5$}}%
    }%
    \gplbacktext
    \put(0,0){\includegraphics{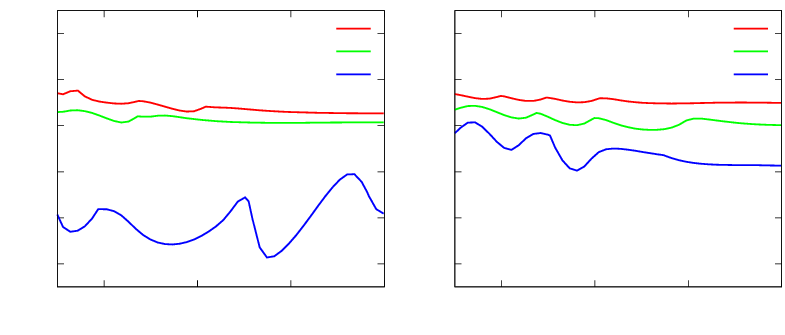}}%
    \gplfronttext
  \end{picture}%
\endgroup

%% file: figure/single-ver-remove-MF.tex
% GNUPLOT: LaTeX picture with Postscript
\begingroup
  \makeatletter
  \providecommand\color[2][]{%
    \GenericError{(gnuplot) \space\space\space\@spaces}{%
      Package color not loaded in conjunction with
      terminal option `colourtext'%
    }{See the gnuplot documentation for explanation.%
    }{Either use 'blacktext' in gnuplot or load the package
      color.sty in LaTeX.}%
    \renewcommand\color[2][]{}%
  }%
  \providecommand\includegraphics[2][]{%
    \GenericError{(gnuplot) \space\space\space\@spaces}{%
      Package graphicx or graphics not loaded%
    }{See the gnuplot documentation for explanation.%
    }{The gnuplot epslatex terminal needs graphicx.sty or graphics.sty.}%
    \renewcommand\includegraphics[2][]{}%
  }%
  \providecommand\rotatebox[2]{#2}%
  \@ifundefined{ifGPcolor}{%
    \newif\ifGPcolor
    \GPcolorfalse
  }{}%
  \@ifundefined{ifGPblacktext}{%
    \newif\ifGPblacktext
    \GPblacktexttrue
  }{}%
  % define a \g@addto@macro without @ in the name:
  \let\gplgaddtomacro\g@addto@macro
  % define empty templates for all commands taking text:
  \gdef\gplbacktext{}%
  \gdef\gplfronttext{}%
  \makeatother
  \ifGPblacktext
    % no textcolor at all
    \def\colorrgb#1{}%
    \def\colorgray#1{}%
  \else
    % gray or color?
    \ifGPcolor
      \def\colorrgb#1{\color[rgb]{#1}}%
      \def\colorgray#1{\color[gray]{#1}}%
      \expandafter\def\csname LTw\endcsname{\color{white}}%
      \expandafter\def\csname LTb\endcsname{\color{black}}%
      \expandafter\def\csname LTa\endcsname{\color{black}}%
      \expandafter\def\csname LT0\endcsname{\color[rgb]{1,0,0}}%
      \expandafter\def\csname LT1\endcsname{\color[rgb]{0,1,0}}%
      \expandafter\def\csname LT2\endcsname{\color[rgb]{0,0,1}}%
      \expandafter\def\csname LT3\endcsname{\color[rgb]{1,0,1}}%
      \expandafter\def\csname LT4\endcsname{\color[rgb]{0,1,1}}%
      \expandafter\def\csname LT5\endcsname{\color[rgb]{1,1,0}}%
      \expandafter\def\csname LT6\endcsname{\color[rgb]{0,0,0}}%
      \expandafter\def\csname LT7\endcsname{\color[rgb]{1,0.3,0}}%
      \expandafter\def\csname LT8\endcsname{\color[rgb]{0.5,0.5,0.5}}%
    \else
      % gray
      \def\colorrgb#1{\color{black}}%
      \def\colorgray#1{\color[gray]{#1}}%
      \expandafter\def\csname LTw\endcsname{\color{white}}%
      \expandafter\def\csname LTb\endcsname{\color{black}}%
      \expandafter\def\csname LTa\endcsname{\color{black}}%
      \expandafter\def\csname LT0\endcsname{\color{black}}%
      \expandafter\def\csname LT1\endcsname{\color{black}}%
      \expandafter\def\csname LT2\endcsname{\color{black}}%
      \expandafter\def\csname LT3\endcsname{\color{black}}%
      \expandafter\def\csname LT4\endcsname{\color{black}}%
      \expandafter\def\csname LT5\endcsname{\color{black}}%
      \expandafter\def\csname LT6\endcsname{\color{black}}%
      \expandafter\def\csname LT7\endcsname{\color{black}}%
      \expandafter\def\csname LT8\endcsname{\color{black}}%
    \fi
  \fi
    \setlength{\unitlength}{0.0500bp}%
    \ifx\gptboxheight\undefined%
      \newlength{\gptboxheight}%
      \newlength{\gptboxwidth}%
      \newsavebox{\gptboxtext}%
    \fi%
    \setlength{\fboxrule}{0.5pt}%
    \setlength{\fboxsep}{1pt}%
\begin{picture}(7560.00,3225.60)%
    \gplgaddtomacro\gplbacktext{%
      \csname LTb\endcsname%%
      \put(584,462){\makebox(0,0)[r]{\strut{}$1$}}%
      \put(584,841){\makebox(0,0)[r]{\strut{}$1.05$}}%
      \put(584,1220){\makebox(0,0)[r]{\strut{}$1.1$}}%
      \put(584,1599){\makebox(0,0)[r]{\strut{}$1.15$}}%
      \put(584,1978){\makebox(0,0)[r]{\strut{}$1.2$}}%
      \put(584,2357){\makebox(0,0)[r]{\strut{}$1.25$}}%
      \put(584,2736){\makebox(0,0)[r]{\strut{}$1.3$}}%
      \put(584,3115){\makebox(0,0)[r]{\strut{}$1.35$}}%
      \put(1134,297){\makebox(0,0){\strut{}$45$}}%
      \put(1885,297){\makebox(0,0){\strut{}$50$}}%
      \put(2637,297){\makebox(0,0){\strut{}$55$}}%
      \put(3388,297){\makebox(0,0){\strut{}$60$}}%
       \put(116,3115){{\makebox(0,0){\strut{}{\large (a)}}}}%
    }%
    \gplgaddtomacro\gplfronttext{%
      \csname LTb\endcsname%%
      \put(116,1788){\rotatebox{-270}{\makebox(0,0){\strut{}$\overline{Nu}$}}}%
      \put(2186,77){\makebox(0,0){\strut{}\large $t$}}%
      \csname LTb\endcsname%%
      \put(3164,2942){\makebox(0,0)[r]{\strut{}$N$=44.5}}%
      \csname LTb\endcsname%%
      \put(3164,2722){\makebox(0,0)[r]{\strut{}Remove MF}}%
    }%
    \gplgaddtomacro\gplbacktext{%
      \csname LTb\endcsname%%
      \put(4334,652){\makebox(0,0)[r]{\strut{}$2.6$}}%
      \put(4334,1031){\makebox(0,0)[r]{\strut{}$2.7$}}%
      \put(4334,1410){\makebox(0,0)[r]{\strut{}$2.8$}}%
      \put(4334,1789){\makebox(0,0)[r]{\strut{}$2.9$}}%
      \put(4334,2168){\makebox(0,0)[r]{\strut{}$3$}}%
      \put(4334,2547){\makebox(0,0)[r]{\strut{}$3.1$}}%
      \put(4334,2926){\makebox(0,0)[r]{\strut{}$3.2$}}%
      \put(4894,297){\makebox(0,0){\strut{}$45$}}%
      \put(5661,297){\makebox(0,0){\strut{}$50$}}%
      \put(6429,297){\makebox(0,0){\strut{}$55$}}%
      \put(7197,297){\makebox(0,0){\strut{}$60$}}%
   \put(3932,3115){{\makebox(0,0){\strut{}{\large (b)}}}}%
    }%
    \gplgaddtomacro\gplfronttext{%
      \csname LTb\endcsname%%
      \put(3932,1788){\rotatebox{-270}{\makebox(0,0){\strut{}$\dot{m}S_{\Gamma}$}}}%
      \put(5968,77){\makebox(0,0){\strut{}\large $t$}}%
      \csname LTb\endcsname%%
      \put(6979,2942){\makebox(0,0)[r]{\strut{}$N$=44.5}}%
      \csname LTb\endcsname%%
      \put(6979,2722){\makebox(0,0)[r]{\strut{}Remove MF}}%
    }%
    \gplbacktext
    \put(0,0){\includegraphics{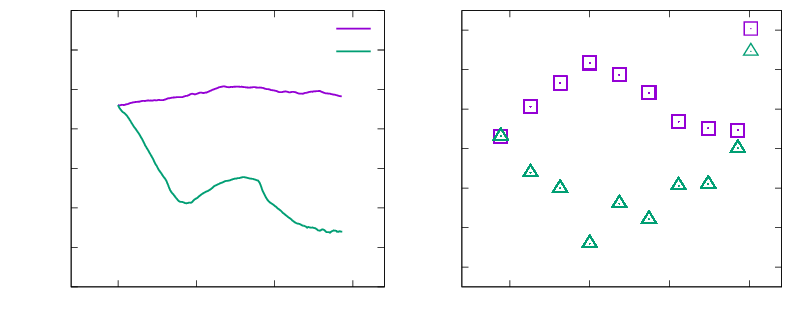}}%
    \gplfronttext
  \end{picture}%
\endgroup

%% file: figure/single-hor-remove-MF.tex
% GNUPLOT: LaTeX picture with Postscript
\begingroup
  \makeatletter
  \providecommand\color[2][]{%
    \GenericError{(gnuplot) \space\space\space\@spaces}{%
      Package color not loaded in conjunction with
      terminal option `colourtext'%
    }{See the gnuplot documentation for explanation.%
    }{Either use 'blacktext' in gnuplot or load the package
      color.sty in LaTeX.}%
    \renewcommand\color[2][]{}%
  }%
  \providecommand\includegraphics[2][]{%
    \GenericError{(gnuplot) \space\space\space\@spaces}{%
      Package graphicx or graphics not loaded%
    }{See the gnuplot documentation for explanation.%
    }{The gnuplot epslatex terminal needs graphicx.sty or graphics.sty.}%
    \renewcommand\includegraphics[2][]{}%
  }%
  \providecommand\rotatebox[2]{#2}%
  \@ifundefined{ifGPcolor}{%
    \newif\ifGPcolor
    \GPcolorfalse
  }{}%
  \@ifundefined{ifGPblacktext}{%
    \newif\ifGPblacktext
    \GPblacktexttrue
  }{}%
  % define a \g@addto@macro without @ in the name:
  \let\gplgaddtomacro\g@addto@macro
  % define empty templates for all commands taking text:
  \gdef\gplbacktext{}%
  \gdef\gplfronttext{}%
  \makeatother
  \ifGPblacktext
    % no textcolor at all
    \def\colorrgb#1{}%
    \def\colorgray#1{}%
  \else
    % gray or color?
    \ifGPcolor
      \def\colorrgb#1{\color[rgb]{#1}}%
      \def\colorgray#1{\color[gray]{#1}}%
      \expandafter\def\csname LTw\endcsname{\color{white}}%
      \expandafter\def\csname LTb\endcsname{\color{black}}%
      \expandafter\def\csname LTa\endcsname{\color{black}}%
      \expandafter\def\csname LT0\endcsname{\color[rgb]{1,0,0}}%
      \expandafter\def\csname LT1\endcsname{\color[rgb]{0,1,0}}%
      \expandafter\def\csname LT2\endcsname{\color[rgb]{0,0,1}}%
      \expandafter\def\csname LT3\endcsname{\color[rgb]{1,0,1}}%
      \expandafter\def\csname LT4\endcsname{\color[rgb]{0,1,1}}%
      \expandafter\def\csname LT5\endcsname{\color[rgb]{1,1,0}}%
      \expandafter\def\csname LT6\endcsname{\color[rgb]{0,0,0}}%
      \expandafter\def\csname LT7\endcsname{\color[rgb]{1,0.3,0}}%
      \expandafter\def\csname LT8\endcsname{\color[rgb]{0.5,0.5,0.5}}%
    \else
      % gray
      \def\colorrgb#1{\color{black}}%
      \def\colorgray#1{\color[gray]{#1}}%
      \expandafter\def\csname LTw\endcsname{\color{white}}%
      \expandafter\def\csname LTb\endcsname{\color{black}}%
      \expandafter\def\csname LTa\endcsname{\color{black}}%
      \expandafter\def\csname LT0\endcsname{\color{black}}%
      \expandafter\def\csname LT1\endcsname{\color{black}}%
      \expandafter\def\csname LT2\endcsname{\color{black}}%
      \expandafter\def\csname LT3\endcsname{\color{black}}%
      \expandafter\def\csname LT4\endcsname{\color{black}}%
      \expandafter\def\csname LT5\endcsname{\color{black}}%
      \expandafter\def\csname LT6\endcsname{\color{black}}%
      \expandafter\def\csname LT7\endcsname{\color{black}}%
      \expandafter\def\csname LT8\endcsname{\color{black}}%
    \fi
  \fi
    \setlength{\unitlength}{0.0500bp}%
    \ifx\gptboxheight\undefined%
      \newlength{\gptboxheight}%
      \newlength{\gptboxwidth}%
      \newsavebox{\gptboxtext}%
    \fi%
    \setlength{\fboxrule}{0.5pt}%
    \setlength{\fboxsep}{1pt}%
\begin{picture}(7560.00,3225.60)%
    \gplgaddtomacro\gplbacktext{%
      \csname LTb\endcsname%%
      \put(584,462){\makebox(0,0)[r]{\strut{}$1.2$}}%
      \put(584,993){\makebox(0,0)[r]{\strut{}$1.22$}}%
      \put(584,1523){\makebox(0,0)[r]{\strut{}$1.24$}}%
      \put(584,2054){\makebox(0,0)[r]{\strut{}$1.26$}}%
      \put(584,2584){\makebox(0,0)[r]{\strut{}$1.28$}}%
      \put(584,3115){\makebox(0,0)[r]{\strut{}$1.3$}}%
      \put(683,297){\makebox(0,0){\strut{}$55$}}%
      \put(1435,297){\makebox(0,0){\strut{}$60$}}%
      \put(2186,297){\makebox(0,0){\strut{}$65$}}%
      \put(2938,297){\makebox(0,0){\strut{}$70$}}%
      \put(3689,297){\makebox(0,0){\strut{}$75$}}%
       \put(116,3115){{\makebox(0,0){\strut{}{\large (a)}}}}%
    }%
    \gplgaddtomacro\gplfronttext{%
      \csname LTb\endcsname%%
      \put(116,1788){\rotatebox{-270}{\makebox(0,0){\strut{}$\overline{Nu}$}}}%
      \put(2186,77){\makebox(0,0){\strut{}\large $t$}}%
      \csname LTb\endcsname%%
      \put(3164,2942){\makebox(0,0)[r]{\strut{}$N$=44.5}}%
      \csname LTb\endcsname%%
      \put(3164,2722){\makebox(0,0)[r]{\strut{}Remove MF}}%
    }%
    \gplgaddtomacro\gplbacktext{%
      \csname LTb\endcsname%%
      \put(4400,462){\makebox(0,0)[r]{\strut{}$1.92$}}%
      \put(4400,944){\makebox(0,0)[r]{\strut{}$1.96$}}%
      \put(4400,1427){\makebox(0,0)[r]{\strut{}$2$}}%
      \put(4400,1909){\makebox(0,0)[r]{\strut{}$2.04$}}%
      \put(4400,2391){\makebox(0,0)[r]{\strut{}$2.08$}}%
      \put(4400,2874){\makebox(0,0)[r]{\strut{}$2.12$}}%
      \put(4499,297){\makebox(0,0){\strut{}$55$}}%
      \put(5250,297){\makebox(0,0){\strut{}$60$}}%
      \put(6002,297){\makebox(0,0){\strut{}$65$}}%
      \put(6753,297){\makebox(0,0){\strut{}$70$}}%
      \put(7504,297){\makebox(0,0){\strut{}$75$}}%
   \put(3932,3115){{\makebox(0,0){\strut{}{\large (b)}}}}%
    }%
    \gplgaddtomacro\gplfronttext{%
      \csname LTb\endcsname%%
      \put(3932,1788){\rotatebox{-270}{\makebox(0,0){\strut{}$\dot{m}S_{\Gamma}$}}}%
      \put(6001,77){\makebox(0,0){\strut{}\large $t$}}%
      \csname LTb\endcsname%%
      \put(6979,2942){\makebox(0,0)[r]{\strut{}$N$=44.5}}%
      \csname LTb\endcsname%%
      \put(6979,2722){\makebox(0,0)[r]{\strut{}Remove MF}}%
    }%
    \gplbacktext
    \put(0,0){\includegraphics{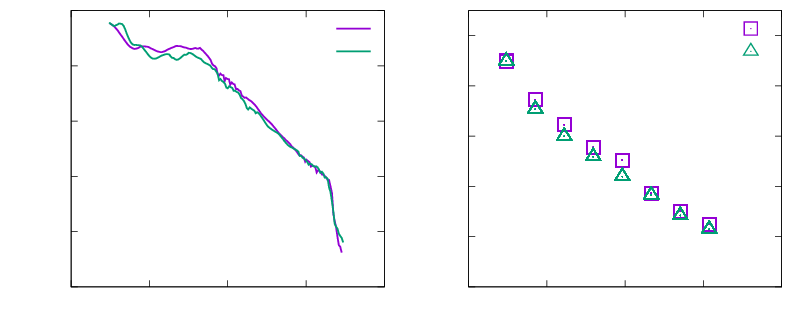}}%
    \gplfronttext
  \end{picture}%
\endgroup

%% file: figure/mut-ver-nu.tex
% GNUPLOT: LaTeX picture with Postscript
\begingroup
  \makeatletter
  \providecommand\color[2][]{%
    \GenericError{(gnuplot) \space\space\space\@spaces}{%
      Package color not loaded in conjunction with
      terminal option `colourtext'%
    }{See the gnuplot documentation for explanation.%
    }{Either use 'blacktext' in gnuplot or load the package
      color.sty in LaTeX.}%
    \renewcommand\color[2][]{}%
  }%
  \providecommand\includegraphics[2][]{%
    \GenericError{(gnuplot) \space\space\space\@spaces}{%
      Package graphicx or graphics not loaded%
    }{See the gnuplot documentation for explanation.%
    }{The gnuplot epslatex terminal needs graphicx.sty or graphics.sty.}%
    \renewcommand\includegraphics[2][]{}%
  }%
  \providecommand\rotatebox[2]{#2}%
  \@ifundefined{ifGPcolor}{%
    \newif\ifGPcolor
    \GPcolorfalse
  }{}%
  \@ifundefined{ifGPblacktext}{%
    \newif\ifGPblacktext
    \GPblacktexttrue
  }{}%
  % define a \g@addto@macro without @ in the name:
  \let\gplgaddtomacro\g@addto@macro
  % define empty templates for all commands taking text:
  \gdef\gplbacktext{}%
  \gdef\gplfronttext{}%
  \makeatother
  \ifGPblacktext
    % no textcolor at all
    \def\colorrgb#1{}%
    \def\colorgray#1{}%
  \else
    % gray or color?
    \ifGPcolor
      \def\colorrgb#1{\color[rgb]{#1}}%
      \def\colorgray#1{\color[gray]{#1}}%
      \expandafter\def\csname LTw\endcsname{\color{white}}%
      \expandafter\def\csname LTb\endcsname{\color{black}}%
      \expandafter\def\csname LTa\endcsname{\color{black}}%
      \expandafter\def\csname LT0\endcsname{\color[rgb]{1,0,0}}%
      \expandafter\def\csname LT1\endcsname{\color[rgb]{0,1,0}}%
      \expandafter\def\csname LT2\endcsname{\color[rgb]{0,0,1}}%
      \expandafter\def\csname LT3\endcsname{\color[rgb]{1,0,1}}%
      \expandafter\def\csname LT4\endcsname{\color[rgb]{0,1,1}}%
      \expandafter\def\csname LT5\endcsname{\color[rgb]{1,1,0}}%
      \expandafter\def\csname LT6\endcsname{\color[rgb]{0,0,0}}%
      \expandafter\def\csname LT7\endcsname{\color[rgb]{1,0.3,0}}%
      \expandafter\def\csname LT8\endcsname{\color[rgb]{0.5,0.5,0.5}}%
    \else
      % gray
      \def\colorrgb#1{\color{black}}%
      \def\colorgray#1{\color[gray]{#1}}%
      \expandafter\def\csname LTw\endcsname{\color{white}}%
      \expandafter\def\csname LTb\endcsname{\color{black}}%
      \expandafter\def\csname LTa\endcsname{\color{black}}%
      \expandafter\def\csname LT0\endcsname{\color{black}}%
      \expandafter\def\csname LT1\endcsname{\color{black}}%
      \expandafter\def\csname LT2\endcsname{\color{black}}%
      \expandafter\def\csname LT3\endcsname{\color{black}}%
      \expandafter\def\csname LT4\endcsname{\color{black}}%
      \expandafter\def\csname LT5\endcsname{\color{black}}%
      \expandafter\def\csname LT6\endcsname{\color{black}}%
      \expandafter\def\csname LT7\endcsname{\color{black}}%
      \expandafter\def\csname LT8\endcsname{\color{black}}%
    \fi
  \fi
    \setlength{\unitlength}{0.0500bp}%
    \ifx\gptboxheight\undefined%
      \newlength{\gptboxheight}%
      \newlength{\gptboxwidth}%
      \newsavebox{\gptboxtext}%
    \fi%
    \setlength{\fboxrule}{0.5pt}%
    \setlength{\fboxsep}{1pt}%
\begin{picture}(9216.00,3225.60)%
    \gplgaddtomacro\gplbacktext{%
      \csname LTb\endcsname%%
      \put(1551,462){\makebox(0,0)[r]{\strut{}$0.6$}}%
      \put(1551,841){\makebox(0,0)[r]{\strut{}$0.8$}}%
      \put(1551,1220){\makebox(0,0)[r]{\strut{}$1$}}%
      \put(1551,1599){\makebox(0,0)[r]{\strut{}$1.2$}}%
      \put(1551,1978){\makebox(0,0)[r]{\strut{}$1.4$}}%
      \put(1551,2357){\makebox(0,0)[r]{\strut{}$1.6$}}%
      \put(1551,2736){\makebox(0,0)[r]{\strut{}$1.8$}}%
      \put(1551,3115){\makebox(0,0)[r]{\strut{}$2$}}%
      \put(1650,297){\makebox(0,0){\strut{}$0$}}%
      \put(2537,297){\makebox(0,0){\strut{}$15$}}%
      \put(3424,297){\makebox(0,0){\strut{}$30$}}%
      \put(4311,297){\makebox(0,0){\strut{}$45$}}%
      \put(5198,297){\makebox(0,0){\strut{}$60$}}%
      \put(6086,297){\makebox(0,0){\strut{}$75$}}%
      \put(6973,297){\makebox(0,0){\strut{}$90$}}%
    }%
    \gplgaddtomacro\gplfronttext{%
      \csname LTb\endcsname%%
      \put(1149,1788){\rotatebox{-270}{\makebox(0,0){\strut{}$Nu$}}}%
      \put(4607,77){\makebox(0,0){\strut{}\large $t$}}%
      \csname LTb\endcsname%%
      \put(8689,3005){\makebox(0,0)[r]{\strut{}$N$=0}}%
      \csname LTb\endcsname%%
      \put(8689,2785){\makebox(0,0)[r]{\strut{}$N$=2.8}}%
      \csname LTb\endcsname%%
      \put(8689,2565){\makebox(0,0)[r]{\strut{}$N$=11.1}}%
      \csname LTb\endcsname%%
      \put(8689,2345){\makebox(0,0)[r]{\strut{}$N$=44.5}}%
      \csname LTb\endcsname%%
      \put(8689,2125){\makebox(0,0)[r]{\strut{}$N$=100.1}}%
      \csname LTb\endcsname%%
      \put(8689,1905){\makebox(0,0)[r]{\strut{}Klimenko}}%
    }%
    \gplbacktext
    \put(0,0){\includegraphics{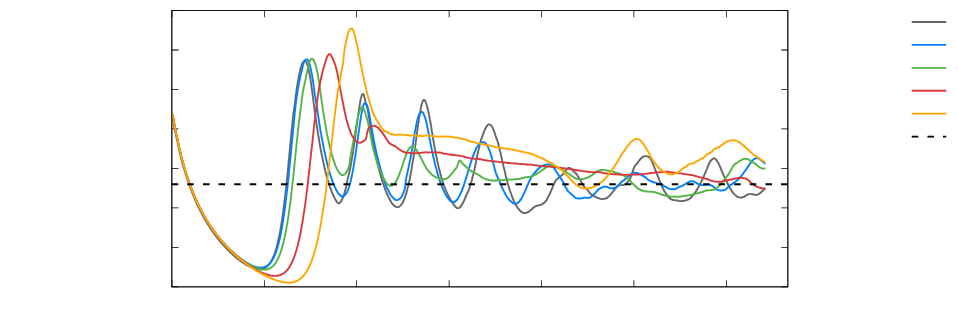}}%
    \gplfronttext
  \end{picture}%
\endgroup

%% file: figure/mut-hor-nu.tex
% GNUPLOT: LaTeX picture with Postscript
\begingroup
  \makeatletter
  \providecommand\color[2][]{%
    \GenericError{(gnuplot) \space\space\space\@spaces}{%
      Package color not loaded in conjunction with
      terminal option `colourtext'%
    }{See the gnuplot documentation for explanation.%
    }{Either use 'blacktext' in gnuplot or load the package
      color.sty in LaTeX.}%
    \renewcommand\color[2][]{}%
  }%
  \providecommand\includegraphics[2][]{%
    \GenericError{(gnuplot) \space\space\space\@spaces}{%
      Package graphicx or graphics not loaded%
    }{See the gnuplot documentation for explanation.%
    }{The gnuplot epslatex terminal needs graphicx.sty or graphics.sty.}%
    \renewcommand\includegraphics[2][]{}%
  }%
  \providecommand\rotatebox[2]{#2}%
  \@ifundefined{ifGPcolor}{%
    \newif\ifGPcolor
    \GPcolorfalse
  }{}%
  \@ifundefined{ifGPblacktext}{%
    \newif\ifGPblacktext
    \GPblacktexttrue
  }{}%
  % define a \g@addto@macro without @ in the name:
  \let\gplgaddtomacro\g@addto@macro
  % define empty templates for all commands taking text:
  \gdef\gplbacktext{}%
  \gdef\gplfronttext{}%
  \makeatother
  \ifGPblacktext
    % no textcolor at all
    \def\colorrgb#1{}%
    \def\colorgray#1{}%
  \else
    % gray or color?
    \ifGPcolor
      \def\colorrgb#1{\color[rgb]{#1}}%
      \def\colorgray#1{\color[gray]{#1}}%
      \expandafter\def\csname LTw\endcsname{\color{white}}%
      \expandafter\def\csname LTb\endcsname{\color{black}}%
      \expandafter\def\csname LTa\endcsname{\color{black}}%
      \expandafter\def\csname LT0\endcsname{\color[rgb]{1,0,0}}%
      \expandafter\def\csname LT1\endcsname{\color[rgb]{0,1,0}}%
      \expandafter\def\csname LT2\endcsname{\color[rgb]{0,0,1}}%
      \expandafter\def\csname LT3\endcsname{\color[rgb]{1,0,1}}%
      \expandafter\def\csname LT4\endcsname{\color[rgb]{0,1,1}}%
      \expandafter\def\csname LT5\endcsname{\color[rgb]{1,1,0}}%
      \expandafter\def\csname LT6\endcsname{\color[rgb]{0,0,0}}%
      \expandafter\def\csname LT7\endcsname{\color[rgb]{1,0.3,0}}%
      \expandafter\def\csname LT8\endcsname{\color[rgb]{0.5,0.5,0.5}}%
    \else
      % gray
      \def\colorrgb#1{\color{black}}%
      \def\colorgray#1{\color[gray]{#1}}%
      \expandafter\def\csname LTw\endcsname{\color{white}}%
      \expandafter\def\csname LTb\endcsname{\color{black}}%
      \expandafter\def\csname LTa\endcsname{\color{black}}%
      \expandafter\def\csname LT0\endcsname{\color{black}}%
      \expandafter\def\csname LT1\endcsname{\color{black}}%
      \expandafter\def\csname LT2\endcsname{\color{black}}%
      \expandafter\def\csname LT3\endcsname{\color{black}}%
      \expandafter\def\csname LT4\endcsname{\color{black}}%
      \expandafter\def\csname LT5\endcsname{\color{black}}%
      \expandafter\def\csname LT6\endcsname{\color{black}}%
      \expandafter\def\csname LT7\endcsname{\color{black}}%
      \expandafter\def\csname LT8\endcsname{\color{black}}%
    \fi
  \fi
    \setlength{\unitlength}{0.0500bp}%
    \ifx\gptboxheight\undefined%
      \newlength{\gptboxheight}%
      \newlength{\gptboxwidth}%
      \newsavebox{\gptboxtext}%
    \fi%
    \setlength{\fboxrule}{0.5pt}%
    \setlength{\fboxsep}{1pt}%
\begin{picture}(9216.00,3225.60)%
    \gplgaddtomacro\gplbacktext{%
      \csname LTb\endcsname%%
      \put(1551,462){\makebox(0,0)[r]{\strut{}$0.6$}}%
      \put(1551,841){\makebox(0,0)[r]{\strut{}$0.8$}}%
      \put(1551,1220){\makebox(0,0)[r]{\strut{}$1$}}%
      \put(1551,1599){\makebox(0,0)[r]{\strut{}$1.2$}}%
      \put(1551,1978){\makebox(0,0)[r]{\strut{}$1.4$}}%
      \put(1551,2357){\makebox(0,0)[r]{\strut{}$1.6$}}%
      \put(1551,2736){\makebox(0,0)[r]{\strut{}$1.8$}}%
      \put(1551,3115){\makebox(0,0)[r]{\strut{}$2$}}%
      \put(1650,297){\makebox(0,0){\strut{}$0$}}%
      \put(2537,297){\makebox(0,0){\strut{}$15$}}%
      \put(3424,297){\makebox(0,0){\strut{}$30$}}%
      \put(4311,297){\makebox(0,0){\strut{}$45$}}%
      \put(5198,297){\makebox(0,0){\strut{}$60$}}%
      \put(6086,297){\makebox(0,0){\strut{}$75$}}%
      \put(6973,297){\makebox(0,0){\strut{}$90$}}%
    }%
    \gplgaddtomacro\gplfronttext{%
      \csname LTb\endcsname%%
      \put(1149,1788){\rotatebox{-270}{\makebox(0,0){\strut{}$Nu$}}}%
      \put(4607,77){\makebox(0,0){\strut{}\large $t$}}%
      \csname LTb\endcsname%%
      \put(8689,3005){\makebox(0,0)[r]{\strut{}$N$=0}}%
      \csname LTb\endcsname%%
      \put(8689,2785){\makebox(0,0)[r]{\strut{}$N$=2.8}}%
      \csname LTb\endcsname%%
      \put(8689,2565){\makebox(0,0)[r]{\strut{}$N$=11.1}}%
      \csname LTb\endcsname%%
      \put(8689,2345){\makebox(0,0)[r]{\strut{}$N$=44.5}}%
      \csname LTb\endcsname%%
      \put(8689,2125){\makebox(0,0)[r]{\strut{}$N$=100.1}}%
      \csname LTb\endcsname%%
      \put(8689,1905){\makebox(0,0)[r]{\strut{}Klimenko}}%
    }%
    \gplbacktext
    \put(0,0){\includegraphics{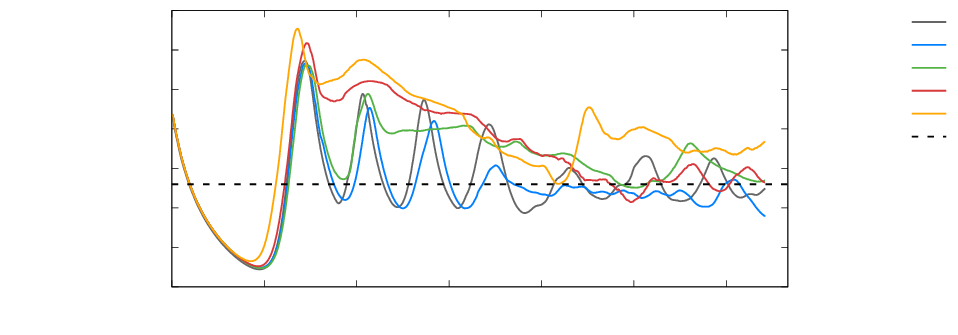}}%
    \gplfronttext
  \end{picture}%
\endgroup

%% file: figure/test-31.tex
% GNUPLOT: LaTeX picture with Postscript
\begingroup
  \makeatletter
  \providecommand\color[2][]{%
    \GenericError{(gnuplot) \space\space\space\@spaces}{%
      Package color not loaded in conjunction with
      terminal option `colourtext'%
    }{See the gnuplot documentation for explanation.%
    }{Either use 'blacktext' in gnuplot or load the package
      color.sty in LaTeX.}%
    \renewcommand\color[2][]{}%
  }%
  \providecommand\includegraphics[2][]{%
    \GenericError{(gnuplot) \space\space\space\@spaces}{%
      Package graphicx or graphics not loaded%
    }{See the gnuplot documentation for explanation.%
    }{The gnuplot epslatex terminal needs graphicx.sty or graphics.sty.}%
    \renewcommand\includegraphics[2][]{}%
  }%
  \providecommand\rotatebox[2]{#2}%
  \@ifundefined{ifGPcolor}{%
    \newif\ifGPcolor
    \GPcolorfalse
  }{}%
  \@ifundefined{ifGPblacktext}{%
    \newif\ifGPblacktext
    \GPblacktexttrue
  }{}%
  % define a \g@addto@macro without @ in the name:
  \let\gplgaddtomacro\g@addto@macro
  % define empty templates for all commands taking text:
  \gdef\gplbacktext{}%
  \gdef\gplfronttext{}%
  \makeatother
  \ifGPblacktext
    % no textcolor at all
    \def\colorrgb#1{}%
    \def\colorgray#1{}%
  \else
    % gray or color?
    \ifGPcolor
      \def\colorrgb#1{\color[rgb]{#1}}%
      \def\colorgray#1{\color[gray]{#1}}%
      \expandafter\def\csname LTw\endcsname{\color{white}}%
      \expandafter\def\csname LTb\endcsname{\color{black}}%
      \expandafter\def\csname LTa\endcsname{\color{black}}%
      \expandafter\def\csname LT0\endcsname{\color[rgb]{1,0,0}}%
      \expandafter\def\csname LT1\endcsname{\color[rgb]{0,1,0}}%
      \expandafter\def\csname LT2\endcsname{\color[rgb]{0,0,1}}%
      \expandafter\def\csname LT3\endcsname{\color[rgb]{1,0,1}}%
      \expandafter\def\csname LT4\endcsname{\color[rgb]{0,1,1}}%
      \expandafter\def\csname LT5\endcsname{\color[rgb]{1,1,0}}%
      \expandafter\def\csname LT6\endcsname{\color[rgb]{0,0,0}}%
      \expandafter\def\csname LT7\endcsname{\color[rgb]{1,0.3,0}}%
      \expandafter\def\csname LT8\endcsname{\color[rgb]{0.5,0.5,0.5}}%
    \else
      % gray
      \def\colorrgb#1{\color{black}}%
      \def\colorgray#1{\color[gray]{#1}}%
      \expandafter\def\csname LTw\endcsname{\color{white}}%
      \expandafter\def\csname LTb\endcsname{\color{black}}%
      \expandafter\def\csname LTa\endcsname{\color{black}}%
      \expandafter\def\csname LT0\endcsname{\color{black}}%
      \expandafter\def\csname LT1\endcsname{\color{black}}%
      \expandafter\def\csname LT2\endcsname{\color{black}}%
      \expandafter\def\csname LT3\endcsname{\color{black}}%
      \expandafter\def\csname LT4\endcsname{\color{black}}%
      \expandafter\def\csname LT5\endcsname{\color{black}}%
      \expandafter\def\csname LT6\endcsname{\color{black}}%
      \expandafter\def\csname LT7\endcsname{\color{black}}%
      \expandafter\def\csname LT8\endcsname{\color{black}}%
    \fi
  \fi
    \setlength{\unitlength}{0.0500bp}%
    \ifx\gptboxheight\undefined%
      \newlength{\gptboxheight}%
      \newlength{\gptboxwidth}%
      \newsavebox{\gptboxtext}%
    \fi%
    \setlength{\fboxrule}{0.5pt}%
    \setlength{\fboxsep}{1pt}%
\begin{picture}(9936.00,4536.00)%
    \gplgaddtomacro\gplbacktext{%
      \csname LTb\endcsname%%
      \put(386,462){\makebox(0,0)[r]{\strut{}$0$}}%
      \put(386,1783){\makebox(0,0)[r]{\strut{}$1$}}%
      \put(386,3104){\makebox(0,0)[r]{\strut{}$2$}}%
      \put(386,4425){\makebox(0,0)[r]{\strut{}$3$}}%
      \put(485,297){\makebox(0,0){\strut{}$-0.5$}}%
      \put(1079,297){\makebox(0,0){\strut{}$0$}}%
      \put(1673,297){\makebox(0,0){\strut{}$0.5$}}%
    }%
    \gplgaddtomacro\gplfronttext{%
      \csname LTb\endcsname%%
      \put(116,2443){\rotatebox{-270}{\makebox(0,0){\strut{}\large $x/ \lambda_d $}}}%
      \put(1079,77){\makebox(0,0){\strut{}\large $y/ \lambda_d $}}%
      \csname LTb\endcsname%%
      \put(1214,3655){\makebox(0,0)[r]{\strut{}$N 0$}}%
      \csname LTb\endcsname%%
      \put(1214,3435){\makebox(0,0)[r]{\strut{} $N 100.1$}}%
       \put(1100,4200){\makebox(0,0)[r]{\strut{} $a-1$}}%
    }%
    \gplgaddtomacro\gplbacktext{%
      \csname LTb\endcsname%%

    }%
    \gplgaddtomacro\gplfronttext{%
      \csname LTb\endcsname%%
       \put(2600,4200){\makebox(0,0)[r]{\strut{} $a-2$}}%
      \put(1628,2443){\rotatebox{-270}{\makebox(0,0){\strut{}}}}%
      \put(2591,77){\makebox(0,0){\strut{}\large $y/ \lambda_d $}}%
    }%
    \gplgaddtomacro\gplbacktext{%
      \csname LTb\endcsname%%

    }%
    \gplgaddtomacro\gplfronttext{%
      \csname LTb\endcsname%%
      \put(3140,2443){\rotatebox{-270}{\makebox(0,0){\strut{}}}}%
      \put(4103,77){\makebox(0,0){\strut{}\large $y/ \lambda_d $}}%
      \csname LTb\endcsname%%
      \put(4238,3655){\makebox(0,0)[r]{\strut{}$N0$}}%
      \csname LTb\endcsname%%
      \put(4238,3435){\makebox(0,0)[r]{\strut{}$N100.1$}}%
      \put(4100,4200){\makebox(0,0)[r]{\strut{} $b-1$}}%
    }%
    \gplgaddtomacro\gplbacktext{%
      \csname LTb\endcsname%%

    }%
    \gplgaddtomacro\gplfronttext{%
      \csname LTb\endcsname%%
      \put(4652,2443){\rotatebox{-270}{\makebox(0,0){\strut{}}}}%
      \put(5615,77){\makebox(0,0){\strut{}\large $y/ \lambda_d $}}%
      \put(5600,4200){\makebox(0,0)[r]{\strut{} $b-2$}}%
    }%
    \gplgaddtomacro\gplbacktext{%
      \csname LTb\endcsname%%

    }%
    \gplgaddtomacro\gplfronttext{%
      \csname LTb\endcsname%%
      \put(6164,2443){\rotatebox{-270}{\makebox(0,0){\strut{}}}}%
      \put(7126,77){\makebox(0,0){\strut{}\large $y/ \lambda_d $}}%
      \put(7100,4200){\makebox(0,0)[r]{\strut{} $c-1$}}%
      \csname LTb\endcsname%%
      \put(7261,3655){\makebox(0,0)[r]{\strut{}$N0$}}%
      \csname LTb\endcsname%%
      \put(7261,3435){\makebox(0,0)[r]{\strut{}$N100.1$}}%
    }%
    \gplgaddtomacro\gplbacktext{%
      \csname LTb\endcsname%%

    }%
    \gplgaddtomacro\gplfronttext{%
      \csname LTb\endcsname%%
      \put(7676,2443){\rotatebox{-270}{\makebox(0,0){\strut{}}}}%
      \put(8638,77){\makebox(0,0){\strut{}\large $y/ \lambda_d $}}%
      \put(8600,4200){\makebox(0,0)[r]{\strut{} $c-2$}}%
    }%
    \gplbacktext
    \put(0,0){\includegraphics{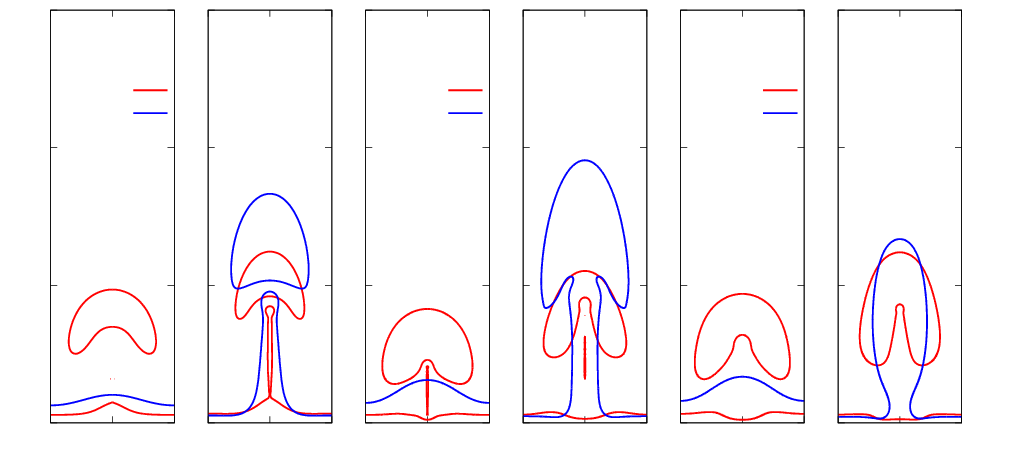}}%
    \gplfronttext
  \end{picture}%
\endgroup